\documentclass[superscriptaddress, onecolumn]{revtex4-2}

\usepackage{graphicx}
\usepackage{geometry}
\usepackage{multirow}
\usepackage{lipsum} 
\usepackage{tcolorbox}
\usepackage{bm}
\usepackage{xr}
\usepackage{float}
\usepackage{amsmath}
\usepackage{amssymb}
\usepackage{amsthm}
\usepackage{amsfonts}
\usepackage{booktabs}
\usepackage{xcolor}
\usepackage{soul}
\usepackage{cleveref}
\usepackage{svg}
\usepackage[normalem]{ulem}
\usepackage{booktabs}
\usepackage{xspace}
\usepackage{algorithm}
\usepackage{algpseudocode}
\usepackage{listings}

\lstdefinelanguage{json}{
    basicstyle=\ttfamily\small,
    numbers=left,
    numberstyle=\tiny,
    stepnumber=1,
    numbersep=5pt,
    showstringspaces=false,
    breaklines=true,
    backgroundcolor=\color{white},
    keywordstyle=\color{blue},
    stringstyle=\color{black},
    morestring=[b]",
    literate=
     *{0}{{{\color{blue}0}}}{1}
      {1}{{{\color{blue}1}}}{1}
      {2}{{{\color{blue}2}}}{1}
      {3}{{{\color{blue}3}}}{1}
      {4}{{{\color{blue}4}}}{1}
      {5}{{{\color{blue}5}}}{1}
      {6}{{{\color{blue}6}}}{1}
      {7}{{{\color{blue}7}}}{1}
      {8}{{{\color{blue}8}}}{1}
      {9}{{{\color{blue}9}}}{1}
}

\begin{document}

\title{Multifaceted polarisation and information reliability in climate change discussions on social media platforms}

\author{Aleix Bassolas}
\affiliation{Eurecat, Centre Tecnològic de Catalunya, Barcelona, Spain}
\author{Joan Massachs}
\affiliation{Departament de F\'isica de la Mat\`eria Condensada and Institute of Complex Systems (UBICS), Universitat de Barcelona, Barcelona, Spain}
\affiliation{Eurecat, Centre Tecnològic de Catalunya, Barcelona, Spain}
\author{Emanuele Cozzo}
\affiliation{Departament de F\'isica de la Mat\`eria Condensada and Institute of Complex Systems (UBICS), Universitat de Barcelona, Barcelona, Spain}
\affiliation{CNSC-IN3 Universitat Oberta de Catalunya, Barcelona, Spain}
\author{Julian Vicens}
\affiliation{Eurecat, Centre Tecnològic de Catalunya, Barcelona, Spain}

\begin{abstract}
Social media platforms like YouTube and Twitter play a key role in disseminating both reliable and unreliable information about climate change. This study analyses the topology of interactions in Twitter and their relation to cross-platform sharing, content discussions and emotional responses. We examined climate change discussions across four topics: the 27th United Nations Climate Change Conference, the Sixth Assessment Report of the United Nations Intergovernmental Panel on Climate Change, climate refugees, and Doñana Natural Park. While retweets reinforce in-group cohesion in the form of echo chambers, inter-group exposure is significant through mentions, suggesting that exposure to opposing views intensifies polarisation, rather than mitigates it. Ideological divides feature content differences accompanied by steeper negative sentiments, especially from right-leaning communities prone to share low-reliability information. We identified a topological alignment between platforms, indicating that ideological communities span multiple sites. Our findings show that climate change polarisation is multifaceted, involving ideological divides, structural isolation, and emotional engagement. These results suggest that effective climate policy discussions must address the emotional and identity-driven nature of public discourse and seek strategies to bridge ideological divides.
\end{abstract}

\maketitle
Polarisation has become a defining feature of contemporary political and social life. While a certain level of conflict and disagreement is essential to the functioning of democracies, an excessively polarised social environment can detrimentally affect their self-governing capacity \cite{McCoy2018}, particularly when it pertains to issues, like climate change, that require broad consensus and coordinated action. In this paper, we examine polarisation in climate change discourse on Twitter (currently X) and YouTube, using a multidimensional approach. To capture different social and geographical contexts, we analyse four datasets related to milestone events: the 2022 UN Climate Change Conference (COP27), the release of the 6th Assessment Report by the Intergovernmental Panel on Climate Change (IPCC), the emergent issue of climate refugees, and a politically charged local debate in Spain over environmental regulation in Doñana.

The term polarisation is used across disciplines to describe the division of a population into two well-separated and opposing groups. It can be understood both as a state – the extent to which opinions or characteristics are opposed at a given time – and as a process – the movement toward greater extremes over time \cite{dimaggio1996have}. Furthermore, polarisation has a specific meaning related to the distribution of opinions, beliefs, attitudes, or other measurable attributes in a population (such as wealth or income) \cite{esteban1994measurement}. Operationalisations of this concept emphasise the formation of distinct groups in the distribution, with high homogeneity within the group and heterogeneity between groups. The core notion being that polarisation is high when a distribution is multi-modal and maximal when it is bimodal (i.e., when there are multiple distinct 'peaks' corresponding to social groups) rather than a continuum \cite{bramson2017understanding}.

Esteban and May \cite{esteban1994measurement} frames polarisation in terms of 'identification' and 'alienation'. Polarisation is conceptualised as the result of people strongly identifying with others in their own group while feeling alienation (distance or hostility) toward those in other groups. A formal model is proposed, where society is more polarised when it contains relatively cohesive subgroups that are far apart from each other. High identification within groups means that each of the groups of individuals has internal similarity or solidarity, and high alienation means there is strong effective distance or antagonism between those groups. This idea leads to well-defined mathematical measures in which polarisation is essentially the weighted sum of pairwise 'antagonisms' between individuals, increasing when people cluster into separated groups. 

The shift of political views towards extreme positions and the formation of separate cohesive groups often is related to the presence of echo chambers, where interactions predominantly occur among individuals who share similar beliefs and ideas \cite{zollo2015emotional,del2016echo,du2017echo}. As interest in measuring \cite{del2016echo,quattrociocchi2016echo,cinelli2021echo,di2021infodemics,grusauskaite2023debating} and modelling  \cite{baumann2020modeling} the dynamics of polarisation and echo chambers grows, efforts to identify the factors influencing the interaction patterns face significant challenges. The recommendation algorithms employed by these platforms influence these dynamics \cite{cho2020search,santos2021link,de2022modelling} by limiting exposure to opposing viewpoints in users' feeds \cite{levy2021social}. Yet the rise of polarisation is not only an effect of homophily as it has been shown that individuals are often exposed to opposed opinions \cite{barbera2020social}.

While polarisation is a natural aspect in pluralistic societies, its intensification over the last decades is a matter of debate both among academics and in public discourse \cite{brady2006polarization,dimaggio1996have,iyengar2019origins,bruggemann2023debates}. Multiple studies point to the fact that the widespread adoption of social media has driven an increase in polarisation as observed across experimental settings \cite{bail2018exposure} and various platforms \cite{quattrociocchi2016echo,levy2021social,iandoli2021impact}. However, its presence and magnitude vary across subjects, can be triggered by external events \cite{Garimella_2017, Waller2021,esquirol2024analyzing,meyer2023between} and is steeper around political discussions than in more general subjects \cite{barbera2015tweeting}. Platforms also play a role, while polarisation has been observed on Twitter, YouTube, or Instagram \cite{bessi2016users,fernandes2020political,wu2021cross,hohmann2023quantifying}, in others, such as WhatsApp, there is a depolarisation effect \cite{yarchi2021political}. Even though the interest in measuring social and political polarisation dates from before the digital era \cite{Krackhardt1988}, the widespread use of online social media and data accessibility sparked the development of frameworks to quantify it. The techniques developed vary depending on the perspective, while the topology of interactions provides insights into the preferences to form social bonds based on political views \citep{salloum2022separating}, textual analysis can capture the divergence in emotions and policies \cite{Fiorina2008}

The combination of a polarised environment \cite{zollo2017debunking} and behavioural factors, such as confirmation bias \cite{del2016spreading,soon2018fake}, has been closely associated with the dissemination of low-reliability information \cite{vicario2019polarization}, influencing the outcome of presidential elections \cite{bovet2019influence} and increasing vaccine hesitancy \cite{Loomba2021}. Despite these challenges, there have been concerted efforts to combat the spread of fake news \cite{van2017inoculating}. Nonetheless, debates persist regarding the distinction between reliable and unreliable news cascades. Although initial studies highlighted structural differences \cite{vosoughi2018spread}, subsequent analyses have suggested that disparities may be more attributable to differences in cascade size \cite{juul2021comparing}.

Polarisation manifests across political issues, but also other kinds of events and subjects as we have witnessed an increase in climate change discussions and related events \cite{falkenberg2022growing, torricelli2023does,bruggemann2023debates,meyer2023between}. Previous research has focused on the origins of this polarisation \cite{fisher2013does}, its relation to political affiliations \cite{Chen_2021}, and the influential roles played by media outlets \cite{Chinn_2020} and corporations \cite{farrell2016corporate}. Recently, an increasing polarisation has been reported in locations affected by extreme weather events or environmental disputes \cite{torricelli2023does}. The sentiment towards climate change mitigation measures varies significantly across regions, shaped by local political and cultural contexts \cite{rivera2024evolution}. While the spread of low-reliability information in these contexts has been noted, its connection with polarised environments remains unclear \cite{lutzke2019priming,treen2020online,sanford2021controversy}. Our study addresses this gap by studying a highly localised and politically charged debate over environmental protection in Spain, Doñana, where the interplay between misinformation, partisanity, and institutional trust is especially salient.

The broad concept of polarisation can then be further specified, with particular relevance to our work, by focusing on the notions of political and affective polarisation. Political polarisation usually refers to the distance between political preferences, identities, or parties within a society \cite{dimaggio1996have}. In relation to this, the analysis of a retweet network relies on the empirical and theoretically supported assumption that the way users share information is related to and a source of their collective political identity  \cite{barandiaran2020defining,van2022polarized}. In the same line, information sharing is related to the practice of networked gatekeeping that put in relation elite and mass polarisation \cite{meraz2013networked} and lead to the formation of affective publics \cite{papacharissi2015affective}. Affective polarisation is also emphasised in political science and is distinct from policy disagreement. It describes the phenomenon of partisans feeling not just different, but actively antagonistic toward each other. \cite{iyengar2019origins,Cota2019}. In other words, it’s the polarisation of feelings or attitudes toward the social groups representing the other, rather than toward policy issues themselves. This is well captured through expressions of antagonism between groups, such as emotional tone in communication. In relation with the structural polarisation metrics we apply in this work, it is central to the notion of how users use retweet and mention as different ways of signalling in- and out-group relations and identities \cite{falkenberg2024patterns}.

The initial metrics to quantify the macroscopic level of structural polarisation primarily concentrate on modularity and the analysis of community boundaries \cite{conover2011political,guerra2013measure,salloum2022separating}. However, those only provide a partial perspective on the organisation of social connections, as they largely disregard the multi-scalar organisation of networks and the role of individual nodes. Metrics designed to capture long-range structural properties include label propagation \cite{morales2015measuring}, centrality measures, and random walk diffusion \cite{Garimella2018}. Even though most assume a dipole organisation, recent developments can capture multipole polarisation \cite{martin2023multipolar}. Structural measures in dealing with the networked organisation of social relations \cite{erikson2018relationalism} provide a measure of polarisation being content-agnostic. Natural language processing techniques are essential for assessing the content of the divergence on policy-related issues \cite{yarchi2021political,falkenberg2022growing} and evaluating the emotions of individuals and their interactions \cite{falkenberg2024patterns}. While our approach focuses on structural and affective dynamics at a group level, recent studies have aimed to integrate these with individual-level opinions \cite{hohmann2023quantifying} providing further context into the structural organisation of different ideological perspectives.

Building on previous literature that tends to isolate structural, ideological, or affective polarisation, this study offers an integrated approach to capture their intersection in climate change discourse. Drawing on Twitter and YouTube data, we examine how community structure, emotional tone, and media reliability interact across platforms and contexts. To do so, we combine network analysis (to assess structural patterns), ideological measures (using political bias and source reliability), and sentiment metrics (focused on toxicity in quote tweets and cross-group mentions). This multidimensional approach allows us to observe how different dimensions of polarisation reinforce one another. Users with similar ideological leanings tend to cluster in retweet networks and express aligned affective reactions, particularly in politically salient debates. Interactions across communities, while rare in the form of retweets, are more frequent through mentions and often carry antagonistic sentiment—suggesting emotional confrontation despite low structural connectivity.

Our main contribution is the development of an integrative framework that captures structural, ideological, and affective polarisation across digital platforms. Applying this framework to four climate-related events, we examine how polarisation manifests differently depending on the topic and geographical context, ranging from global summits to locally contentious environmental debates. This multidimensional analysis reveals consistent links between the structure of online communities, the emotional tone of user interactions, and the reliability of the content being shared. Notably, we observe that polarisation is not confined to a single platform: ideologically aligned groups on Twitter also engage with similar content on YouTube, indicating a broader cross-platform coherence. These findings underscore the importance of moving beyond single-platform or single-dimension studies in order to understand the full complexity of polarisation in digital climate discourse.

\section{Data Description}

We extracted four datasets related to climate change topics from Twitter through the Twitter Academic API. These datasets correspond to the 2022 United Nations Climate Change Conference (COP27), the release of the 6th Assessment Report by the Intergovernmental Panel on Climate Change (IPCC), the climate refugee crisis resulting from extreme weather conditions (Climate Refugees), and to legislation affecting the natural reserve of Do\~{n}ana in the south of Spain (Do\~{n}ana). Tweets were collected using Twarc \cite{Summers2023} by searching for keywords related to each discussion. The keywords selected were closely related to each event. For COP27 the keywords included \textit{COP27} and related concepts such as \textit{LossAndDamage}. For IPCC the keywords were \textit{IPCC} and its official account (\textit{@IPCC\_CH}). For climate refugees the keywords were related combinations included \textit{climate refugee(s)} and \textit{climate migration}. For Do\~{n}ana only the keyword \textit{Doñana} was used. The datasets for specific events (COP27 and IPCC) were filtered within the span of the pre and post-event dates. The Climate Refugees and Doñana datasets span a longer period compared to the IPCC and COP27 datasets. The IPCC dataset, in particular, covers only nine days, coinciding with the peak discussion surrounding the release of the 6th Assessment Report (See Table \ref{summary_table_twitter}). We extracted the corresponding YouTube dataset by analysing the URLs referenced in each Twitter dataset and downloading the posts and comments using the YouTube Data API. To ensure relevance, we filtered the posts by searching for occurrences of specific keywords in each video title or description (Section S1).

\begin{table*}
\centering
\begin{tabular}{lccccccc}
& \textbf{Min. date} & \textbf{Max. date} & \textbf{Users} & \textbf{Tweets} & \textbf{Retweets} & \textbf{Replies} & \textbf{Quotes} \\
\textbf{COP27} & 2022-09-01 & 2022-11-27 & 1,351,903& 866,753& 4,977,874& 205,973& 175,078\\
\textbf{IPCC} & 2023-03-18 & 2023-03-26 & 157,056& 31,138& 267,971& 45,863& 7,751\\
\textbf{C. Refugees} & 2008-03-10 & 2022-12-31 & 841,454& 384,267& 1,376,057& 139,699& 38,909\\
\textbf{Do\~{n}ana}  & 2019-01-01 & 2023-04-30 & 290,782& 139,478& 1,187,646& 135,245& 25,056\\\end{tabular}
\caption{Summary table of the Twitter datasets analysed. Minimum and maximum dates, observed number of users, and number of tweets by typology.}\label{summary_table_twitter} 
\end{table*}

\begin{table*} 
\centering 
\begin{tabular}{lcccccc} 
 & \textbf{Videos} & \textbf{Min. Date} & \textbf{Max. Date} & \textbf{Avg. views} & \textbf{Avg. likes} & \textbf{Avg. comments} \\
 \textbf{COP27} & 624 & 2008-02-11 & 2023-05-12 & 571,964.35& 5,860.32& 306.99 \\ 
 \textbf{IPCC} & 145 & 2007-11-03 & 2023-06-09 & 350,178.66& 8,991.67& 1,455.16\\ 
 \textbf{C. Refugees} & 215 & 2009-11-22 & 2023-11-06 & 243,720.27& 3,604.81& 527.40 \\ 
 \textbf{Do\~{n}ana} & 191 & 2009-01-11 & 2023-06-01 & 101,098.32& 3,622.68& 246.25 \\ \end{tabular} 
 \caption{Summary table of the YouTube datasets analysed. Number of videos in each community, minimum and maximum dates observed, and average number of views, likes, and comments.}\label{summary_table_youtube} 
 \end{table*}

Unlike the Twitter dataset, the period of the comments is longer due to users frequently referencing older videos. Each dataset exhibits peaks at different times based on its unique characteristics. In particular, we observe a strong alignment in the IPCC and COP27 datasets with the event dates (Table \ref{summary_table_youtube}). Additional insights, including the data collection procedure can be found in Section S1 of the Supplementary Material.

\section{Results}
\subsection{Network analysis and structural polarisation}
\begin{figure}[!htbp] 
\begin{center}
\includegraphics[width=0.9\textwidth]{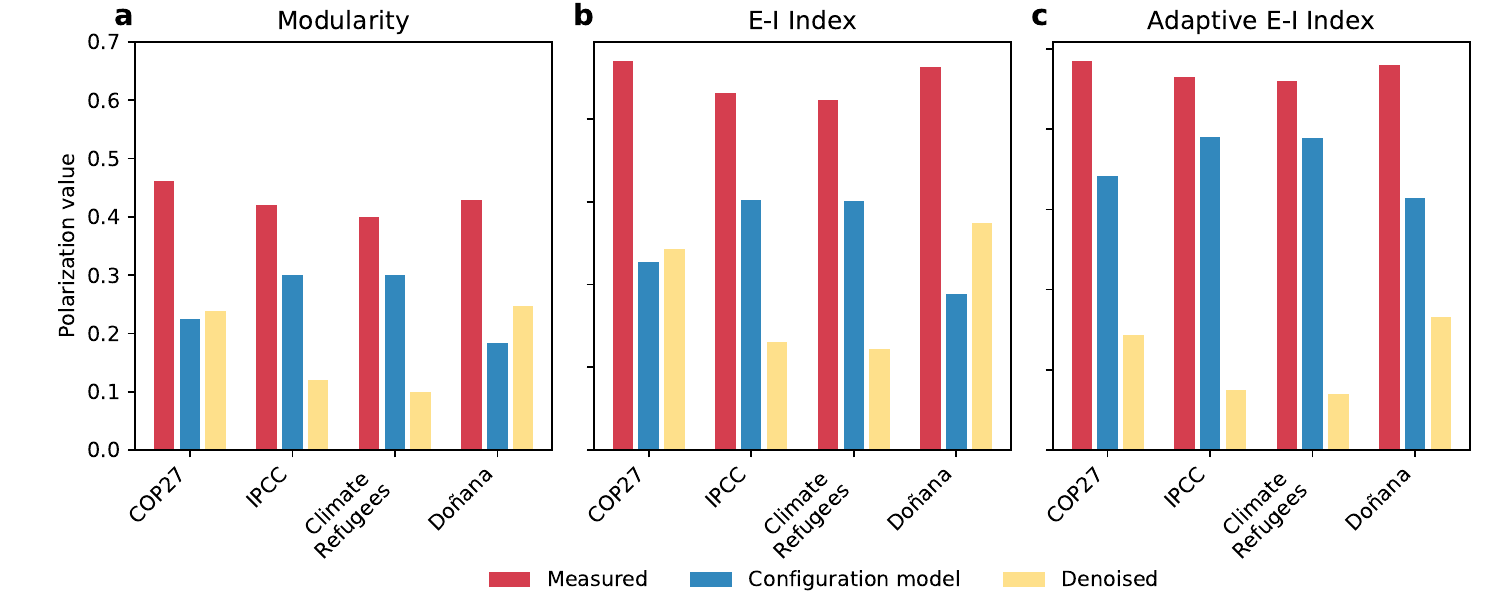} 
\end{center} 
\caption{\textbf{Polarization in the Twitter dataset.} Quantification of the polarization in Twitter: \textbf{(a)} modularity, \textbf{(b)} E-I index, and \textbf{(c)} adaptive E-I index. We provide the measures for the observed networks in red, across 100 realisations of the configuration model with preserved degree sequences in blue, and the difference between them (denoised value) in yellow.} \label{global_polarization} 
\end{figure}

We evaluate the structural polarisation on Twitter by considering the unweighted network of retweets with undirected links indicating a retweet interaction between two users. Previous studies have used retweets as a proxy for the influence between individuals allowing for the computation of their latent ideology \cite{morales2015measuring,falkenberg2022growing}. Similarly, we aim to measure how the influence between individuals shapes network organisation. We applied the Metis partition algorithm \cite{karypis1998fast} to split the graph into two clusters, and we calculated the modularity \cite{newman2006modularity}, the E-I index \cite{krackhardt1988informal} and the adaptive E-I index \cite{chen2021polarization} that adjusts for the cluster sizes. These metrics evaluate the topological interactions between users, the mesoscale structure and the extent to which users in different groups interact with one another without considering the content. The results for the four datasets analysed are presented in Fig. \ref{global_polarization}. Following \cite{salloum2022separating}, we computed polarisation metrics in the observed network $\Phi(G)$, as well as in the ensemble of configuration models with equivalent degree distribution $\Phi(G_{CM})$ \cite{fosdick2018configuring}, and calculated the denoised value $\hat{\Phi}(G)=\Phi(G)-\Phi(G_{CM})$. For completeness, we report the standardised values \cite{salloum2022separating} in Fig. S3.

All networks exhibit structural polarisation according to the denoised value, suggesting the presence of two groups of strongly connected individuals. In particular, the COP27 and Do\~{n}ana datasets displayed larger denoised values, suggesting a stronger polarisation around the topics. The higher structural polarisation observed in the COP27 dataset may be attributed to transnational spheres using different languages that do not interact with each other \cite{neff2024transnational}. The first and second largest communities, in terms of the number of messages, are dominated by English. The third and fourth communities primarily consist of Portuguese messages, followed by others where French and Spanish are more prevalent. Further details on this distribution per dataset and community can be found in Section S2.2. The Do\~{n}ana dataset captures the discussion around a new law related to natural park preservation, which involves political parties and users with strong political affiliations like politicians, journalists, or activists. The lower polarisation observed in the IPCC dataset may be due to its nature as a scientific report release rather than a mainstream  international conference like COP, potentially attracting a more scientifically educated audience and inducing a lower level of polarisation. Although the largest community (labelled as 0) is composed mainly of scientists, primarily from the UK, other communities consist of activists (community 3) or politicians (community 6), such as @GretaThunberg and @BarackObama, respectively. 

\begin{figure}[!htbp] 
\begin{center} 
\includegraphics[width=0.9\textwidth]{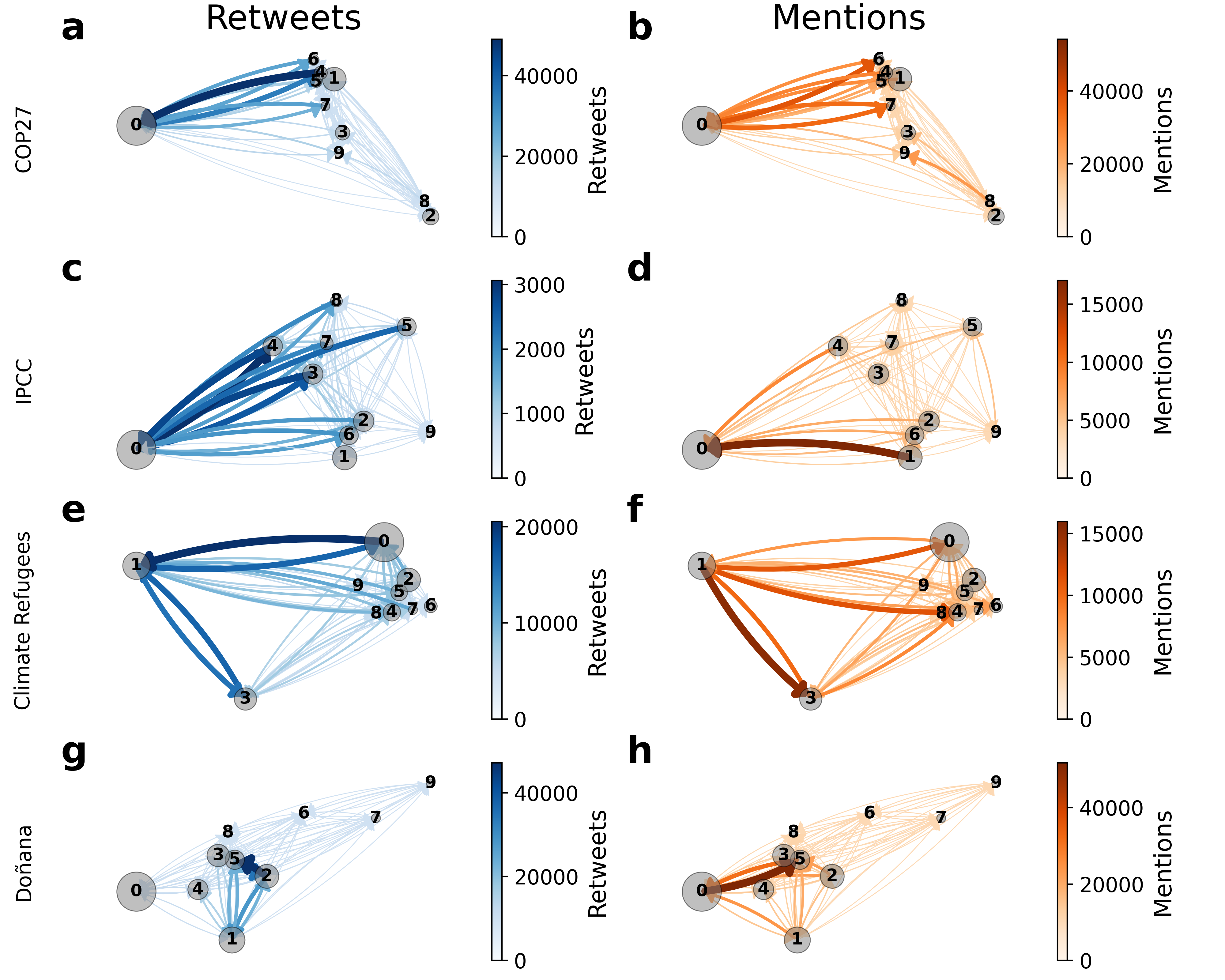} \end{center} 
\caption{\textbf{Community interaction network.} Interaction between the ten largest communities in \textbf{(a, b)} COP27, \textbf{(c, d)} IPCC, \textbf{(e, f)} Climate Refugees, \textbf{(g, h)} Do\~{n}ana. We report the interactions via \textbf{(a, c, e, g)} retweets and \textbf{(b, d, f, h)} mentions. Arrows indicate the direction of the interaction, the colour intensity and width, the volume. The size of the dots represents community size. In blue, we show the interaction through retweets and, in orange, through mentions.} \label{community_network} 
\end{figure}

While the retweet network has been previously interpreted as the influence between social media users, other types of interactions, such as mentions, often feature different structural properties \cite{conover2011political}. We examine the network of retweets and mentions among the $10$ largest communities of the networks in Fig. \ref{community_network}. We identified these communities using modularity optimisation on the directed weighted graph of retweets only \cite{newman2006modularity}, and then we incorporate mentions. This approach allows us to identify clusters based on influence and later analyse how they interact through mentions and retweets. The interaction patterns among the largest community (labelled as 0) and the remaining communities vary across datasets and types of interactions. In the COP27 and IPCC networks, the smaller communities mostly retweet content from the largest community, which is more isolated and engages in fewer interactions with the rest of the network. The largest communities include official accounts and major figures, such as the Secretary-General of the United Nations (@antonioguterres) or the Intergovernmental Panel on Climate Change (@IPCC\_CH), representing COP27 and the IPCC, respectively. Conversely, the Do\~{n}ana network exhibits higher flow heterogeneity between communities, with community 5 playing a central role in attracting most of the flow. The detailed overview of the users in the community reveals the presence of the president of Spain (@sanchezcastejon), the ruling political party (@PSOE), and sympathiser journalists (@eldiarioes, @iescolar). There are significant structural differences depending on the type of interaction. For example, in the IPCC network, there are relatively few retweets from community 1, with accounts concentrated flow such as @uksciencechiefUK (Government Office for Science) and @BernieSpofforth (the anti-climate movement figure) to community 0, formed by well-know climate science accounts such as @ed\_hawkins or @MrMatthewTodd. However, there is a high volume of mentions between these communities (see details in Section S2.1).

\begin{figure}[!htbp] 
\begin{center} 
\includegraphics[width=0.9\textwidth]{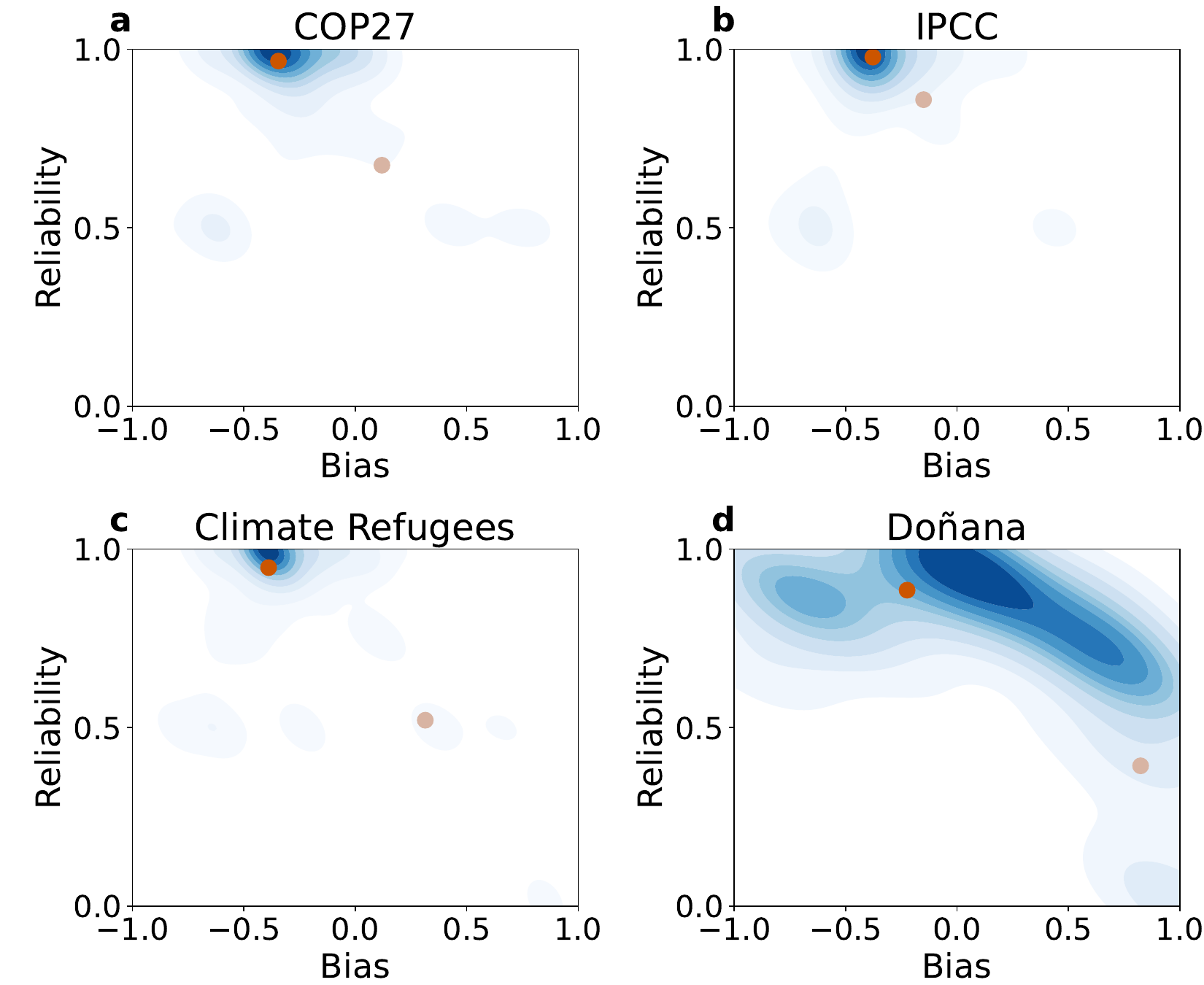} \end{center} \caption{\textbf{Reliability as a function of political bias in the studied networks.}  
Density plot showing the distribution of reliability as a function of bias for the communities detected in \textbf{(a)} COP27, \textbf{(b)} IPCC, \textbf{(c)} Climate Refugees, and \textbf{(d)} Doñana. For each dataset, the two communities with the largest and most divergent political views are identified. In dark brown the left-biased community and in light brown the right-biased community}
\label{biasreliability} 
\end{figure}

\subsubsection{Isolation and entropy}

We further evaluate the connectivity patterns between communities by quantifying the structural properties of their inter- and intra-group interactions. To quantify the isolation of communities, we calculated the directed interactions between them, and we computed the normalised outflow difference given by
\begin{equation} 
\Delta F ^{out}_i=\frac{\sum_{j, j\neq i}T_{ij}-T_{ii}}{\sum_{\forall j} T_{ij}}, 
\end{equation} 
where $T_{ij}$ is the number of interactions from community $i$ to community $j$. $\Delta F ^{out}_i$ provides a measure of a community's isolation. A value of $-1$ indicates that a community has only internal flow, whereas a value close to $1$ suggests that most of the flow is directed toward other communities. A value of 0 signifies that half of the flow is external to the community, while the other half remains internal.

To understand the diversity of interactions among communities, we compute the entropy of flows as
\begin{equation} 
H^{out}_i=-\sum_{\forall j} p_{ij} log p_{ij}, 
\end{equation} 

where $p_{ij}$ is the probability of having an interaction from $i$ to $j$ given by $T_{ij}/\sum_{\forall j} T_{ij}$. While $\Delta F ^{out}_i$ quantifies the ratio between external and internal links and goes from $-1$ (when all flows are internal) to $1$ (when all flows are external), the entropy ($H^{out}_i$) provides information on the variety of those flows. The entropy is $0$ when there is no variety of flows and $1$ when it is maximum.
Most of the largest communities exhibit a certain degree of isolation, indicated by negative values. In the COP27 network, the most isolated communities are 2 and 3, both Brazilian ($\Delta F ^{out}_2=-0.97$ and $\Delta F ^{out}_3=-0.94$), followed by community 1 ($\Delta F ^{out}_1=-0.88$), which is the most isolated and has the lowest entropy ($H^{out}_1=0.05$) among English-speaking groups. This community consists of sceptics with influential accounts such as @JamesMelville and @BernieSpofforth. Following the same pattern, in the IPCC network, community 1 has the highest isolation value ($\Delta F ^{out}_1=-0.98$) and the lowest entropy ($H^{out}_1=0.014$). This community also includes, among other accounts, @BernieSpofforth. Similarly, in the Climate Refugees network, community 6 is the most isolated ($\Delta F ^{out}_6=-0.93$) and has the lowest entropy ($H^{out}_6=0.03$), featuring accounts such as @MaximeBernier and @PrisonPlanet. In the Spanish Doñana network, community 0 is the most isolated ($\Delta F ^{out}_0=-0.95$) and has the lowest entropy ($H^{out}_0=0.02$). Across all datasets, communities with right-wing alignment demonstrate strong isolation and low entropy, implying that their messages have limited spread across the network.

However, the pattern changes when considering the mention graph. The largest mainstream communities show far less mentions of other communities, for instance, community 0 on IPCC and COP27 networks has high isolation ($\Delta F ^{out}_0=-0.5$ and $\Delta F ^{out}_0=-0.57$, respectively). Whereas the low-reliability communities display lower isolation and more external mentions, its the case of community 1 in COP27 ($\Delta F ^{out}_1=0.37$ and $H^{out}_1 = 0.25$). These results reinforce, first, the heterogeneity in group connectivity depending on the interaction types and, second, different levels of isolation by communities.

\subsection{Political polarisation}

\subsubsection{Bias and reliability}
We assess if the structural polarisation is related to the divergence in political views and information diet by calculating the bias and reliability of the URLs shared by the users. Denser connections between users with similar ideological biases could indicate a closer influence between them and a greater disagreement on climate change discussions. Information on the political bias and reliability of news outlets was obtained from MediaBias fact-check \cite{mediabias}, where positive and negative political bias correspond to right-wing and left-wing media, respectively. Outlets with higher reliability are considered more trustworthy and tend to share more verified news. For the Do\~{n}ana dataset, the reliability and bias were based on a separate dataset \cite{politicalwatch}, given that most media sources are Spanish. To facilitate the analysis, we have normalised the bias within the range [-1, 1] and the reliability within the range [0, 1]. The minimum and maximum biases are set according to the higher absolute value observed. In Fig. \ref{biasreliability}\textbf{(a)-(d)}, we show the density plot of community reliability as a function of their political bias according to the shared URLs in the IPCC, COP27, Climate Refugees, and Do\~{n}ana networks. We computed user-level values by averaging over the sources they shared and then calculated community-level values by averaging over the users. All datasets exhibit a left-wing bias, as most communities show negative values, especially in the IPCC and Climate Refugee networks. Possible explanations include a systemic bias in media coverage of these topics, the tendency of climate change sceptics to share information from unconventional sources that are not explicitly flagged by media bias indicators, or an over-representation of left-wing users in climate change discussions, as they may be more likely to engage due to greater awareness of the issue. The left-wing bias is particularly pronounced in the IPCC and Climate Refugee networks, where the cluster of large communities is centred around negative values. However, the bias is more dispersed in the Do\~{n}ana dataset, possibly due to differences in the shared media sources or a wider ideological variety. 

The communities with right-wing bias we have identified are: community 1 in COP27 (with key actors as @JamesMelville) and IPCC (with the presence of @BernieSpofforth), community 6 in Climate Refugees (with @MaximeBernier or @PrisonPlanet), and community 0 in Doñana (with @JuanMa\_Moreno or @alfonso\_ussia). We performed significance tests to compare the bias and reliability of communities, revealing several significant differences. In terms of bias, the most notable differences (p-value < 0.001) are observed in the previously mentioned right-wing biased communities compared to the other communities. Regarding reliability, community 4 in COP27 (with main actors such as  @CarolineLucas), community 3 in IPCC (with @GretaThunberg or  @CarolineLucas), community 5 in Climate Refugees (@jeremycorbyn), and community 1 in Doñana (@WWFespana and other conservation organisations) show the most significant higher reliability relative to the other communities (see Figs. S6-S7).

Based on the previous findings and manual inspection of the communities, we identified the two largest communities with opposing political leanings. In the IPCC and COP27 datasets, community 0 represents the largest community aligned with the mainstream climate change narrative, while community 1 is the primary advocate of a counter-narrative message. However, in the Do\~{n}ana network, the labels are swapped, as the right-wing biased community has a larger user base. For the Climate Refugees network, most of the top 10 communities exhibit similar bias-reliability values, except for community 6, which stands out with low reliability and a right-wing bias. In this dataset, we designate community 1 as the primary left-wing biased community, as it includes well-known climate change activists such as @MikeHudema or @GretaThunberg and engages in more interactions with the low-reliability community. Our results point out that most communities in climate change discussions share reliable information and have a left-biased alignment. However, there is at least one large community in each dataset on the right-side alignment that shares less reliable information.

These results indicate differences in the network topology depending on the type of interactions. Retweets, which are linked to influence, connect individuals in communities with similar biases. Instead, a higher number of mentions across communities suggests interactions between individuals from opposing sides of the ideological spectrum.

Given that many networks include users from different countries speaking various languages, we also measured polarisation by focusing on the two largest communities that speak the same language but have opposing biases, comparing the retweet and mention networks. The results still indicate polarisation in the retweet network, with modularity values ranging from $0.07$ in the IPCC network to $0.21$ in the Doñana network. Even in the mention networks—where structural scores are lower (modularity is $0.05$ in COP27, $0.16$ in Climate Refugees, and $0.15$ in Doñana)—only the IPCC network shows no polarisation (modularity = $-0.23$). These findings further support the observed disparity in network organisation based on interaction typology (see Section S2.3 for more details).

\subsubsection{Echo chambers}

Throughout this section, we assess the presence of echo chambers with a focus on the two largest communities exhibiting opposing political biases. Our objective is to assess the interaction patterns between users as a function of their bias. We employed an approach developed in \cite{cinelli2021echo} to evaluate the existence of echo chambers. For a given user $i$ with degree $k_i$, we computed its average bias $x_i$ and reliability $y_i$, along with the average bias and reliability of its neighbourhood, denoted as $\frac{1}{k_i}\sum_{\forall j\neq i}A_{ij}x_j$ and $\frac{1}{k_i}\sum_{\forall j\neq i}A_{ij}y_j$, respectively.

\begin{figure}[!htbp]
 \begin{center}
  \includegraphics[width=0.9\textwidth]{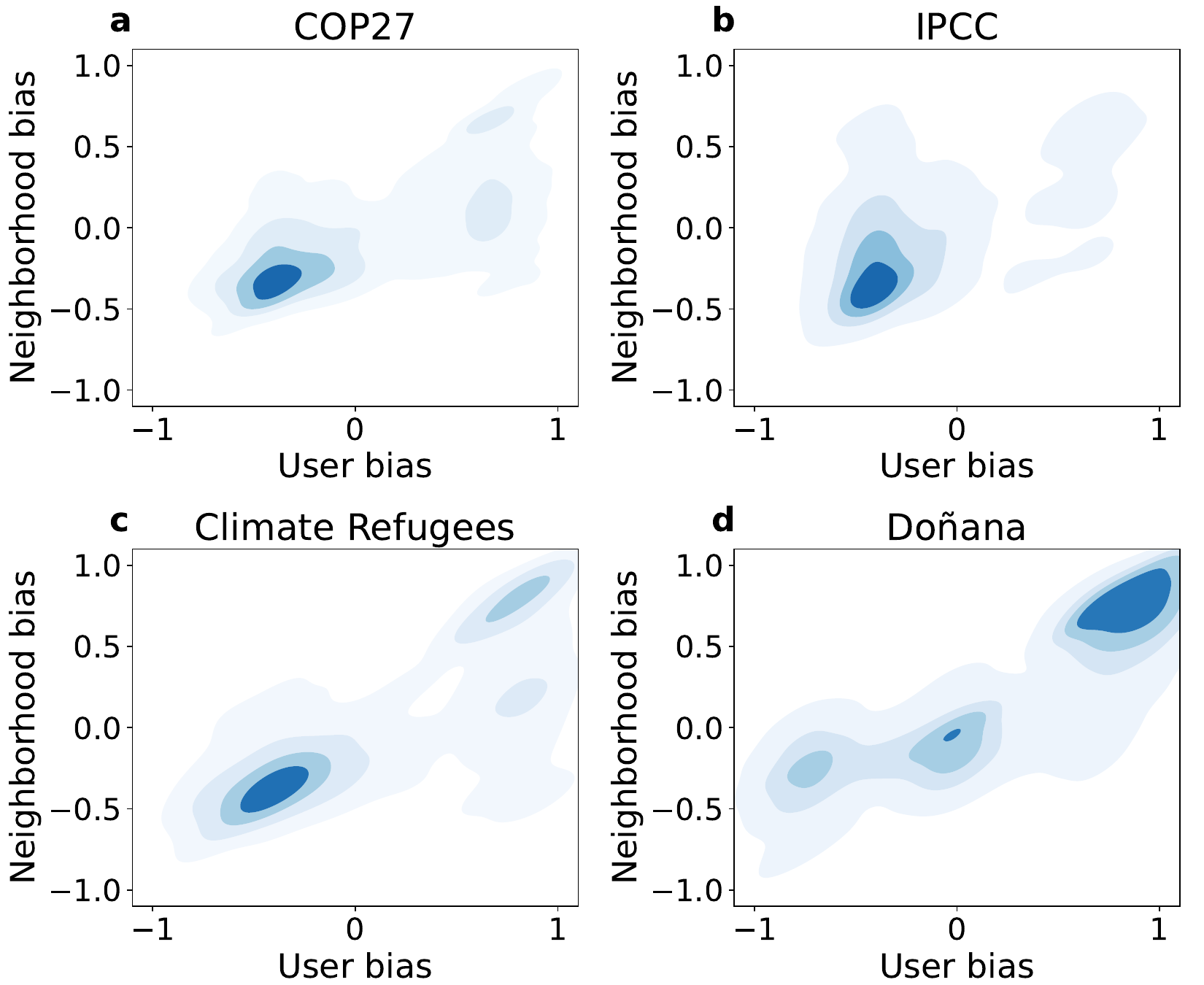}
   \end{center} \caption{\textbf{Detection of echo chambers in ideological biases for the retweet networks.} Neighbourhood political bias as a function of the user bias in the retweet network for \textbf{(a)} COP27, \textbf{(b)} IPCC, \textbf{(c)} Climate Refugees, and \textbf{(d)} Do\~{n}ana.}\label{neighboor_bias_retweet} 
   \end{figure}
   
We show the density plot for the bias of the neighbourhood of users as a function of their values when considering the interactions through retweets in Fig.\ref{neighboor_bias_retweet}. As expected from the overall ideological positioning of the users, there is a hotspot on the left of the ideological spectrum and a weaker one on the right side. We have identified a diagonal trend that could be related to the presence of echo chambers. Right-wing biased users interact more with left-wing biased users, as there are two hotspots for positive values in some cases. The presence of echo chambers is more evident in the Do\~{n}ana dataset than in the rest of the networks, mainly due to the presence of more users with right-leaning discourse. There are three hotspots of comparable magnitude across the diagonal with right-, centre-, and left-leaning discourses. Still, the hotspot on the left side of the spectrum displays a lower ideological isolation, since it interacts with users featuring a bias close to 0. The right-wing biased users interact mainly with each other and not with users in the centre. When analysing other types of interactions, such as the mentions, the presence of echo chambers is weaker with nearly horizontal hotspots in the COP27 and IPCC datasets (see details in Figs. S11). This observation is consistent with the results on polarisation from the previous section. The existence of echo chambers is less pronounced when considering different types of interactions, and mentions could serve as a bridge between users with different political biases.

We also examine the echo chambers based on the reliability of the content shared by users and their neighbourhoods, where we find three hotspots for high, medium, and small values. The hotspots in high reliability suggest that users who share highly reliable sources interact with each other, while low-reliability users interact with users who spread highly reliable sources (see Fig. S12).
  
We inspect in detail the user interaction similarity by quantifying their echo chamber \cite{kolic2022quantifying}, without considering the ideology. Given a set of leading users $i$, which have an audience $A_i$ composed of the set of users that retweeted it, their chamber $C_i$ is the set of users retweeted by audience $A_i$. We focus on the two communities with opposing political biases, specifically the top twenty users with the most retweets in each of those communities. The first quantity we will analyse is the chamber overlap between a user $i$ and a user $j$ given by
\begin{equation} 
q_{ij}=\frac{C_i \cap C_j}{C_i \cup C_j}, 
\end{equation} 
where $C_i$ and $C_j$ are the chambers of users $i$ and $j$. It is worth noting that the quantity $q_{ij}$ is symmetric by definition and is equal to $1$ when chambers are identical and $0$ when they are completely different.

We compute the chamber overlap distribution $P(q_{ij})$, where a notable disparity between networks is observed (see Fig.S13). A strong bimodal distribution suggests the presence of echo chambers, as each user has low values of $q_{ij}$ with users outside its echo chamber and high values with those inside it. The results from the COP27 and Climate Refugees datasets suggest that leaders have more heterogeneous chambers compared to the IPCC and Do\~{n}ana datasets, where we observe a peak for high chamber overlap. The Do\~{n}ana dataset features a strong echo chamber effect with two separated peaks at low and high values of $q_{ij}$. We have performed hierarchical clustering of the top 20 users per community based on the similarity of their chambers, showing a strong alignment between the communities detected (for more details see Section S2.6).

\begin{figure}[!htbp] 
\begin{center} 
\includegraphics[width=\textwidth]{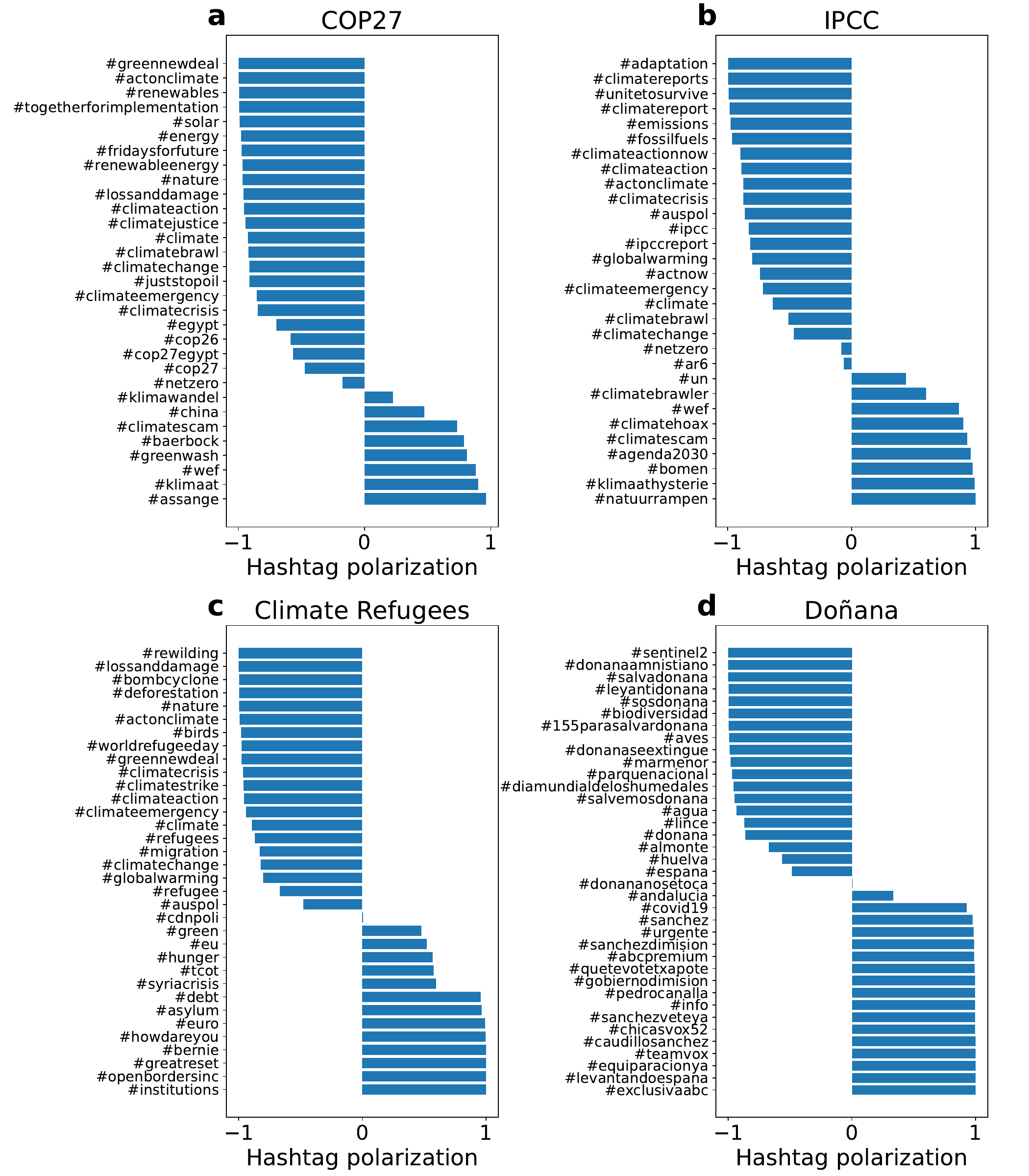} 
\end{center} 
\caption{\textbf{Hashtag polarisation.} Polarisation around hashtags for \textbf{(a)} COP27, \textbf{(b)} IPCC, \textbf{(c)} Climate Refugees, and \textbf{(d)} Do\~{n}ana. Negative and positive values correspond to hashtags more abundant in the left-biased and right-biased communities, respectively.} \label{hashtag_polarization} 
\end{figure}

\subsection{Policy and affective polarisation}

On account of the structural polarisation observed, with groups of users sharing political bias, we further investigate if there are perceptible differences in their climate change views. Structural metrics are purely based on user interactions and, thus, provide limited information on the reasons and motivations of the groups. To get a first assessment of the content discussed in each of the two opposed communities and evaluate if they have confronted perspectives on climate change, we inspect the patterns of hashtags shared. Despite their ideological confrontation, they share most of them, likely due to the common subject of discussion (Fig. S15). The mainstream communities use rather generalist hashtags such as $\#COP27$, $\#IPCC$, $\#climatechange$, $\#ar6$, or $\#climatereport$. In Do\~{n}ana, common hashtags between communities are less frequent, which could indicate a stronger political polarisation. We observe hashtags related to the environment as $\#donana$, $\#donanaseextingue$, or $\#salvemosdonana$ for the mainstream community and hashtags related to right-wing parties as $\#teamvox$, or against the left-wing government as $\#gobiernodimision$, in the right-wing biased community.

To compute the hashtag polarisation, we calculated the fraction of tweets with each hashtag by community, given by 
\begin{equation}\label{eq:hashtag_polarization}
P_i^{\#}=\frac{S^{l}N_i^{r}-S^{r}N_i^{l}}{S^{r}N^{l}_i+S^{l}N_i^{r}}, 
\end{equation}
where $N_i^{l}$ and $N_i^r$ are the number of tweets with the hashtag $i$ in the main left- and right-wing communities, respectively, and $S^l$ and $S^r$ the number of users in each community. This formulation takes into account the uneven sizes of the communities. We display the polarisation of the top hashtags in Fig. \ref{hashtag_polarization}. The number of hashtags in each network varies since we have considered the $20$ most shared in each community, and some of them can overlap. Interestingly, we observe only a few hashtags with values close to $0$, suggesting a strong polarisation in the use of hashtags. In the IPCC and COP27 datasets, the mainstream community dominates most of the hashtags, while the community spreading a counter-narrative dominates only a few hashtags related to the denial of climate change that include $\#climatehoax$ or $\#agenda2030$. In COP27, we also observe hashtags related to environmental hypocrisy such as $\#greenwash$. In the Do\~{n}ana network, we also observe only a few hashtags close to zero and hashtags almost equally divided between $+1$ and $-1$. We performed a Z-test to assess the significance of hashtag frequency differences between the most polarised communities. Among the hashtags with highest absolute Z-scores, $\#climatescam$ is predominantly used by right-leaning communities in both the COP27 and IPCC networks. In the Climate Refugees network, one of the hashtags with a high Z-score is $\#howdareyou$, again primarily used by the right-wing community. In the case of the Spanish discussion about Doñana, the hashtags with the highest Z-scores are those criticising the Spanish government, as previously mentioned (see Section S3.1 for details).

Whereas hashtags can provide a glimpse of the ideas discussed within each community, they offer a limited perspective to the opinion of individuals as they can be used in different contexts. To better understand the divergence in climate change views, we conducted a topic analysis on tweets from the two largest communities with opposing political perspectives. We assess topic polarisation by quantifying the number of users within each community who discuss a given topic, using Eq. \ref{eq:hashtag_polarization}. Most topics are either dominated by the left-biased community or the low-reliability community. In the COP27 network, the right-leaning discourse reveals corruption discussions with money-related terms, and political references to the UK, China and India. In contrast, the left-biased community focuses on essential resources such as water and food, as well as the urgency of taking action. In the IPCC network, topics within the right-biased community express scepticism about the scientific basis of climate change predictions, reference COVID-19, and include discussions that are often unrelated or misleading (for more details about the topics see Section S3.2). The results confirm that it could be composed of individuals posting low-reliability information. The left-biased community emphasises the need to reduce emissions and the urgency of taking action, using language similar to that found in COP27 discussions. In the Climate Refugees network, the right-biased community focuses on jobs and border security, whereas the left-biased community highlights the risks posed by natural disasters. The greatest divergence in discussions is observed in the Do\~{n}ana network. The right-biased community focuses on the political handling of the situation, with an emphasis on the Spanish president, the ruling party (PSOE), and the European Union. In contrast, the mainstream community centres its discussion on the Do\~{n}ana National Park and its conservation.

\begin{figure}[!htbp] 
\begin{center} 
\includegraphics[width=0.9\textwidth]{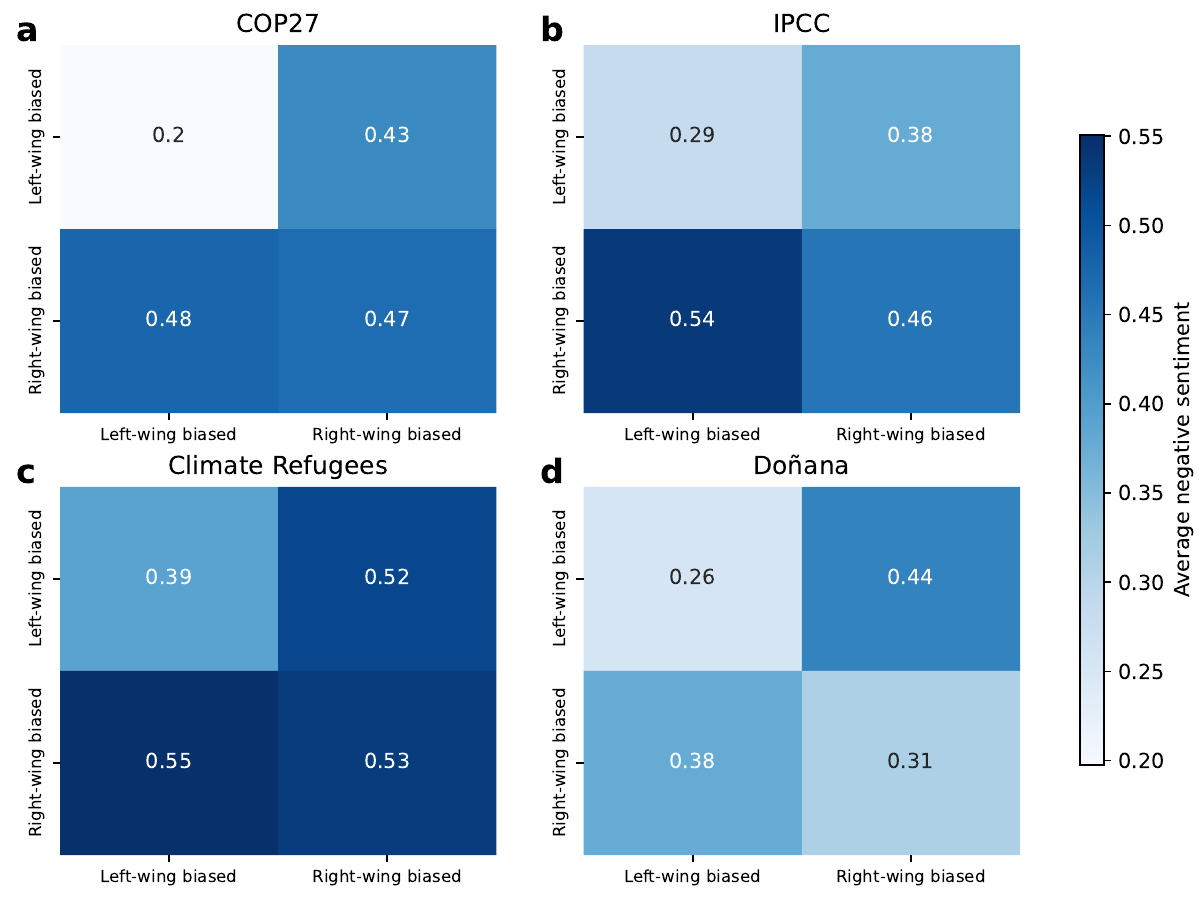} 
\end{center} 
\caption{\textbf{Average negative sentiment in messages in community interactions.} Average negative sentiment of the messages as a function of the poster community (y-axis) and quoted (x-axis). The results correspond to \textbf{(a)} COP27, \textbf{(b)} IPCC,\textbf{(c)} Climate Refugees, \textbf{(d)} Do\~{n}ana.} \label{sentiment_analysis} 
\end{figure}

Besides the disparity in the content shared, via hashtags and topics, we want to adress whether those groups are aware of each other and if they are emotionally aligned. Therefore, We evaluate emotional repulsion in relation to political views by analysing the sentiment of quote messages. Specifically, we assess the average negative sentiment based on the community of both the sender and the receiver (Fig. \ref{sentiment_analysis}). We have implemented a sentiment analysis on the four Twitter datasets using a RoBERTa-based model fine-tuned on around 58 million of tweets \cite{huggingface, barbieri2020tweeteval}. Overall, messages from right-wing biased communities exhibit a higher degree of negative sentiment, particularly when directed at left-wing biased communities (0.48 for COP27, 0.54 for IPCC, and 0.55 for Climate Refugees). The difference between intra- and inter-community messages is most pronounced in the IPCC and Doñana networks. In the IPCC network, the negativity score for right-leaning to left-leaning interactions is 0.54, while for right-leaning to right-leaning interactions, it is 0.46 ($\Delta=0.08$). Similarly, in the Doñana network, right-leaning to left-leaning interactions have a negativity score of 0.38, compared to 0.31 for right-leaning to right-leaning interactions ($\Delta=0.07$). Interestingly, the most pronounced difference in sentiment between intra- and inter-community interactions occurs among users in the left-wing biased community. In the extreme case of the COP27 network, left-leaning users display a $\Delta=0.23$ difference, while also being associated with the dissemination of more reliable information. Across all networks, the lowest levels of negative sentiment are observed within the left-wing biased communities, ranging from 0.20 in COP27 to 0.39 in Climate Refugees. In contrast, the right-wing biased community exhibits consistently higher negativity, even within its own interactions, with values ranging from 0.31 in Doñana to 0.53 in Climate Refugees.

We also conducted a Mann-Whitney U test to evaluate the statistical significance of negativity differences between communities. A consistent pattern emerges in the COP27, IPCC, and Climate Refugees networks: right-leaning to left-leaning interactions display significantly greater negativity than all other combinations. For example, in the IPCC network, the negativity of right-leaning to left-leaning interactions is significantly greater than that of left-wing biased to right-wing biased interactions ($p<.001$), left-wing biased to left-wing biased interactions ($p<.001$), and right-wing biased to right-wing biased interactions ($p<.001$) (see Fig. S19 for additional comparisons). Furthermore, across all datasets, right-wing biased to right-wing biased interactions exhibit significantly higher negativity compared to left-wing biased to left-wing biased interactions (COP27: $p<.001$; IPCC: $p<.001$; Climate Refugees: $p<.001$; and Doñana: $p<.001$). A similar pattern is observed when comparing left-wing biased to right-wing biased interactions against left-wing biased to left-wing biased interactions (COP27: $p<.001$; IPCC: $p<.001$; Climate Refugees: $p<.001$; and Doñana: $p<.001$). Hence, the structural communities identified are aligned with a divergence on content discussion related to climate change and stronger emotional responses. The results suggest that polarisation on the subject takes multiple forms.

\subsection{Cross-platform analysis}

We have analysed the dataset created from the YouTube videos quoted and referenced on Twitter to assess if the observed polarisation facets are aligned with content consumption in video platforms as partisan media affects the dynamics of social networks \cite{guess2021consequences}. The information gathered for each post includes the transcription and description of each video, the corresponding channel, and the user comments. The comment information allows us to create a bipartite network between users and posts since we know the users who have commented on each video. We can project the bipartite network into a post network or a user network. In the post network, the weight corresponds to the number of users in common and, in the user network, to the posts in common. The graphs are undirected by construction in both cases. We performed a community analysis using the greedy modularity optimisation on the weighted network \cite{newman2006modularity}. The organisation of communities do not display a clear separation between communities with several interconnections except for the Do\~{n}ana network, where communities zero and one have a clear separation (see Section S4). 

\begin{figure}[!htbp] 
\begin{center} 
\includegraphics[width=0.9\textwidth]{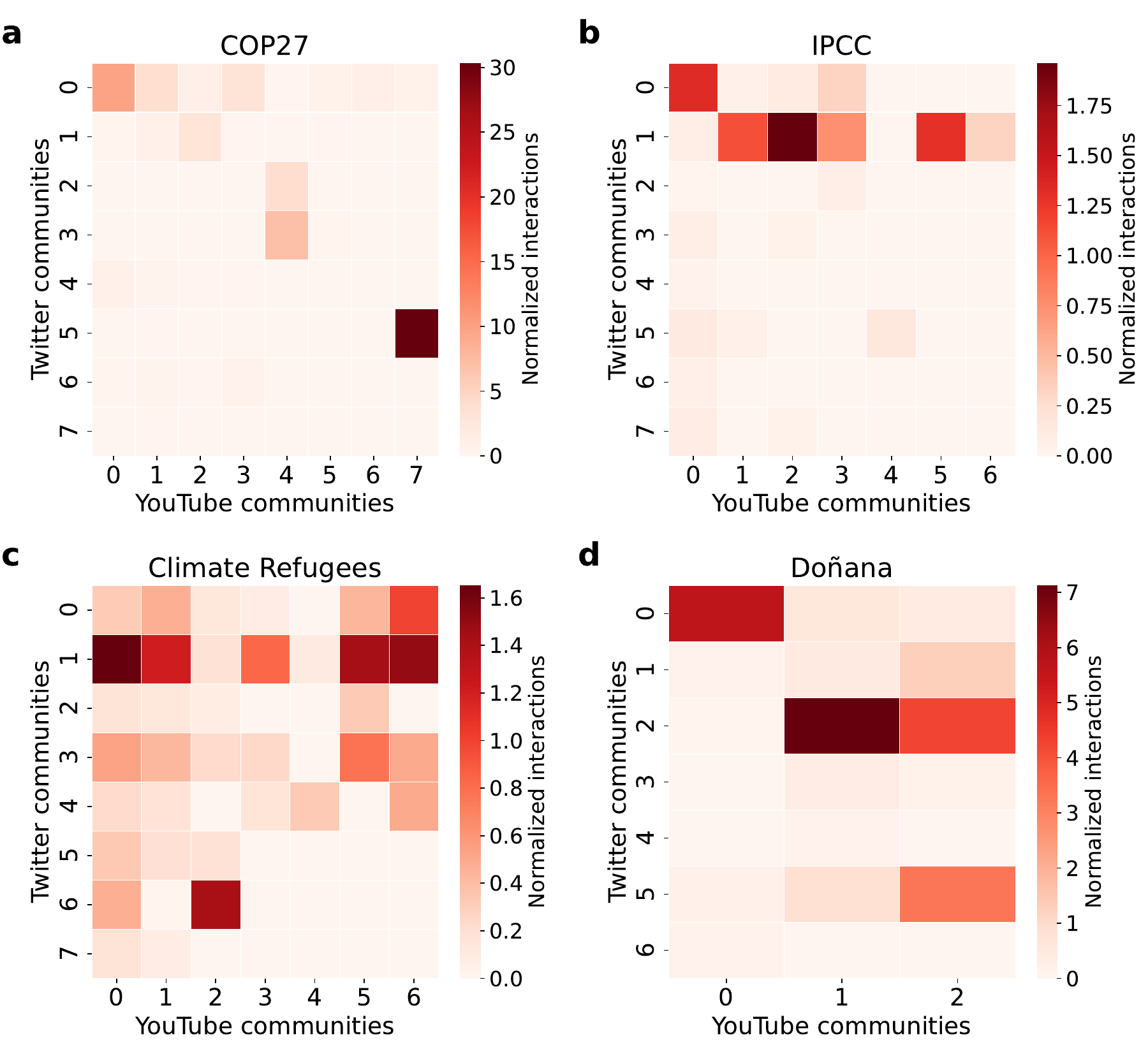} 
\end{center} 
\caption{\textbf{Interaction between Twitter and YouTube communities.} Number of references to YouTube communities by the top six Twitter communities in \textbf{(a)} COP27, \textbf{(b)} IPCC, \textbf{(c)} Climate Refugees, and \textbf{(d)} Do\~{n}ana. To account for the uneven size of YouTube communities, we have divided the counts by the YouTube community size.} \label{twitter_youtube_interaction} 
\end{figure}

We conducted an analysis linking Twitter and YouTube social media platforms to study the cross-platforms effects. We create the Twitter subnetwork of users who shared YouTube links related to climate change, tagged according to the YouTube community of the videos. The connection between network organisation and the shared YouTube content is clearer in the Doñana and IPCC networks. In both cases, users tweeting videos from YouTube community zero are more tightly connected than other users (see the subnetwork in Figs. S22 and S23).

Furthermore, we computed the frequency with which each Twitter community references YouTube posts and their corresponding communities. To account for the heterogeneity in community sizes observed in the YouTube dataset, we normalised the number of references to each YouTube community by its size. A strong connection exists between Twitter and YouTube communities, as most references within each Twitter community align with a specific YouTube community. Additionally, each Twitter community exhibits a unique pattern of references to YouTube.

In the COP27 network, the larger left-wing biased Twitter community 0 predominantly references YouTube community 0 ($35.8\%$), while the larger right-wing biased and low-reliability community 1 primarily references posts in YouTube community 2 ($57.3\%$). Similarly, in the Climate Refugees network, Twitter community 6, known for spreading low-reliability information, interacts significantly with YouTube community 2 ($74.6\%$). Finally, in the Doñana network, community 0 from both social networks interacts intensively ($84.3\%$). Conversely, in the IPCC network, the larger right-wing biased community 1 references multiple YouTube communities, specifically communities 1 ($20.2\%$), 2 ($35.3\%$), and 5 ($23.4\%$), suggesting a tendency to share diverse typologies of YouTube content. With the exception of the IPCC network, posts shared by the larger right-wing biased community on Twitter predominantly align with a single YouTube community (Fig. \ref{twitter_youtube_interaction}).
\section{Discussion}
Our results confirm that polarisation in climate change discussions manifests in different forms through structural, ideological, and emotional components. We also observed distinct views on climate change and patterns of content consumption. The structural polarisation around the subject is aligned with distinct political biases and information reliability consumption. Not only that, the topological communities also show an alignment on perspective towards the environmental crisis and a stronger emotional response. Overall, polarisation takes multiple aligned forms. While previous work often emphasised the role of ideological segregation, our cross-platform analysis shows that inter-community exposure remains significant—especially via mentions—but is frequently accompanied by affective repulsion and asymmetric negativity, particularly from low-reliability, right-leaning communities.

This suggests that polarisation is not merely a consequence of echo chambers or algorithmic filtering, but also reflects active processes of antagonism and identity signalling, in line with theories of affective polarisation. Even in multilingual and transnational contexts, such as the COP27 dataset, we observe stable polarisation structures shaped more by political affiliations and emotional tone than by language or geography. Moreover, the structural signature of interactions varies across platforms and interaction types: while retweets act as endorsement signals and tend to reinforce in-group cohesion, mentions frequently bridge ideological divides, albeit often with antagonistic or negative sentiment.

The interplay between information reliability and political alignment also emerges as a key factor. Right-leaning communities consistently share lower-reliability content and exhibit higher isolation and negativity, yet they remain engaged with mainstream communities—albeit often in confrontational terms. This pattern challenges the notion that low-reliability information spreads in isolation and points instead to a contested media space where political identity, trust, and emotion are tightly intertwined. Our results align with other studies showing that, despite higher engagement within ideologically similar groups, users are still exposed to opposing views \cite{barbera2020social}. However, such exposure does not necessarily mitigate polarisation. On the contrary, it has been shown that confrontation with opposing views can intensify polarisation instead of reducing it \cite{bail2018exposure}. This is reflected in our sentiment analysis, where messages from right-biased communities consistently show higher levels of negativity—especially when directed toward left-leaning users \cite{falkenberg2024patterns}. Overall, the sentiment in the low-reliability community was more negative. Our results are in line with recent research \cite{meyer2023between}, which finds that climate change dissenters are closely related and tend to engage with a left-biased mainstream community \cite{meyer2023between,falkenberg2024patterns,meyer2025disruptive}. Furthermore, toxicity levels are not evenly distributed across interaction types: replies and mentions are significantly more toxic than other forms of engagement \cite{bruggemann2023debates}. This observation suggests that low-reliability individuals may reference the mainstream community to engage in discussions but get little attention in return \cite{bruggemann2023debates}, highlighting an asymmetry between left- and right-leaning communities \cite{eady2019many}.

The cross-platform alignment between Twitter and YouTube reinforces the idea that ideological communities are transversal across platforms, with users not only engaging with like-minded individuals but also selecting and amplifying content from ideologically aligned sources elsewhere. This calls for a broader, integrated approach to studying online polarisation, moving beyond single-platform analyses. Additionally, our results indicate that the degree and manifestation of polarisation can vary substantially across topics and events—suggesting that it is not static, but shaped by contextual factors such as the political salience of the issue or the type of event. This may reflect a dynamic of partisan sorting, whereby individuals increasingly align their social media behaviour with broader political identities, as already observed in the Finnish Twittersphere \cite{chen2021polarization}. Prior studies, such as \cite{sanford2021controversy}, observed that the 2019 IPCC Special Report on Climate Change and Land sparked significant public discourse on social media, with dietary recommendations becoming a highly polarised topic. Similar to our findings on localised polarisation patterns (e.g., Doñana), that report showed that climate solutions affecting lifestyle choices may trigger elevated levels of contention on platforms like Twitter. The alignment between platforms could be amplified by recommendation algorithms, which can function as filter bubbles that reinforce users' beliefs \cite{cho2020search,park2024filter}. However, recent studies have downplayed the contribution of algorithms to polarisation \cite{liu2025short}. Our results support the idea that polarisation is ubiquitous across platforms, but not driven solely by lack of exposure to opposing viewpoints. 

These findings have important implications for the broader climate change debate. The persistence of polarised, emotionally charged interactions—even in the presence of exposure to opposing views—suggests that efforts to foster consensus around climate policy must account for the affective and identity-driven nature of public discourse. Moreover, the asymmetric structure of negativity and the association between political alignment and information reliability can contribute to public confusion, distrust in science, and a fragmentation of climate narratives, making it harder to establish the shared understanding necessary for effective action. Addressing polarisation in this context requires more than correcting low-reliability information—it demands strategies that bridge ideological divides and rebuild trust in both institutions and science. 

Our static analysis of the polarisation around climate change could be extended to assess how it evolves with time as a function of internal dynamics, external sources, and the unfolding of events. For instance, coordinated campaigns have been observed to impact social media dynamics \cite{pacheco2021uncovering}, and could play a role in shaping the emergence or intensification of polarisation. It is unclear whether message negativity is intrinsic to the topic or evolves with the discussion, potentially triggered by specific interactions. The asymmetric spread of emotionally polarised content suggests that by targeting specific small groups, polarisation could be significantly decreased \cite{meyer2025disruptive}. However, the opacity of most online social networks makes content moderation a major challenge. It would be essential to assess whether content moderation or fact-checking could have a positive impact on the reduction of polarisation.  Since our results suggest that individuals sharing low-reliability information are exposed to opposing views, the development of intervention strategies becomes a major challenge. The role of exposure in increasing polarisation could be further addressed by investigating the temporal evolution of user ideology and reliability as a function of their progressive interactions with others. In a similar fashion, it opens the door to the development of models to unveil how the exposure to opposing views can further exacerbate polarisation.

\section*{Data availability statement}

The Twitter data are made available in accordance with Twitter’s terms of service at \url{https://github.com/clint-project/clint-data} (https//doi.org:/10.5281/zenodo.14095218) and the codes to calculate the metrics at \url{https://github.com/clint-project/polarization_metrics} (https://doi.org/10.5281/zenodo.14095108). The YouTube data is available at \newline https://doi.org:/10.5281/zenodo.15046657.

\section*{Author contribution}

A.B: data collection, data curation, investigation, methodology, visualization; J.M.: data collection, data curation,  investigation, methodology. E.C.: conceptualization, investigation, methodology, writing; J.V.: conceptualization, investigation, methodology, writing. All authors gave final approval for publication and agreed to be held accountable for the work performed therein.

\section*{Acknowledgements}

J.M. was a fellow of Eurecat’s “Vicente López” PhD grant program. This work was financially supported by the Catalan Government through the funding grant ACCIÓ-Eurecat (Project TRAÇA 2023 - CLINT). E.C. acknowledges support from the Spanish grants PGC2018-094754-B-C22 and PID2021-128005NB-C22, funded by MCIN/AEL 10.13039/501100011033 and "ERDF A way of making Europe", and from Generalitat de Catalunya under project 2021-SGR-00856.

\bibliography{sample}

\begin{thebibliography}{84}%
\makeatletter
\providecommand \@ifxundefined [1]{%
 \@ifx{#1\undefined}
}%
\providecommand \@ifnum [1]{%
 \ifnum #1\expandafter \@firstoftwo
 \else \expandafter \@secondoftwo
 \fi
}%
\providecommand \@ifx [1]{%
 \ifx #1\expandafter \@firstoftwo
 \else \expandafter \@secondoftwo
 \fi
}%
\providecommand \natexlab [1]{#1}%
\providecommand \enquote  [1]{``#1''}%
\providecommand \bibnamefont  [1]{#1}%
\providecommand \bibfnamefont [1]{#1}%
\providecommand \citenamefont [1]{#1}%
\providecommand \href@noop [0]{\@secondoftwo}%
\providecommand \href [0]{\begingroup \@sanitize@url \@href}%
\providecommand \@href[1]{\@@startlink{#1}\@@href}%
\providecommand \@@href[1]{\endgroup#1\@@endlink}%
\providecommand \@sanitize@url [0]{\catcode `\\12\catcode `\$12\catcode
  `\&12\catcode `\#12\catcode `\^12\catcode `\_12\catcode `\%12\relax}%
\providecommand \@@startlink[1]{}%
\providecommand \@@endlink[0]{}%
\providecommand \url  [0]{\begingroup\@sanitize@url \@url }%
\providecommand \@url [1]{\endgroup\@href {#1}{\urlprefix }}%
\providecommand \urlprefix  [0]{URL }%
\providecommand \Eprint [0]{\href }%
\providecommand \doibase [0]{https://doi.org/}%
\providecommand \selectlanguage [0]{\@gobble}%
\providecommand \bibinfo  [0]{\@secondoftwo}%
\providecommand \bibfield  [0]{\@secondoftwo}%
\providecommand \translation [1]{[#1]}%
\providecommand \BibitemOpen [0]{}%
\providecommand \bibitemStop [0]{}%
\providecommand \bibitemNoStop [0]{.\EOS\space}%
\providecommand \EOS [0]{\spacefactor3000\relax}%
\providecommand \BibitemShut  [1]{\csname bibitem#1\endcsname}%
\let\auto@bib@innerbib\@empty
\bibitem [{\citenamefont {McCoy}\ \emph {et~al.}(2018)\citenamefont {McCoy},
  \citenamefont {Rahman},\ and\ \citenamefont {Somer}}]{McCoy2018}%
  \BibitemOpen
  \bibfield  {author} {\bibinfo {author} {\bibfnamefont {J.}~\bibnamefont
  {McCoy}}, \bibinfo {author} {\bibfnamefont {T.}~\bibnamefont {Rahman}},\ and\
  \bibinfo {author} {\bibfnamefont {M.}~\bibnamefont {Somer}},\ }\bibfield
  {title} {\bibinfo {title} {Polarization and the global crisis of democracy:
  Common patterns, dynamics, and pernicious consequences for democratic
  polities},\ }\href {https://doi.org/10.1177/0002764218759576} {\bibfield
  {journal} {\bibinfo  {journal} {American Behavioral Scientist}\ }\textbf
  {\bibinfo {volume} {62}},\ \bibinfo {pages} {16–42} (\bibinfo {year}
  {2018})}\BibitemShut {NoStop}%
\bibitem [{\citenamefont {DiMaggio}\ \emph {et~al.}(1996)\citenamefont
  {DiMaggio}, \citenamefont {Evans},\ and\ \citenamefont
  {Bryson}}]{dimaggio1996have}%
  \BibitemOpen
  \bibfield  {author} {\bibinfo {author} {\bibfnamefont {P.}~\bibnamefont
  {DiMaggio}}, \bibinfo {author} {\bibfnamefont {J.}~\bibnamefont {Evans}},\
  and\ \bibinfo {author} {\bibfnamefont {B.}~\bibnamefont {Bryson}},\
  }\bibfield  {title} {\bibinfo {title} {Have american's social attitudes
  become more polarized?},\ }\href@noop {} {\bibfield  {journal} {\bibinfo
  {journal} {American journal of Sociology}\ }\textbf {\bibinfo {volume}
  {102}},\ \bibinfo {pages} {690} (\bibinfo {year} {1996})}\BibitemShut
  {NoStop}%
\bibitem [{\citenamefont {Esteban}\ and\ \citenamefont
  {Ray}(1994)}]{esteban1994measurement}%
  \BibitemOpen
  \bibfield  {author} {\bibinfo {author} {\bibfnamefont {J.-M.}\ \bibnamefont
  {Esteban}}\ and\ \bibinfo {author} {\bibfnamefont {D.}~\bibnamefont {Ray}},\
  }\bibfield  {title} {\bibinfo {title} {On the measurement of polarization},\
  }\href@noop {} {\bibfield  {journal} {\bibinfo  {journal} {Econometrica:
  Journal of the Econometric Society}\ ,\ \bibinfo {pages} {819}} (\bibinfo
  {year} {1994})}\BibitemShut {NoStop}%
\bibitem [{\citenamefont {Bramson}\ \emph {et~al.}(2017)\citenamefont
  {Bramson}, \citenamefont {Grim}, \citenamefont {Singer}, \citenamefont
  {Berger}, \citenamefont {Sack}, \citenamefont {Fisher}, \citenamefont
  {Flocken},\ and\ \citenamefont {Holman}}]{bramson2017understanding}%
  \BibitemOpen
  \bibfield  {author} {\bibinfo {author} {\bibfnamefont {A.}~\bibnamefont
  {Bramson}}, \bibinfo {author} {\bibfnamefont {P.}~\bibnamefont {Grim}},
  \bibinfo {author} {\bibfnamefont {D.~J.}\ \bibnamefont {Singer}}, \bibinfo
  {author} {\bibfnamefont {W.~J.}\ \bibnamefont {Berger}}, \bibinfo {author}
  {\bibfnamefont {G.}~\bibnamefont {Sack}}, \bibinfo {author} {\bibfnamefont
  {S.}~\bibnamefont {Fisher}}, \bibinfo {author} {\bibfnamefont
  {C.}~\bibnamefont {Flocken}},\ and\ \bibinfo {author} {\bibfnamefont
  {B.}~\bibnamefont {Holman}},\ }\bibfield  {title} {\bibinfo {title}
  {Understanding polarization: Meanings, measures, and model evaluation},\
  }\href@noop {} {\bibfield  {journal} {\bibinfo  {journal} {Philosophy of
  science}\ }\textbf {\bibinfo {volume} {84}},\ \bibinfo {pages} {115}
  (\bibinfo {year} {2017})}\BibitemShut {NoStop}%
\bibitem [{\citenamefont {Zollo}\ \emph {et~al.}(2015)\citenamefont {Zollo},
  \citenamefont {Novak}, \citenamefont {Del~Vicario}, \citenamefont {Bessi},
  \citenamefont {Mozeti{\v{c}}}, \citenamefont {Scala}, \citenamefont
  {Caldarelli},\ and\ \citenamefont {Quattrociocchi}}]{zollo2015emotional}%
  \BibitemOpen
  \bibfield  {author} {\bibinfo {author} {\bibfnamefont {F.}~\bibnamefont
  {Zollo}}, \bibinfo {author} {\bibfnamefont {P.~K.}\ \bibnamefont {Novak}},
  \bibinfo {author} {\bibfnamefont {M.}~\bibnamefont {Del~Vicario}}, \bibinfo
  {author} {\bibfnamefont {A.}~\bibnamefont {Bessi}}, \bibinfo {author}
  {\bibfnamefont {I.}~\bibnamefont {Mozeti{\v{c}}}}, \bibinfo {author}
  {\bibfnamefont {A.}~\bibnamefont {Scala}}, \bibinfo {author} {\bibfnamefont
  {G.}~\bibnamefont {Caldarelli}},\ and\ \bibinfo {author} {\bibfnamefont
  {W.}~\bibnamefont {Quattrociocchi}},\ }\bibfield  {title} {\bibinfo {title}
  {Emotional dynamics in the age of misinformation},\ }\href@noop {} {\bibfield
   {journal} {\bibinfo  {journal} {PloS one}\ }\textbf {\bibinfo {volume}
  {10}},\ \bibinfo {pages} {e0138740} (\bibinfo {year} {2015})}\BibitemShut
  {NoStop}%
\bibitem [{\citenamefont {Del~Vicario}\ \emph
  {et~al.}(2016{\natexlab{a}})\citenamefont {Del~Vicario}, \citenamefont
  {Vivaldo}, \citenamefont {Bessi}, \citenamefont {Zollo}, \citenamefont
  {Scala}, \citenamefont {Caldarelli},\ and\ \citenamefont
  {Quattrociocchi}}]{del2016echo}%
  \BibitemOpen
  \bibfield  {author} {\bibinfo {author} {\bibfnamefont {M.}~\bibnamefont
  {Del~Vicario}}, \bibinfo {author} {\bibfnamefont {G.}~\bibnamefont
  {Vivaldo}}, \bibinfo {author} {\bibfnamefont {A.}~\bibnamefont {Bessi}},
  \bibinfo {author} {\bibfnamefont {F.}~\bibnamefont {Zollo}}, \bibinfo
  {author} {\bibfnamefont {A.}~\bibnamefont {Scala}}, \bibinfo {author}
  {\bibfnamefont {G.}~\bibnamefont {Caldarelli}},\ and\ \bibinfo {author}
  {\bibfnamefont {W.}~\bibnamefont {Quattrociocchi}},\ }\bibfield  {title}
  {\bibinfo {title} {Echo chambers: Emotional contagion and group polarization
  on facebook},\ }\href@noop {} {\bibfield  {journal} {\bibinfo  {journal}
  {Scientific reports}\ }\textbf {\bibinfo {volume} {6}},\ \bibinfo {pages}
  {37825} (\bibinfo {year} {2016}{\natexlab{a}})}\BibitemShut {NoStop}%
\bibitem [{\citenamefont {Du}\ and\ \citenamefont
  {Gregory}(2017)}]{du2017echo}%
  \BibitemOpen
  \bibfield  {author} {\bibinfo {author} {\bibfnamefont {S.}~\bibnamefont
  {Du}}\ and\ \bibinfo {author} {\bibfnamefont {S.}~\bibnamefont {Gregory}},\
  }\bibfield  {title} {\bibinfo {title} {The echo chamber effect in twitter:
  does community polarization increase?},\ }in\ \href@noop {} {\emph {\bibinfo
  {booktitle} {Complex Networks \& Their Applications V: Proceedings of the 5th
  International Workshop on Complex Networks and their Applications (COMPLEX
  NETWORKS 2016)}}}\ (\bibinfo {organization} {Springer},\ \bibinfo {year}
  {2017})\ pp.\ \bibinfo {pages} {373--378}\BibitemShut {NoStop}%
\bibitem [{\citenamefont {Quattrociocchi}\ \emph {et~al.}(2016)\citenamefont
  {Quattrociocchi}, \citenamefont {Scala},\ and\ \citenamefont
  {Sunstein}}]{quattrociocchi2016echo}%
  \BibitemOpen
  \bibfield  {author} {\bibinfo {author} {\bibfnamefont {W.}~\bibnamefont
  {Quattrociocchi}}, \bibinfo {author} {\bibfnamefont {A.}~\bibnamefont
  {Scala}},\ and\ \bibinfo {author} {\bibfnamefont {C.~R.}\ \bibnamefont
  {Sunstein}},\ }\bibfield  {title} {\bibinfo {title} {Echo chambers on
  facebook},\ }\href@noop {} {\bibfield  {journal} {\bibinfo  {journal}
  {Available at SSRN 2795110}\ } (\bibinfo {year} {2016})}\BibitemShut
  {NoStop}%
\bibitem [{\citenamefont {Cinelli}\ \emph {et~al.}(2021)\citenamefont
  {Cinelli}, \citenamefont {De~Francisci~Morales}, \citenamefont {Galeazzi},
  \citenamefont {Quattrociocchi},\ and\ \citenamefont
  {Starnini}}]{cinelli2021echo}%
  \BibitemOpen
  \bibfield  {author} {\bibinfo {author} {\bibfnamefont {M.}~\bibnamefont
  {Cinelli}}, \bibinfo {author} {\bibfnamefont {G.}~\bibnamefont
  {De~Francisci~Morales}}, \bibinfo {author} {\bibfnamefont {A.}~\bibnamefont
  {Galeazzi}}, \bibinfo {author} {\bibfnamefont {W.}~\bibnamefont
  {Quattrociocchi}},\ and\ \bibinfo {author} {\bibfnamefont {M.}~\bibnamefont
  {Starnini}},\ }\bibfield  {title} {\bibinfo {title} {The echo chamber effect
  on social media},\ }\href@noop {} {\bibfield  {journal} {\bibinfo  {journal}
  {Proceedings of the National Academy of Sciences}\ }\textbf {\bibinfo
  {volume} {118}},\ \bibinfo {pages} {e2023301118} (\bibinfo {year}
  {2021})}\BibitemShut {NoStop}%
\bibitem [{\citenamefont {Di~Marco}\ \emph {et~al.}(2021)\citenamefont
  {Di~Marco}, \citenamefont {Cinelli},\ and\ \citenamefont
  {Quattrociocchi}}]{di2021infodemics}%
  \BibitemOpen
  \bibfield  {author} {\bibinfo {author} {\bibfnamefont {N.}~\bibnamefont
  {Di~Marco}}, \bibinfo {author} {\bibfnamefont {M.}~\bibnamefont {Cinelli}},\
  and\ \bibinfo {author} {\bibfnamefont {W.}~\bibnamefont {Quattrociocchi}},\
  }\bibfield  {title} {\bibinfo {title} {Infodemics on youtube: Reliability of
  content and echo chambers on covid-19},\ }\href@noop {} {\bibfield  {journal}
  {\bibinfo  {journal} {arXiv e-prints}\ ,\ \bibinfo {pages} {arXiv}} (\bibinfo
  {year} {2021})}\BibitemShut {NoStop}%
\bibitem [{\citenamefont {Grusauskaite}\ \emph {et~al.}(2023)\citenamefont
  {Grusauskaite}, \citenamefont {Carbone}, \citenamefont {Harambam},\ and\
  \citenamefont {Aupers}}]{grusauskaite2023debating}%
  \BibitemOpen
  \bibfield  {author} {\bibinfo {author} {\bibfnamefont {K.}~\bibnamefont
  {Grusauskaite}}, \bibinfo {author} {\bibfnamefont {L.}~\bibnamefont
  {Carbone}}, \bibinfo {author} {\bibfnamefont {J.}~\bibnamefont {Harambam}},\
  and\ \bibinfo {author} {\bibfnamefont {S.}~\bibnamefont {Aupers}},\
  }\bibfield  {title} {\bibinfo {title} {Debating (in) echo chambers: How
  culture shapes communication in conspiracy theory networks on youtube},\
  }\href@noop {} {\bibfield  {journal} {\bibinfo  {journal} {New Media \&
  Society}\ ,\ \bibinfo {pages} {14614448231162585}} (\bibinfo {year}
  {2023})}\BibitemShut {NoStop}%
\bibitem [{\citenamefont {Baumann}\ \emph {et~al.}(2020)\citenamefont
  {Baumann}, \citenamefont {Lorenz-Spreen}, \citenamefont {Sokolov},\ and\
  \citenamefont {Starnini}}]{baumann2020modeling}%
  \BibitemOpen
  \bibfield  {author} {\bibinfo {author} {\bibfnamefont {F.}~\bibnamefont
  {Baumann}}, \bibinfo {author} {\bibfnamefont {P.}~\bibnamefont
  {Lorenz-Spreen}}, \bibinfo {author} {\bibfnamefont {I.~M.}\ \bibnamefont
  {Sokolov}},\ and\ \bibinfo {author} {\bibfnamefont {M.}~\bibnamefont
  {Starnini}},\ }\bibfield  {title} {\bibinfo {title} {Modeling echo chambers
  and polarization dynamics in social networks},\ }\href@noop {} {\bibfield
  {journal} {\bibinfo  {journal} {Physical Review Letters}\ }\textbf {\bibinfo
  {volume} {124}},\ \bibinfo {pages} {048301} (\bibinfo {year}
  {2020})}\BibitemShut {NoStop}%
\bibitem [{\citenamefont {Cho}\ \emph {et~al.}(2020)\citenamefont {Cho},
  \citenamefont {Ahmed}, \citenamefont {Hilbert}, \citenamefont {Liu},\ and\
  \citenamefont {Luu}}]{cho2020search}%
  \BibitemOpen
  \bibfield  {author} {\bibinfo {author} {\bibfnamefont {J.}~\bibnamefont
  {Cho}}, \bibinfo {author} {\bibfnamefont {S.}~\bibnamefont {Ahmed}}, \bibinfo
  {author} {\bibfnamefont {M.}~\bibnamefont {Hilbert}}, \bibinfo {author}
  {\bibfnamefont {B.}~\bibnamefont {Liu}},\ and\ \bibinfo {author}
  {\bibfnamefont {J.}~\bibnamefont {Luu}},\ }\bibfield  {title} {\bibinfo
  {title} {Do search algorithms endanger democracy? an experimental
  investigation of algorithm effects on political polarization},\ }\href@noop
  {} {\bibfield  {journal} {\bibinfo  {journal} {Journal of Broadcasting \&
  Electronic Media}\ }\textbf {\bibinfo {volume} {64}},\ \bibinfo {pages} {150}
  (\bibinfo {year} {2020})}\BibitemShut {NoStop}%
\bibitem [{\citenamefont {Santos}\ \emph {et~al.}(2021)\citenamefont {Santos},
  \citenamefont {Lelkes},\ and\ \citenamefont {Levin}}]{santos2021link}%
  \BibitemOpen
  \bibfield  {author} {\bibinfo {author} {\bibfnamefont {F.~P.}\ \bibnamefont
  {Santos}}, \bibinfo {author} {\bibfnamefont {Y.}~\bibnamefont {Lelkes}},\
  and\ \bibinfo {author} {\bibfnamefont {S.~A.}\ \bibnamefont {Levin}},\
  }\bibfield  {title} {\bibinfo {title} {Link recommendation algorithms and
  dynamics of polarization in online social networks},\ }\href@noop {}
  {\bibfield  {journal} {\bibinfo  {journal} {Proceedings of the National
  Academy of Sciences}\ }\textbf {\bibinfo {volume} {118}},\ \bibinfo {pages}
  {e2102141118} (\bibinfo {year} {2021})}\BibitemShut {NoStop}%
\bibitem [{\citenamefont {de~Arruda}\ \emph {et~al.}(2022)\citenamefont
  {de~Arruda}, \citenamefont {Cardoso}, \citenamefont {de~Arruda},
  \citenamefont {Hern{\'a}ndez}, \citenamefont {da~Fontoura~Costa},\ and\
  \citenamefont {Moreno}}]{de2022modelling}%
  \BibitemOpen
  \bibfield  {author} {\bibinfo {author} {\bibfnamefont {H.~F.}\ \bibnamefont
  {de~Arruda}}, \bibinfo {author} {\bibfnamefont {F.~M.}\ \bibnamefont
  {Cardoso}}, \bibinfo {author} {\bibfnamefont {G.~F.}\ \bibnamefont
  {de~Arruda}}, \bibinfo {author} {\bibfnamefont {A.~R.}\ \bibnamefont
  {Hern{\'a}ndez}}, \bibinfo {author} {\bibfnamefont {L.}~\bibnamefont
  {da~Fontoura~Costa}},\ and\ \bibinfo {author} {\bibfnamefont
  {Y.}~\bibnamefont {Moreno}},\ }\bibfield  {title} {\bibinfo {title}
  {Modelling how social network algorithms can influence opinion
  polarization},\ }\href@noop {} {\bibfield  {journal} {\bibinfo  {journal}
  {Information Sciences}\ }\textbf {\bibinfo {volume} {588}},\ \bibinfo {pages}
  {265} (\bibinfo {year} {2022})}\BibitemShut {NoStop}%
\bibitem [{\citenamefont {Levy}(2021)}]{levy2021social}%
  \BibitemOpen
  \bibfield  {author} {\bibinfo {author} {\bibfnamefont {R.}~\bibnamefont
  {Levy}},\ }\bibfield  {title} {\bibinfo {title} {Social media, news
  consumption, and polarization: Evidence from a field experiment},\
  }\href@noop {} {\bibfield  {journal} {\bibinfo  {journal} {American economic
  review}\ }\textbf {\bibinfo {volume} {111}},\ \bibinfo {pages} {831}
  (\bibinfo {year} {2021})}\BibitemShut {NoStop}%
\bibitem [{\citenamefont {Barber{\'a}}(2020)}]{barbera2020social}%
  \BibitemOpen
  \bibfield  {author} {\bibinfo {author} {\bibfnamefont {P.}~\bibnamefont
  {Barber{\'a}}},\ }\bibfield  {title} {\bibinfo {title} {Social media, echo
  chambers, and political polarization},\ }\href@noop {} {\bibfield  {journal}
  {\bibinfo  {journal} {Social media and democracy: The state of the field,
  prospects for reform}\ ,\ \bibinfo {pages} {34}} (\bibinfo {year}
  {2020})}\BibitemShut {NoStop}%
\bibitem [{\citenamefont {Brady}\ and\ \citenamefont
  {Han}(2006)}]{brady2006polarization}%
  \BibitemOpen
  \bibfield  {author} {\bibinfo {author} {\bibfnamefont {D.~W.}\ \bibnamefont
  {Brady}}\ and\ \bibinfo {author} {\bibfnamefont {H.~C.}\ \bibnamefont
  {Han}},\ }\bibfield  {title} {\bibinfo {title} {Polarization then and now: A
  historical perspective},\ }\href@noop {} {\bibfield  {journal} {\bibinfo
  {journal} {Red and Blue Nation? Characteristics and Causes of America’s
  Polarized Politics}\ }\textbf {\bibinfo {volume} {1}},\ \bibinfo {pages}
  {119} (\bibinfo {year} {2006})}\BibitemShut {NoStop}%
\bibitem [{\citenamefont {Iyengar}\ \emph {et~al.}(2019)\citenamefont
  {Iyengar}, \citenamefont {Lelkes}, \citenamefont {Levendusky}, \citenamefont
  {Malhotra},\ and\ \citenamefont {Westwood}}]{iyengar2019origins}%
  \BibitemOpen
  \bibfield  {author} {\bibinfo {author} {\bibfnamefont {S.}~\bibnamefont
  {Iyengar}}, \bibinfo {author} {\bibfnamefont {Y.}~\bibnamefont {Lelkes}},
  \bibinfo {author} {\bibfnamefont {M.}~\bibnamefont {Levendusky}}, \bibinfo
  {author} {\bibfnamefont {N.}~\bibnamefont {Malhotra}},\ and\ \bibinfo
  {author} {\bibfnamefont {S.~J.}\ \bibnamefont {Westwood}},\ }\bibfield
  {title} {\bibinfo {title} {The origins and consequences of affective
  polarization in the united states},\ }\href@noop {} {\bibfield  {journal}
  {\bibinfo  {journal} {Annual review of political science}\ }\textbf {\bibinfo
  {volume} {22}},\ \bibinfo {pages} {129} (\bibinfo {year} {2019})}\BibitemShut
  {NoStop}%
\bibitem [{\citenamefont {Br{\"u}ggemann}\ and\ \citenamefont
  {Meyer}(2023)}]{bruggemann2023debates}%
  \BibitemOpen
  \bibfield  {author} {\bibinfo {author} {\bibfnamefont {M.}~\bibnamefont
  {Br{\"u}ggemann}}\ and\ \bibinfo {author} {\bibfnamefont {H.}~\bibnamefont
  {Meyer}},\ }\bibfield  {title} {\bibinfo {title} {When debates break apart:
  Discursive polarization as a multi-dimensional divergence emerging in and
  through communication},\ }\href@noop {} {\bibfield  {journal} {\bibinfo
  {journal} {Communication Theory}\ }\textbf {\bibinfo {volume} {33}},\
  \bibinfo {pages} {132} (\bibinfo {year} {2023})}\BibitemShut {NoStop}%
\bibitem [{\citenamefont {Bail}\ \emph {et~al.}(2018)\citenamefont {Bail},
  \citenamefont {Argyle}, \citenamefont {Brown}, \citenamefont {Bumpus},
  \citenamefont {Chen}, \citenamefont {Hunzaker}, \citenamefont {Lee},
  \citenamefont {Mann}, \citenamefont {Merhout},\ and\ \citenamefont
  {Volfovsky}}]{bail2018exposure}%
  \BibitemOpen
  \bibfield  {author} {\bibinfo {author} {\bibfnamefont {C.~A.}\ \bibnamefont
  {Bail}}, \bibinfo {author} {\bibfnamefont {L.~P.}\ \bibnamefont {Argyle}},
  \bibinfo {author} {\bibfnamefont {T.~W.}\ \bibnamefont {Brown}}, \bibinfo
  {author} {\bibfnamefont {J.~P.}\ \bibnamefont {Bumpus}}, \bibinfo {author}
  {\bibfnamefont {H.}~\bibnamefont {Chen}}, \bibinfo {author} {\bibfnamefont
  {M.~F.}\ \bibnamefont {Hunzaker}}, \bibinfo {author} {\bibfnamefont
  {J.}~\bibnamefont {Lee}}, \bibinfo {author} {\bibfnamefont {M.}~\bibnamefont
  {Mann}}, \bibinfo {author} {\bibfnamefont {F.}~\bibnamefont {Merhout}},\ and\
  \bibinfo {author} {\bibfnamefont {A.}~\bibnamefont {Volfovsky}},\ }\bibfield
  {title} {\bibinfo {title} {Exposure to opposing views on social media can
  increase political polarization},\ }\href@noop {} {\bibfield  {journal}
  {\bibinfo  {journal} {Proceedings of the National Academy of Sciences}\
  }\textbf {\bibinfo {volume} {115}},\ \bibinfo {pages} {9216} (\bibinfo {year}
  {2018})}\BibitemShut {NoStop}%
\bibitem [{\citenamefont {Iandoli}\ \emph {et~al.}(2021)\citenamefont
  {Iandoli}, \citenamefont {Primario},\ and\ \citenamefont
  {Zollo}}]{iandoli2021impact}%
  \BibitemOpen
  \bibfield  {author} {\bibinfo {author} {\bibfnamefont {L.}~\bibnamefont
  {Iandoli}}, \bibinfo {author} {\bibfnamefont {S.}~\bibnamefont {Primario}},\
  and\ \bibinfo {author} {\bibfnamefont {G.}~\bibnamefont {Zollo}},\ }\bibfield
   {title} {\bibinfo {title} {The impact of group polarization on the quality
  of online debate in social media: A systematic literature review},\
  }\href@noop {} {\bibfield  {journal} {\bibinfo  {journal} {Technological
  Forecasting and Social Change}\ }\textbf {\bibinfo {volume} {170}},\ \bibinfo
  {pages} {120924} (\bibinfo {year} {2021})}\BibitemShut {NoStop}%
\bibitem [{\citenamefont {Garimella}\ and\ \citenamefont
  {Weber}(2017)}]{Garimella_2017}%
  \BibitemOpen
  \bibfield  {author} {\bibinfo {author} {\bibfnamefont {V.~R.~K.}\
  \bibnamefont {Garimella}}\ and\ \bibinfo {author} {\bibfnamefont
  {I.}~\bibnamefont {Weber}},\ }\bibfield  {title} {\bibinfo {title} {A
  long-term analysis of polarization on twitter},\ }\href
  {https://doi.org/10.1609/icwsm.v11i1.14918} {\bibfield  {journal} {\bibinfo
  {journal} {Proceedings of the International AAAI Conference on Web and Social
  Media}\ }\textbf {\bibinfo {volume} {11}},\ \bibinfo {pages} {528–531}
  (\bibinfo {year} {2017})}\BibitemShut {NoStop}%
\bibitem [{\citenamefont {Waller}\ and\ \citenamefont
  {Anderson}(2021)}]{Waller2021}%
  \BibitemOpen
  \bibfield  {author} {\bibinfo {author} {\bibfnamefont {I.}~\bibnamefont
  {Waller}}\ and\ \bibinfo {author} {\bibfnamefont {A.}~\bibnamefont
  {Anderson}},\ }\bibfield  {title} {\bibinfo {title} {Quantifying social
  organization and political polarization in online platforms},\ }\href
  {https://doi.org/10.1038/s41586-021-04167-x} {\bibfield  {journal} {\bibinfo
  {journal} {Nature}\ }\textbf {\bibinfo {volume} {600}},\ \bibinfo {pages}
  {264–268} (\bibinfo {year} {2021})}\BibitemShut {NoStop}%
\bibitem [{\citenamefont {Esquirol}\ \emph {et~al.}(2024)\citenamefont
  {Esquirol}, \citenamefont {Prignano}, \citenamefont {D{\'\i}az-Guilera},\
  and\ \citenamefont {Cozzo}}]{esquirol2024analyzing}%
  \BibitemOpen
  \bibfield  {author} {\bibinfo {author} {\bibfnamefont {B.}~\bibnamefont
  {Esquirol}}, \bibinfo {author} {\bibfnamefont {L.}~\bibnamefont {Prignano}},
  \bibinfo {author} {\bibfnamefont {A.}~\bibnamefont {D{\'\i}az-Guilera}},\
  and\ \bibinfo {author} {\bibfnamefont {E.}~\bibnamefont {Cozzo}},\ }\bibfield
   {title} {\bibinfo {title} {Analyzing user activity on twitter during
  long-lasting crisis events: a case study of the covid-19 crisis in spain},\
  }\href@noop {} {\bibfield  {journal} {\bibinfo  {journal} {Social Network
  Analysis and Mining}\ }\textbf {\bibinfo {volume} {14}},\ \bibinfo {pages}
  {73} (\bibinfo {year} {2024})}\BibitemShut {NoStop}%
\bibitem [{\citenamefont {Meyer}\ \emph {et~al.}(2023)\citenamefont {Meyer},
  \citenamefont {Peach}, \citenamefont {Guenther}, \citenamefont {Kedar},\ and\
  \citenamefont {Br{\"u}ggemann}}]{meyer2023between}%
  \BibitemOpen
  \bibfield  {author} {\bibinfo {author} {\bibfnamefont {H.}~\bibnamefont
  {Meyer}}, \bibinfo {author} {\bibfnamefont {A.~K.}\ \bibnamefont {Peach}},
  \bibinfo {author} {\bibfnamefont {L.}~\bibnamefont {Guenther}}, \bibinfo
  {author} {\bibfnamefont {H.~E.}\ \bibnamefont {Kedar}},\ and\ \bibinfo
  {author} {\bibfnamefont {M.}~\bibnamefont {Br{\"u}ggemann}},\ }\bibfield
  {title} {\bibinfo {title} {Between calls for action and narratives of denial:
  Climate change attention structures on twitter},\ }\href@noop {} {\bibfield
  {journal} {\bibinfo  {journal} {Media and Communication}\ }\textbf {\bibinfo
  {volume} {11}},\ \bibinfo {pages} {278} (\bibinfo {year} {2023})}\BibitemShut
  {NoStop}%
\bibitem [{\citenamefont {Barber{\'a}}\ \emph {et~al.}(2015)\citenamefont
  {Barber{\'a}}, \citenamefont {Jost}, \citenamefont {Nagler}, \citenamefont
  {Tucker},\ and\ \citenamefont {Bonneau}}]{barbera2015tweeting}%
  \BibitemOpen
  \bibfield  {author} {\bibinfo {author} {\bibfnamefont {P.}~\bibnamefont
  {Barber{\'a}}}, \bibinfo {author} {\bibfnamefont {J.~T.}\ \bibnamefont
  {Jost}}, \bibinfo {author} {\bibfnamefont {J.}~\bibnamefont {Nagler}},
  \bibinfo {author} {\bibfnamefont {J.~A.}\ \bibnamefont {Tucker}},\ and\
  \bibinfo {author} {\bibfnamefont {R.}~\bibnamefont {Bonneau}},\ }\bibfield
  {title} {\bibinfo {title} {Tweeting from left to right: Is online political
  communication more than an echo chamber?},\ }\href@noop {} {\bibfield
  {journal} {\bibinfo  {journal} {Psychological science}\ }\textbf {\bibinfo
  {volume} {26}},\ \bibinfo {pages} {1531} (\bibinfo {year}
  {2015})}\BibitemShut {NoStop}%
\bibitem [{\citenamefont {Bessi}\ \emph {et~al.}(2016)\citenamefont {Bessi},
  \citenamefont {Zollo}, \citenamefont {Del~Vicario}, \citenamefont {Puliga},
  \citenamefont {Scala}, \citenamefont {Caldarelli}, \citenamefont {Uzzi},\
  and\ \citenamefont {Quattrociocchi}}]{bessi2016users}%
  \BibitemOpen
  \bibfield  {author} {\bibinfo {author} {\bibfnamefont {A.}~\bibnamefont
  {Bessi}}, \bibinfo {author} {\bibfnamefont {F.}~\bibnamefont {Zollo}},
  \bibinfo {author} {\bibfnamefont {M.}~\bibnamefont {Del~Vicario}}, \bibinfo
  {author} {\bibfnamefont {M.}~\bibnamefont {Puliga}}, \bibinfo {author}
  {\bibfnamefont {A.}~\bibnamefont {Scala}}, \bibinfo {author} {\bibfnamefont
  {G.}~\bibnamefont {Caldarelli}}, \bibinfo {author} {\bibfnamefont
  {B.}~\bibnamefont {Uzzi}},\ and\ \bibinfo {author} {\bibfnamefont
  {W.}~\bibnamefont {Quattrociocchi}},\ }\bibfield  {title} {\bibinfo {title}
  {Users polarization on facebook and youtube},\ }\href@noop {} {\bibfield
  {journal} {\bibinfo  {journal} {PloS one}\ }\textbf {\bibinfo {volume}
  {11}},\ \bibinfo {pages} {e0159641} (\bibinfo {year} {2016})}\BibitemShut
  {NoStop}%
\bibitem [{\citenamefont {Fernandes}\ \emph {et~al.}(2020)\citenamefont
  {Fernandes}, \citenamefont {Ademir~de Oliveira}, \citenamefont {Motta~de
  Campos},\ and\ \citenamefont {Gomes}}]{fernandes2020political}%
  \BibitemOpen
  \bibfield  {author} {\bibinfo {author} {\bibfnamefont {C.~M.}\ \bibnamefont
  {Fernandes}}, \bibinfo {author} {\bibfnamefont {L.}~\bibnamefont {Ademir~de
  Oliveira}}, \bibinfo {author} {\bibfnamefont {M.}~\bibnamefont {Motta~de
  Campos}},\ and\ \bibinfo {author} {\bibfnamefont {V.~B.}\ \bibnamefont
  {Gomes}},\ }\bibfield  {title} {\bibinfo {title} {Political polarization in
  the brazilian election campaign for the presidency of brazil in 2018: an
  analysis of the social network instagram},\ }\href@noop {} {\bibfield
  {journal} {\bibinfo  {journal} {Int'l J. Soc. Sci. Stud.}\ }\textbf {\bibinfo
  {volume} {8}},\ \bibinfo {pages} {119} (\bibinfo {year} {2020})}\BibitemShut
  {NoStop}%
\bibitem [{\citenamefont {Wu}\ and\ \citenamefont
  {Resnick}(2021)}]{wu2021cross}%
  \BibitemOpen
  \bibfield  {author} {\bibinfo {author} {\bibfnamefont {S.}~\bibnamefont
  {Wu}}\ and\ \bibinfo {author} {\bibfnamefont {P.}~\bibnamefont {Resnick}},\
  }\bibfield  {title} {\bibinfo {title} {Cross-partisan discussions on youtube:
  Conservatives talk to liberals but liberals don't talk to conservatives},\
  }in\ \href@noop {} {\emph {\bibinfo {booktitle} {Proceedings of the
  International AAAI Conference on Web and Social Media}}},\ Vol.~\bibinfo
  {volume} {15}\ (\bibinfo {year} {2021})\ pp.\ \bibinfo {pages}
  {808--819}\BibitemShut {NoStop}%
\bibitem [{\citenamefont {Hohmann}\ \emph {et~al.}(2023)\citenamefont
  {Hohmann}, \citenamefont {Devriendt},\ and\ \citenamefont
  {Coscia}}]{hohmann2023quantifying}%
  \BibitemOpen
  \bibfield  {author} {\bibinfo {author} {\bibfnamefont {M.}~\bibnamefont
  {Hohmann}}, \bibinfo {author} {\bibfnamefont {K.}~\bibnamefont {Devriendt}},\
  and\ \bibinfo {author} {\bibfnamefont {M.}~\bibnamefont {Coscia}},\
  }\bibfield  {title} {\bibinfo {title} {Quantifying ideological polarization
  on a network using generalized euclidean distance},\ }\href@noop {}
  {\bibfield  {journal} {\bibinfo  {journal} {Science Advances}\ }\textbf
  {\bibinfo {volume} {9}},\ \bibinfo {pages} {eabq2044} (\bibinfo {year}
  {2023})}\BibitemShut {NoStop}%
\bibitem [{\citenamefont {Yarchi}\ \emph {et~al.}(2021)\citenamefont {Yarchi},
  \citenamefont {Baden},\ and\ \citenamefont
  {Kligler-Vilenchik}}]{yarchi2021political}%
  \BibitemOpen
  \bibfield  {author} {\bibinfo {author} {\bibfnamefont {M.}~\bibnamefont
  {Yarchi}}, \bibinfo {author} {\bibfnamefont {C.}~\bibnamefont {Baden}},\ and\
  \bibinfo {author} {\bibfnamefont {N.}~\bibnamefont {Kligler-Vilenchik}},\
  }\bibfield  {title} {\bibinfo {title} {Political polarization on the digital
  sphere: A cross-platform, over-time analysis of interactional, positional,
  and affective polarization on social media},\ }\href@noop {} {\bibfield
  {journal} {\bibinfo  {journal} {Political Communication}\ }\textbf {\bibinfo
  {volume} {38}},\ \bibinfo {pages} {98} (\bibinfo {year} {2021})}\BibitemShut
  {NoStop}%
\bibitem [{\citenamefont {Krackhardt}\ and\ \citenamefont
  {Stern}(1988{\natexlab{a}})}]{Krackhardt1988}%
  \BibitemOpen
  \bibfield  {author} {\bibinfo {author} {\bibfnamefont {D.}~\bibnamefont
  {Krackhardt}}\ and\ \bibinfo {author} {\bibfnamefont {R.~N.}\ \bibnamefont
  {Stern}},\ }\bibfield  {title} {\bibinfo {title} {Informal networks and
  organizational crises: An experimental simulation},\ }\href
  {https://doi.org/10.2307/2786835} {\bibfield  {journal} {\bibinfo  {journal}
  {Social Psychology Quarterly}\ }\textbf {\bibinfo {volume} {51}},\ \bibinfo
  {pages} {123} (\bibinfo {year} {1988}{\natexlab{a}})}\BibitemShut {NoStop}%
\bibitem [{\citenamefont {Salloum}\ \emph {et~al.}(2022)\citenamefont
  {Salloum}, \citenamefont {Chen},\ and\ \citenamefont
  {Kivel{\"a}}}]{salloum2022separating}%
  \BibitemOpen
  \bibfield  {author} {\bibinfo {author} {\bibfnamefont {A.}~\bibnamefont
  {Salloum}}, \bibinfo {author} {\bibfnamefont {T.~H.~Y.}\ \bibnamefont
  {Chen}},\ and\ \bibinfo {author} {\bibfnamefont {M.}~\bibnamefont
  {Kivel{\"a}}},\ }\bibfield  {title} {\bibinfo {title} {Separating
  polarization from noise: comparison and normalization of structural
  polarization measures},\ }\href@noop {} {\bibfield  {journal} {\bibinfo
  {journal} {Proceedings of the ACM on human-computer interaction}\ }\textbf
  {\bibinfo {volume} {6}},\ \bibinfo {pages} {1} (\bibinfo {year}
  {2022})}\BibitemShut {NoStop}%
\bibitem [{\citenamefont {Fiorina}\ and\ \citenamefont
  {Abrams}(2008)}]{Fiorina2008}%
  \BibitemOpen
  \bibfield  {author} {\bibinfo {author} {\bibfnamefont {M.~P.}\ \bibnamefont
  {Fiorina}}\ and\ \bibinfo {author} {\bibfnamefont {S.~J.}\ \bibnamefont
  {Abrams}},\ }\bibfield  {title} {\bibinfo {title} {Political polarization in
  the american public},\ }\href
  {https://doi.org/10.1146/annurev.polisci.11.053106.153836} {\bibfield
  {journal} {\bibinfo  {journal} {Annual Review of Political Science}\ }\textbf
  {\bibinfo {volume} {11}},\ \bibinfo {pages} {563–588} (\bibinfo {year}
  {2008})}\BibitemShut {NoStop}%
\bibitem [{\citenamefont {Zollo}\ \emph {et~al.}(2017)\citenamefont {Zollo},
  \citenamefont {Bessi}, \citenamefont {Del~Vicario}, \citenamefont {Scala},
  \citenamefont {Caldarelli}, \citenamefont {Shekhtman}, \citenamefont
  {Havlin},\ and\ \citenamefont {Quattrociocchi}}]{zollo2017debunking}%
  \BibitemOpen
  \bibfield  {author} {\bibinfo {author} {\bibfnamefont {F.}~\bibnamefont
  {Zollo}}, \bibinfo {author} {\bibfnamefont {A.}~\bibnamefont {Bessi}},
  \bibinfo {author} {\bibfnamefont {M.}~\bibnamefont {Del~Vicario}}, \bibinfo
  {author} {\bibfnamefont {A.}~\bibnamefont {Scala}}, \bibinfo {author}
  {\bibfnamefont {G.}~\bibnamefont {Caldarelli}}, \bibinfo {author}
  {\bibfnamefont {L.}~\bibnamefont {Shekhtman}}, \bibinfo {author}
  {\bibfnamefont {S.}~\bibnamefont {Havlin}},\ and\ \bibinfo {author}
  {\bibfnamefont {W.}~\bibnamefont {Quattrociocchi}},\ }\bibfield  {title}
  {\bibinfo {title} {Debunking in a world of tribes},\ }\href@noop {}
  {\bibfield  {journal} {\bibinfo  {journal} {PloS one}\ }\textbf {\bibinfo
  {volume} {12}},\ \bibinfo {pages} {e0181821} (\bibinfo {year}
  {2017})}\BibitemShut {NoStop}%
\bibitem [{\citenamefont {Del~Vicario}\ \emph
  {et~al.}(2016{\natexlab{b}})\citenamefont {Del~Vicario}, \citenamefont
  {Bessi}, \citenamefont {Zollo}, \citenamefont {Petroni}, \citenamefont
  {Scala}, \citenamefont {Caldarelli}, \citenamefont {Stanley},\ and\
  \citenamefont {Quattrociocchi}}]{del2016spreading}%
  \BibitemOpen
  \bibfield  {author} {\bibinfo {author} {\bibfnamefont {M.}~\bibnamefont
  {Del~Vicario}}, \bibinfo {author} {\bibfnamefont {A.}~\bibnamefont {Bessi}},
  \bibinfo {author} {\bibfnamefont {F.}~\bibnamefont {Zollo}}, \bibinfo
  {author} {\bibfnamefont {F.}~\bibnamefont {Petroni}}, \bibinfo {author}
  {\bibfnamefont {A.}~\bibnamefont {Scala}}, \bibinfo {author} {\bibfnamefont
  {G.}~\bibnamefont {Caldarelli}}, \bibinfo {author} {\bibfnamefont {H.~E.}\
  \bibnamefont {Stanley}},\ and\ \bibinfo {author} {\bibfnamefont
  {W.}~\bibnamefont {Quattrociocchi}},\ }\bibfield  {title} {\bibinfo {title}
  {The spreading of misinformation online},\ }\href@noop {} {\bibfield
  {journal} {\bibinfo  {journal} {Proceedings of the national academy of
  Sciences}\ }\textbf {\bibinfo {volume} {113}},\ \bibinfo {pages} {554}
  (\bibinfo {year} {2016}{\natexlab{b}})}\BibitemShut {NoStop}%
\bibitem [{\citenamefont {Soon}\ and\ \citenamefont
  {Goh}(2018)}]{soon2018fake}%
  \BibitemOpen
  \bibfield  {author} {\bibinfo {author} {\bibfnamefont {C.}~\bibnamefont
  {Soon}}\ and\ \bibinfo {author} {\bibfnamefont {S.}~\bibnamefont {Goh}},\
  }\bibfield  {title} {\bibinfo {title} {Fake news, false information and more:
  countering human biases},\ }\href@noop {} {\bibfield  {journal} {\bibinfo
  {journal} {Institute of Policy Studies (IPS) Working Papers}\ }\textbf
  {\bibinfo {volume} {31}} (\bibinfo {year} {2018})}\BibitemShut {NoStop}%
\bibitem [{\citenamefont {Vicario}\ \emph {et~al.}(2019)\citenamefont
  {Vicario}, \citenamefont {Quattrociocchi}, \citenamefont {Scala},\ and\
  \citenamefont {Zollo}}]{vicario2019polarization}%
  \BibitemOpen
  \bibfield  {author} {\bibinfo {author} {\bibfnamefont {M.~D.}\ \bibnamefont
  {Vicario}}, \bibinfo {author} {\bibfnamefont {W.}~\bibnamefont
  {Quattrociocchi}}, \bibinfo {author} {\bibfnamefont {A.}~\bibnamefont
  {Scala}},\ and\ \bibinfo {author} {\bibfnamefont {F.}~\bibnamefont {Zollo}},\
  }\bibfield  {title} {\bibinfo {title} {Polarization and fake news: Early
  warning of potential misinformation targets},\ }\href@noop {} {\bibfield
  {journal} {\bibinfo  {journal} {ACM Transactions on the Web (TWEB)}\ }\textbf
  {\bibinfo {volume} {13}},\ \bibinfo {pages} {1} (\bibinfo {year}
  {2019})}\BibitemShut {NoStop}%
\bibitem [{\citenamefont {Bovet}\ and\ \citenamefont
  {Makse}(2019)}]{bovet2019influence}%
  \BibitemOpen
  \bibfield  {author} {\bibinfo {author} {\bibfnamefont {A.}~\bibnamefont
  {Bovet}}\ and\ \bibinfo {author} {\bibfnamefont {H.~A.}\ \bibnamefont
  {Makse}},\ }\bibfield  {title} {\bibinfo {title} {Influence of fake news in
  twitter during the 2016 us presidential election},\ }\href@noop {} {\bibfield
   {journal} {\bibinfo  {journal} {Nature communications}\ }\textbf {\bibinfo
  {volume} {10}},\ \bibinfo {pages} {7} (\bibinfo {year} {2019})}\BibitemShut
  {NoStop}%
\bibitem [{\citenamefont {Loomba}\ \emph {et~al.}(2021)\citenamefont {Loomba},
  \citenamefont {de~Figueiredo}, \citenamefont {Piatek}, \citenamefont
  {de~Graaf},\ and\ \citenamefont {Larson}}]{Loomba2021}%
  \BibitemOpen
  \bibfield  {author} {\bibinfo {author} {\bibfnamefont {S.}~\bibnamefont
  {Loomba}}, \bibinfo {author} {\bibfnamefont {A.}~\bibnamefont
  {de~Figueiredo}}, \bibinfo {author} {\bibfnamefont {S.~J.}\ \bibnamefont
  {Piatek}}, \bibinfo {author} {\bibfnamefont {K.}~\bibnamefont {de~Graaf}},\
  and\ \bibinfo {author} {\bibfnamefont {H.~J.}\ \bibnamefont {Larson}},\
  }\bibfield  {title} {\bibinfo {title} {Measuring the impact of covid-19
  vaccine misinformation on vaccination intent in the uk and usa},\ }\href
  {https://doi.org/10.1038/s41562-021-01056-1} {\bibfield  {journal} {\bibinfo
  {journal} {Nature Human Behaviour}\ }\textbf {\bibinfo {volume} {5}},\
  \bibinfo {pages} {337–348} (\bibinfo {year} {2021})}\BibitemShut {NoStop}%
\bibitem [{\citenamefont {Van~der Linden}\ \emph {et~al.}(2017)\citenamefont
  {Van~der Linden}, \citenamefont {Leiserowitz}, \citenamefont {Rosenthal},\
  and\ \citenamefont {Maibach}}]{van2017inoculating}%
  \BibitemOpen
  \bibfield  {author} {\bibinfo {author} {\bibfnamefont {S.}~\bibnamefont
  {Van~der Linden}}, \bibinfo {author} {\bibfnamefont {A.}~\bibnamefont
  {Leiserowitz}}, \bibinfo {author} {\bibfnamefont {S.}~\bibnamefont
  {Rosenthal}},\ and\ \bibinfo {author} {\bibfnamefont {E.}~\bibnamefont
  {Maibach}},\ }\bibfield  {title} {\bibinfo {title} {Inoculating the public
  against misinformation about climate change},\ }\href@noop {} {\bibfield
  {journal} {\bibinfo  {journal} {Global challenges}\ }\textbf {\bibinfo
  {volume} {1}},\ \bibinfo {pages} {1600008} (\bibinfo {year}
  {2017})}\BibitemShut {NoStop}%
\bibitem [{\citenamefont {Vosoughi}\ \emph {et~al.}(2018)\citenamefont
  {Vosoughi}, \citenamefont {Roy},\ and\ \citenamefont
  {Aral}}]{vosoughi2018spread}%
  \BibitemOpen
  \bibfield  {author} {\bibinfo {author} {\bibfnamefont {S.}~\bibnamefont
  {Vosoughi}}, \bibinfo {author} {\bibfnamefont {D.}~\bibnamefont {Roy}},\ and\
  \bibinfo {author} {\bibfnamefont {S.}~\bibnamefont {Aral}},\ }\bibfield
  {title} {\bibinfo {title} {The spread of true and false news online},\
  }\href@noop {} {\bibfield  {journal} {\bibinfo  {journal} {science}\ }\textbf
  {\bibinfo {volume} {359}},\ \bibinfo {pages} {1146} (\bibinfo {year}
  {2018})}\BibitemShut {NoStop}%
\bibitem [{\citenamefont {Juul}\ and\ \citenamefont
  {Ugander}(2021)}]{juul2021comparing}%
  \BibitemOpen
  \bibfield  {author} {\bibinfo {author} {\bibfnamefont {J.~L.}\ \bibnamefont
  {Juul}}\ and\ \bibinfo {author} {\bibfnamefont {J.}~\bibnamefont {Ugander}},\
  }\bibfield  {title} {\bibinfo {title} {Comparing information diffusion
  mechanisms by matching on cascade size},\ }\href@noop {} {\bibfield
  {journal} {\bibinfo  {journal} {Proceedings of the National Academy of
  Sciences}\ }\textbf {\bibinfo {volume} {118}},\ \bibinfo {pages}
  {e2100786118} (\bibinfo {year} {2021})}\BibitemShut {NoStop}%
\bibitem [{\citenamefont {Falkenberg}\ \emph {et~al.}(2022)\citenamefont
  {Falkenberg}, \citenamefont {Galeazzi}, \citenamefont {Torricelli},
  \citenamefont {Di~Marco}, \citenamefont {Larosa}, \citenamefont {Sas},
  \citenamefont {Mekacher}, \citenamefont {Pearce}, \citenamefont {Zollo},
  \citenamefont {Quattrociocchi} \emph {et~al.}}]{falkenberg2022growing}%
  \BibitemOpen
  \bibfield  {author} {\bibinfo {author} {\bibfnamefont {M.}~\bibnamefont
  {Falkenberg}}, \bibinfo {author} {\bibfnamefont {A.}~\bibnamefont
  {Galeazzi}}, \bibinfo {author} {\bibfnamefont {M.}~\bibnamefont
  {Torricelli}}, \bibinfo {author} {\bibfnamefont {N.}~\bibnamefont
  {Di~Marco}}, \bibinfo {author} {\bibfnamefont {F.}~\bibnamefont {Larosa}},
  \bibinfo {author} {\bibfnamefont {M.}~\bibnamefont {Sas}}, \bibinfo {author}
  {\bibfnamefont {A.}~\bibnamefont {Mekacher}}, \bibinfo {author}
  {\bibfnamefont {W.}~\bibnamefont {Pearce}}, \bibinfo {author} {\bibfnamefont
  {F.}~\bibnamefont {Zollo}}, \bibinfo {author} {\bibfnamefont
  {W.}~\bibnamefont {Quattrociocchi}}, \emph {et~al.},\ }\bibfield  {title}
  {\bibinfo {title} {Growing polarization around climate change on social
  media},\ }\href@noop {} {\bibfield  {journal} {\bibinfo  {journal} {Nature
  Climate Change}\ }\textbf {\bibinfo {volume} {12}},\ \bibinfo {pages} {1114}
  (\bibinfo {year} {2022})}\BibitemShut {NoStop}%
\bibitem [{\citenamefont {Torricelli}\ \emph {et~al.}(2023)\citenamefont
  {Torricelli}, \citenamefont {Falkenberg}, \citenamefont {Galeazzi},
  \citenamefont {Zollo}, \citenamefont {Quattrociocchi},\ and\ \citenamefont
  {Baronchelli}}]{torricelli2023does}%
  \BibitemOpen
  \bibfield  {author} {\bibinfo {author} {\bibfnamefont {M.}~\bibnamefont
  {Torricelli}}, \bibinfo {author} {\bibfnamefont {M.}~\bibnamefont
  {Falkenberg}}, \bibinfo {author} {\bibfnamefont {A.}~\bibnamefont
  {Galeazzi}}, \bibinfo {author} {\bibfnamefont {F.}~\bibnamefont {Zollo}},
  \bibinfo {author} {\bibfnamefont {W.}~\bibnamefont {Quattrociocchi}},\ and\
  \bibinfo {author} {\bibfnamefont {A.}~\bibnamefont {Baronchelli}},\
  }\bibfield  {title} {\bibinfo {title} {How does extreme weather impact the
  climate change discourse? insights from the twitter discussion on
  hurricanes},\ }\href@noop {} {\bibfield  {journal} {\bibinfo  {journal} {Plos
  Climate}\ }\textbf {\bibinfo {volume} {2}},\ \bibinfo {pages} {e0000277}
  (\bibinfo {year} {2023})}\BibitemShut {NoStop}%
\bibitem [{\citenamefont {Fisher}\ \emph {et~al.}(2013)\citenamefont {Fisher},
  \citenamefont {Waggle},\ and\ \citenamefont {Leifeld}}]{fisher2013does}%
  \BibitemOpen
  \bibfield  {author} {\bibinfo {author} {\bibfnamefont {D.~R.}\ \bibnamefont
  {Fisher}}, \bibinfo {author} {\bibfnamefont {J.}~\bibnamefont {Waggle}},\
  and\ \bibinfo {author} {\bibfnamefont {P.}~\bibnamefont {Leifeld}},\
  }\bibfield  {title} {\bibinfo {title} {Where does political polarization come
  from? locating polarization within the us climate change debate},\
  }\href@noop {} {\bibfield  {journal} {\bibinfo  {journal} {American
  Behavioral Scientist}\ }\textbf {\bibinfo {volume} {57}},\ \bibinfo {pages}
  {70} (\bibinfo {year} {2013})}\BibitemShut {NoStop}%
\bibitem [{\citenamefont {Chen}\ \emph
  {et~al.}(2021{\natexlab{a}})\citenamefont {Chen}, \citenamefont {Salloum},
  \citenamefont {Gronow}, \citenamefont {Ylä-Anttila},\ and\ \citenamefont
  {Kivelä}}]{Chen_2021}%
  \BibitemOpen
  \bibfield  {author} {\bibinfo {author} {\bibfnamefont {T.~H.~Y.}\
  \bibnamefont {Chen}}, \bibinfo {author} {\bibfnamefont {A.}~\bibnamefont
  {Salloum}}, \bibinfo {author} {\bibfnamefont {A.}~\bibnamefont {Gronow}},
  \bibinfo {author} {\bibfnamefont {T.}~\bibnamefont {Ylä-Anttila}},\ and\
  \bibinfo {author} {\bibfnamefont {M.}~\bibnamefont {Kivelä}},\ }\bibfield
  {title} {\bibinfo {title} {Polarization of climate politics results from
  partisan sorting: Evidence from finnish twittersphere},\ }\href
  {https://doi.org/10.1016/j.gloenvcha.2021.102348} {\bibfield  {journal}
  {\bibinfo  {journal} {Global Environmental Change}\ }\textbf {\bibinfo
  {volume} {71}},\ \bibinfo {pages} {102348} (\bibinfo {year}
  {2021}{\natexlab{a}})}\BibitemShut {NoStop}%
\bibitem [{\citenamefont {Chinn}\ \emph {et~al.}(2020)\citenamefont {Chinn},
  \citenamefont {Hart},\ and\ \citenamefont {Soroka}}]{Chinn_2020}%
  \BibitemOpen
  \bibfield  {author} {\bibinfo {author} {\bibfnamefont {S.}~\bibnamefont
  {Chinn}}, \bibinfo {author} {\bibfnamefont {P.~S.}\ \bibnamefont {Hart}},\
  and\ \bibinfo {author} {\bibfnamefont {S.}~\bibnamefont {Soroka}},\
  }\bibfield  {title} {\bibinfo {title} {Politicization and polarization in
  climate change news content, 1985-2017},\ }\href
  {https://doi.org/10.1177/1075547019900290} {\bibfield  {journal} {\bibinfo
  {journal} {Science Communication}\ }\textbf {\bibinfo {volume} {42}},\
  \bibinfo {pages} {112–129} (\bibinfo {year} {2020})}\BibitemShut {NoStop}%
\bibitem [{\citenamefont {Farrell}(2016)}]{farrell2016corporate}%
  \BibitemOpen
  \bibfield  {author} {\bibinfo {author} {\bibfnamefont {J.}~\bibnamefont
  {Farrell}},\ }\bibfield  {title} {\bibinfo {title} {Corporate funding and
  ideological polarization about climate change},\ }\href@noop {} {\bibfield
  {journal} {\bibinfo  {journal} {Proceedings of the National Academy of
  Sciences}\ }\textbf {\bibinfo {volume} {113}},\ \bibinfo {pages} {92}
  (\bibinfo {year} {2016})}\BibitemShut {NoStop}%
\bibitem [{\citenamefont {Rivera}\ and\ \citenamefont
  {Jemielniak}(2024)}]{rivera2024evolution}%
  \BibitemOpen
  \bibfield  {author} {\bibinfo {author} {\bibfnamefont {A.~G.}\ \bibnamefont
  {Rivera}}\ and\ \bibinfo {author} {\bibfnamefont {D.}~\bibnamefont
  {Jemielniak}},\ }\bibfield  {title} {\bibinfo {title} {Evolution of
  international sentiment towards climate change on twitter},\ }\href@noop {}
  {\bibfield  {journal} {\bibinfo  {journal} {Discover Sustainability}\
  }\textbf {\bibinfo {volume} {5}},\ \bibinfo {pages} {163} (\bibinfo {year}
  {2024})}\BibitemShut {NoStop}%
\bibitem [{\citenamefont {Lutzke}\ \emph {et~al.}(2019)\citenamefont {Lutzke},
  \citenamefont {Drummond}, \citenamefont {Slovic},\ and\ \citenamefont
  {{\'A}rvai}}]{lutzke2019priming}%
  \BibitemOpen
  \bibfield  {author} {\bibinfo {author} {\bibfnamefont {L.}~\bibnamefont
  {Lutzke}}, \bibinfo {author} {\bibfnamefont {C.}~\bibnamefont {Drummond}},
  \bibinfo {author} {\bibfnamefont {P.}~\bibnamefont {Slovic}},\ and\ \bibinfo
  {author} {\bibfnamefont {J.}~\bibnamefont {{\'A}rvai}},\ }\bibfield  {title}
  {\bibinfo {title} {Priming critical thinking: Simple interventions limit the
  influence of fake news about climate change on facebook},\ }\href@noop {}
  {\bibfield  {journal} {\bibinfo  {journal} {Global environmental change}\
  }\textbf {\bibinfo {volume} {58}},\ \bibinfo {pages} {101964} (\bibinfo
  {year} {2019})}\BibitemShut {NoStop}%
\bibitem [{\citenamefont {Treen}\ \emph {et~al.}(2020)\citenamefont {Treen},
  \citenamefont {Williams},\ and\ \citenamefont {O'Neill}}]{treen2020online}%
  \BibitemOpen
  \bibfield  {author} {\bibinfo {author} {\bibfnamefont {K.~M.~d.}\
  \bibnamefont {Treen}}, \bibinfo {author} {\bibfnamefont {H.~T.}\ \bibnamefont
  {Williams}},\ and\ \bibinfo {author} {\bibfnamefont {S.~J.}\ \bibnamefont
  {O'Neill}},\ }\bibfield  {title} {\bibinfo {title} {Online misinformation
  about climate change},\ }\href@noop {} {\bibfield  {journal} {\bibinfo
  {journal} {Wiley Interdisciplinary Reviews: Climate Change}\ }\textbf
  {\bibinfo {volume} {11}},\ \bibinfo {pages} {e665} (\bibinfo {year}
  {2020})}\BibitemShut {NoStop}%
\bibitem [{\citenamefont {Sanford}\ \emph {et~al.}(2021)\citenamefont
  {Sanford}, \citenamefont {Painter}, \citenamefont {Yasseri},\ and\
  \citenamefont {Lorimer}}]{sanford2021controversy}%
  \BibitemOpen
  \bibfield  {author} {\bibinfo {author} {\bibfnamefont {M.}~\bibnamefont
  {Sanford}}, \bibinfo {author} {\bibfnamefont {J.}~\bibnamefont {Painter}},
  \bibinfo {author} {\bibfnamefont {T.}~\bibnamefont {Yasseri}},\ and\ \bibinfo
  {author} {\bibfnamefont {J.}~\bibnamefont {Lorimer}},\ }\bibfield  {title}
  {\bibinfo {title} {Controversy around climate change reports: a case study of
  twitter responses to the 2019 ipcc report on land},\ }\href@noop {}
  {\bibfield  {journal} {\bibinfo  {journal} {Climatic change}\ }\textbf
  {\bibinfo {volume} {167}},\ \bibinfo {pages} {59} (\bibinfo {year}
  {2021})}\BibitemShut {NoStop}%
\bibitem [{\citenamefont {Barandiaran}\ \emph {et~al.}(2020)\citenamefont
  {Barandiaran}, \citenamefont {Calleja-L{\'o}pez},\ and\ \citenamefont
  {Cozzo}}]{barandiaran2020defining}%
  \BibitemOpen
  \bibfield  {author} {\bibinfo {author} {\bibfnamefont {X.~E.}\ \bibnamefont
  {Barandiaran}}, \bibinfo {author} {\bibfnamefont {A.}~\bibnamefont
  {Calleja-L{\'o}pez}},\ and\ \bibinfo {author} {\bibfnamefont
  {E.}~\bibnamefont {Cozzo}},\ }\bibfield  {title} {\bibinfo {title} {Defining
  collective identities in technopolitical interaction networks},\ }\href@noop
  {} {\bibfield  {journal} {\bibinfo  {journal} {Frontiers in Psychology}\
  }\textbf {\bibinfo {volume} {11}},\ \bibinfo {pages} {1549} (\bibinfo {year}
  {2020})}\BibitemShut {NoStop}%
\bibitem [{\citenamefont {Van~Baar}\ and\ \citenamefont
  {FeldmanHall}(2022)}]{van2022polarized}%
  \BibitemOpen
  \bibfield  {author} {\bibinfo {author} {\bibfnamefont {J.~M.}\ \bibnamefont
  {Van~Baar}}\ and\ \bibinfo {author} {\bibfnamefont {O.}~\bibnamefont
  {FeldmanHall}},\ }\bibfield  {title} {\bibinfo {title} {The polarized mind in
  context: Interdisciplinary approaches to the psychology of political
  polarization.},\ }\href@noop {} {\bibfield  {journal} {\bibinfo  {journal}
  {American Psychologist}\ }\textbf {\bibinfo {volume} {77}},\ \bibinfo {pages}
  {394} (\bibinfo {year} {2022})}\BibitemShut {NoStop}%
\bibitem [{\citenamefont {Meraz}\ and\ \citenamefont
  {Papacharissi}(2013)}]{meraz2013networked}%
  \BibitemOpen
  \bibfield  {author} {\bibinfo {author} {\bibfnamefont {S.}~\bibnamefont
  {Meraz}}\ and\ \bibinfo {author} {\bibfnamefont {Z.}~\bibnamefont
  {Papacharissi}},\ }\bibfield  {title} {\bibinfo {title} {Networked
  gatekeeping and networked framing on \#egypt},\ }\href@noop {} {\bibfield
  {journal} {\bibinfo  {journal} {The international journal of press/politics}\
  }\textbf {\bibinfo {volume} {18}},\ \bibinfo {pages} {138} (\bibinfo {year}
  {2013})}\BibitemShut {NoStop}%
\bibitem [{\citenamefont {Papacharissi}(2015)}]{papacharissi2015affective}%
  \BibitemOpen
  \bibfield  {author} {\bibinfo {author} {\bibfnamefont {Z.}~\bibnamefont
  {Papacharissi}},\ }\href@noop {} {\emph {\bibinfo {title} {Affective publics:
  Sentiment, technology, and politics}}}\ (\bibinfo  {publisher} {Oxford
  University Press},\ \bibinfo {year} {2015})\BibitemShut {NoStop}%
\bibitem [{\citenamefont {Cota}\ \emph {et~al.}(2019)\citenamefont {Cota},
  \citenamefont {Ferreira}, \citenamefont {Pastor-Satorras},\ and\
  \citenamefont {Starnini}}]{Cota2019}%
  \BibitemOpen
  \bibfield  {author} {\bibinfo {author} {\bibfnamefont {W.}~\bibnamefont
  {Cota}}, \bibinfo {author} {\bibfnamefont {S.~C.}\ \bibnamefont {Ferreira}},
  \bibinfo {author} {\bibfnamefont {R.}~\bibnamefont {Pastor-Satorras}},\ and\
  \bibinfo {author} {\bibfnamefont {M.}~\bibnamefont {Starnini}},\ }\bibfield
  {title} {\bibinfo {title} {Quantifying echo chamber effects in information
  spreading over political communication networks},\ }\href@noop {} {\bibfield
  {journal} {\bibinfo  {journal} {EPJ Data Sci.}\ }\textbf {\bibinfo {volume}
  {8}} (\bibinfo {year} {2019})}\BibitemShut {NoStop}%
\bibitem [{\citenamefont {Falkenberg}\ \emph {et~al.}(2024)\citenamefont
  {Falkenberg}, \citenamefont {Zollo}, \citenamefont {Quattrociocchi},
  \citenamefont {Pfeffer},\ and\ \citenamefont
  {Baronchelli}}]{falkenberg2024patterns}%
  \BibitemOpen
  \bibfield  {author} {\bibinfo {author} {\bibfnamefont {M.}~\bibnamefont
  {Falkenberg}}, \bibinfo {author} {\bibfnamefont {F.}~\bibnamefont {Zollo}},
  \bibinfo {author} {\bibfnamefont {W.}~\bibnamefont {Quattrociocchi}},
  \bibinfo {author} {\bibfnamefont {J.}~\bibnamefont {Pfeffer}},\ and\ \bibinfo
  {author} {\bibfnamefont {A.}~\bibnamefont {Baronchelli}},\ }\bibfield
  {title} {\bibinfo {title} {Patterns of partisan toxicity and engagement
  reveal the common structure of online political communication across
  countries},\ }\href@noop {} {\bibfield  {journal} {\bibinfo  {journal}
  {Nature Communications}\ }\textbf {\bibinfo {volume} {15}},\ \bibinfo {pages}
  {9560} (\bibinfo {year} {2024})}\BibitemShut {NoStop}%
\bibitem [{\citenamefont {Conover}\ \emph {et~al.}(2011)\citenamefont
  {Conover}, \citenamefont {Ratkiewicz}, \citenamefont {Francisco},
  \citenamefont {Gon{\c{c}}alves}, \citenamefont {Menczer},\ and\ \citenamefont
  {Flammini}}]{conover2011political}%
  \BibitemOpen
  \bibfield  {author} {\bibinfo {author} {\bibfnamefont {M.}~\bibnamefont
  {Conover}}, \bibinfo {author} {\bibfnamefont {J.}~\bibnamefont {Ratkiewicz}},
  \bibinfo {author} {\bibfnamefont {M.}~\bibnamefont {Francisco}}, \bibinfo
  {author} {\bibfnamefont {B.}~\bibnamefont {Gon{\c{c}}alves}}, \bibinfo
  {author} {\bibfnamefont {F.}~\bibnamefont {Menczer}},\ and\ \bibinfo {author}
  {\bibfnamefont {A.}~\bibnamefont {Flammini}},\ }\bibfield  {title} {\bibinfo
  {title} {Political polarization on twitter},\ }in\ \href@noop {} {\emph
  {\bibinfo {booktitle} {Proceedings of the international aaai conference on
  web and social media}}},\ Vol.~\bibinfo {volume} {5}\ (\bibinfo {year}
  {2011})\ pp.\ \bibinfo {pages} {89--96}\BibitemShut {NoStop}%
\bibitem [{\citenamefont {Guerra}\ \emph {et~al.}(2013)\citenamefont {Guerra},
  \citenamefont {Meira~Jr}, \citenamefont {Cardie},\ and\ \citenamefont
  {Kleinberg}}]{guerra2013measure}%
  \BibitemOpen
  \bibfield  {author} {\bibinfo {author} {\bibfnamefont {P.}~\bibnamefont
  {Guerra}}, \bibinfo {author} {\bibfnamefont {W.}~\bibnamefont {Meira~Jr}},
  \bibinfo {author} {\bibfnamefont {C.}~\bibnamefont {Cardie}},\ and\ \bibinfo
  {author} {\bibfnamefont {R.}~\bibnamefont {Kleinberg}},\ }\bibfield  {title}
  {\bibinfo {title} {A measure of polarization on social media networks based
  on community boundaries},\ }in\ \href@noop {} {\emph {\bibinfo {booktitle}
  {Proceedings of the international AAAI conference on web and social
  media}}},\ Vol.~\bibinfo {volume} {7}\ (\bibinfo {year} {2013})\ pp.\
  \bibinfo {pages} {215--224}\BibitemShut {NoStop}%
\bibitem [{\citenamefont {Morales}\ \emph {et~al.}(2015)\citenamefont
  {Morales}, \citenamefont {Borondo}, \citenamefont {Losada},\ and\
  \citenamefont {Benito}}]{morales2015measuring}%
  \BibitemOpen
  \bibfield  {author} {\bibinfo {author} {\bibfnamefont {A.~J.}\ \bibnamefont
  {Morales}}, \bibinfo {author} {\bibfnamefont {J.}~\bibnamefont {Borondo}},
  \bibinfo {author} {\bibfnamefont {J.~C.}\ \bibnamefont {Losada}},\ and\
  \bibinfo {author} {\bibfnamefont {R.~M.}\ \bibnamefont {Benito}},\ }\bibfield
   {title} {\bibinfo {title} {Measuring political polarization: Twitter shows
  the two sides of venezuela},\ }\href@noop {} {\bibfield  {journal} {\bibinfo
  {journal} {Chaos: An Interdisciplinary Journal of Nonlinear Science}\
  }\textbf {\bibinfo {volume} {25}} (\bibinfo {year} {2015})}\BibitemShut
  {NoStop}%
\bibitem [{\citenamefont {Garimella}\ \emph {et~al.}(2018)\citenamefont
  {Garimella}, \citenamefont {Morales}, \citenamefont {Gionis},\ and\
  \citenamefont {Mathioudakis}}]{Garimella2018}%
  \BibitemOpen
  \bibfield  {author} {\bibinfo {author} {\bibfnamefont {K.}~\bibnamefont
  {Garimella}}, \bibinfo {author} {\bibfnamefont {G.~D.~F.}\ \bibnamefont
  {Morales}}, \bibinfo {author} {\bibfnamefont {A.}~\bibnamefont {Gionis}},\
  and\ \bibinfo {author} {\bibfnamefont {M.}~\bibnamefont {Mathioudakis}},\
  }\bibfield  {title} {\bibinfo {title} {Quantifying controversy on social
  media},\ }\href {https://doi.org/10.1145/3140565} {\bibfield  {journal}
  {\bibinfo  {journal} {ACM Transactions on Social Computing}\ }\textbf
  {\bibinfo {volume} {1}},\ \bibinfo {pages} {1–27} (\bibinfo {year}
  {2018})}\BibitemShut {NoStop}%
\bibitem [{\citenamefont {Martin-Gutierrez}\ \emph {et~al.}(2023)\citenamefont
  {Martin-Gutierrez}, \citenamefont {Losada},\ and\ \citenamefont
  {Benito}}]{martin2023multipolar}%
  \BibitemOpen
  \bibfield  {author} {\bibinfo {author} {\bibfnamefont {S.}~\bibnamefont
  {Martin-Gutierrez}}, \bibinfo {author} {\bibfnamefont {J.~C.}\ \bibnamefont
  {Losada}},\ and\ \bibinfo {author} {\bibfnamefont {R.~M.}\ \bibnamefont
  {Benito}},\ }\bibfield  {title} {\bibinfo {title} {Multipolar social systems:
  Measuring polarization beyond dichotomous contexts},\ }\href@noop {}
  {\bibfield  {journal} {\bibinfo  {journal} {Chaos, Solitons \& Fractals}\
  }\textbf {\bibinfo {volume} {169}},\ \bibinfo {pages} {113244} (\bibinfo
  {year} {2023})}\BibitemShut {NoStop}%
\bibitem [{\citenamefont {Erikson}(2018)}]{erikson2018relationalism}%
  \BibitemOpen
  \bibfield  {author} {\bibinfo {author} {\bibfnamefont {E.}~\bibnamefont
  {Erikson}},\ }\bibfield  {title} {\bibinfo {title} {Relationalism and social
  networks},\ }\href@noop {} {\bibfield  {journal} {\bibinfo  {journal} {The
  Palgrave handbook of relational sociology}\ ,\ \bibinfo {pages} {271}}
  (\bibinfo {year} {2018})}\BibitemShut {NoStop}%
\bibitem [{\citenamefont {Summers}\ \emph {et~al.}(2023)\citenamefont
  {Summers}, \citenamefont {Brigadir}, \citenamefont {Hames}, \citenamefont
  {Kemenade}, \citenamefont {Binkley}, \citenamefont {Tinafigueroa},
  \citenamefont {Ruest}, \citenamefont {Walmir}, \citenamefont {Chudnov},
  \citenamefont {Thiel},\ and\ \citenamefont {et~al.}}]{Summers2023}%
  \BibitemOpen
  \bibfield  {author} {\bibinfo {author} {\bibfnamefont {E.}~\bibnamefont
  {Summers}}, \bibinfo {author} {\bibfnamefont {I.}~\bibnamefont {Brigadir}},
  \bibinfo {author} {\bibfnamefont {S.}~\bibnamefont {Hames}}, \bibinfo
  {author} {\bibfnamefont {H.~v.}\ \bibnamefont {Kemenade}}, \bibinfo {author}
  {\bibfnamefont {P.}~\bibnamefont {Binkley}}, \bibinfo {author} {\bibnamefont
  {Tinafigueroa}}, \bibinfo {author} {\bibfnamefont {N.}~\bibnamefont {Ruest}},
  \bibinfo {author} {\bibnamefont {Walmir}}, \bibinfo {author} {\bibfnamefont
  {D.}~\bibnamefont {Chudnov}}, \bibinfo {author} {\bibfnamefont
  {D.}~\bibnamefont {Thiel}},\ and\ \bibinfo {author} {\bibnamefont {et~al.}},\
  }\href {https://zenodo.org/records/7799050} {\bibinfo {title} {Docnow/twarc:
  V2.14.0}} (\bibinfo {year} {2023})\BibitemShut {NoStop}%
\bibitem [{\citenamefont {Karypis}\ and\ \citenamefont
  {Kumar}(1998)}]{karypis1998fast}%
  \BibitemOpen
  \bibfield  {author} {\bibinfo {author} {\bibfnamefont {G.}~\bibnamefont
  {Karypis}}\ and\ \bibinfo {author} {\bibfnamefont {V.}~\bibnamefont
  {Kumar}},\ }\bibfield  {title} {\bibinfo {title} {A fast and high quality
  multilevel scheme for partitioning irregular graphs},\ }\href@noop {}
  {\bibfield  {journal} {\bibinfo  {journal} {SIAM Journal on scientific
  Computing}\ }\textbf {\bibinfo {volume} {20}},\ \bibinfo {pages} {359}
  (\bibinfo {year} {1998})}\BibitemShut {NoStop}%
\bibitem [{\citenamefont {Newman}(2006)}]{newman2006modularity}%
  \BibitemOpen
  \bibfield  {author} {\bibinfo {author} {\bibfnamefont {M.~E.}\ \bibnamefont
  {Newman}},\ }\bibfield  {title} {\bibinfo {title} {Modularity and community
  structure in networks},\ }\href@noop {} {\bibfield  {journal} {\bibinfo
  {journal} {Proceedings of the national academy of sciences}\ }\textbf
  {\bibinfo {volume} {103}},\ \bibinfo {pages} {8577} (\bibinfo {year}
  {2006})}\BibitemShut {NoStop}%
\bibitem [{\citenamefont {Krackhardt}\ and\ \citenamefont
  {Stern}(1988{\natexlab{b}})}]{krackhardt1988informal}%
  \BibitemOpen
  \bibfield  {author} {\bibinfo {author} {\bibfnamefont {D.}~\bibnamefont
  {Krackhardt}}\ and\ \bibinfo {author} {\bibfnamefont {R.~N.}\ \bibnamefont
  {Stern}},\ }\bibfield  {title} {\bibinfo {title} {Informal networks and
  organizational crises: An experimental simulation},\ }\href@noop {}
  {\bibfield  {journal} {\bibinfo  {journal} {Social psychology quarterly}\ ,\
  \bibinfo {pages} {123}} (\bibinfo {year} {1988}{\natexlab{b}})}\BibitemShut
  {NoStop}%
\bibitem [{\citenamefont {Chen}\ \emph
  {et~al.}(2021{\natexlab{b}})\citenamefont {Chen}, \citenamefont {Salloum},
  \citenamefont {Gronow}, \citenamefont {Yl{\"a}-Anttila},\ and\ \citenamefont
  {Kivel{\"a}}}]{chen2021polarization}%
  \BibitemOpen
  \bibfield  {author} {\bibinfo {author} {\bibfnamefont {T.~H.~Y.}\
  \bibnamefont {Chen}}, \bibinfo {author} {\bibfnamefont {A.}~\bibnamefont
  {Salloum}}, \bibinfo {author} {\bibfnamefont {A.}~\bibnamefont {Gronow}},
  \bibinfo {author} {\bibfnamefont {T.}~\bibnamefont {Yl{\"a}-Anttila}},\ and\
  \bibinfo {author} {\bibfnamefont {M.}~\bibnamefont {Kivel{\"a}}},\ }\bibfield
   {title} {\bibinfo {title} {Polarization of climate politics results from
  partisan sorting: Evidence from finnish twittersphere},\ }\href@noop {}
  {\bibfield  {journal} {\bibinfo  {journal} {Global Environmental Change}\
  }\textbf {\bibinfo {volume} {71}},\ \bibinfo {pages} {102348} (\bibinfo
  {year} {2021}{\natexlab{b}})}\BibitemShut {NoStop}%
\bibitem [{\citenamefont {Fosdick}\ \emph {et~al.}(2018)\citenamefont
  {Fosdick}, \citenamefont {Larremore}, \citenamefont {Nishimura},\ and\
  \citenamefont {Ugander}}]{fosdick2018configuring}%
  \BibitemOpen
  \bibfield  {author} {\bibinfo {author} {\bibfnamefont {B.~K.}\ \bibnamefont
  {Fosdick}}, \bibinfo {author} {\bibfnamefont {D.~B.}\ \bibnamefont
  {Larremore}}, \bibinfo {author} {\bibfnamefont {J.}~\bibnamefont
  {Nishimura}},\ and\ \bibinfo {author} {\bibfnamefont {J.}~\bibnamefont
  {Ugander}},\ }\bibfield  {title} {\bibinfo {title} {Configuring random graph
  models with fixed degree sequences},\ }\href@noop {} {\bibfield  {journal}
  {\bibinfo  {journal} {Siam Review}\ }\textbf {\bibinfo {volume} {60}},\
  \bibinfo {pages} {315} (\bibinfo {year} {2018})}\BibitemShut {NoStop}%
\bibitem [{\citenamefont {Neff}\ and\ \citenamefont
  {Jemielniak}(2024)}]{neff2024transnational}%
  \BibitemOpen
  \bibfield  {author} {\bibinfo {author} {\bibfnamefont {T.}~\bibnamefont
  {Neff}}\ and\ \bibinfo {author} {\bibfnamefont {D.}~\bibnamefont
  {Jemielniak}},\ }\bibfield  {title} {\bibinfo {title} {How do transnational
  public spheres emerge? comparing news and social media networks during the
  madrid climate talks},\ }\href@noop {} {\bibfield  {journal} {\bibinfo
  {journal} {new media \& society}\ }\textbf {\bibinfo {volume} {26}},\
  \bibinfo {pages} {2066} (\bibinfo {year} {2024})}\BibitemShut {NoStop}%
\bibitem [{med()}]{mediabias}%
  \BibitemOpen
  \href@noop {} {\bibinfo {title} {Media bias/fact check}},\ \bibinfo
  {howpublished} {https://mediabiasfactcheck.com/},\ \bibinfo {note} {accessed:
  2024-06-30}\BibitemShut {NoStop}%
\bibitem [{pol()}]{politicalwatch}%
  \BibitemOpen
  \href@noop {} {\bibinfo {title} {Political watch}},\ \bibinfo {howpublished}
  {https://politicalwatch.es/},\ \bibinfo {note} {accessed:
  2024-06-30}\BibitemShut {NoStop}%
\bibitem [{\citenamefont {Kolic}\ \emph {et~al.}(2022)\citenamefont {Kolic},
  \citenamefont {Aguirre-L{\'o}pez}, \citenamefont {Hern{\'a}ndez-Williams},\
  and\ \citenamefont {Gardu{\~n}o-Hern{\'a}ndez}}]{kolic2022quantifying}%
  \BibitemOpen
  \bibfield  {author} {\bibinfo {author} {\bibfnamefont {B.}~\bibnamefont
  {Kolic}}, \bibinfo {author} {\bibfnamefont {F.}~\bibnamefont
  {Aguirre-L{\'o}pez}}, \bibinfo {author} {\bibfnamefont {S.}~\bibnamefont
  {Hern{\'a}ndez-Williams}},\ and\ \bibinfo {author} {\bibfnamefont
  {G.}~\bibnamefont {Gardu{\~n}o-Hern{\'a}ndez}},\ }\bibfield  {title}
  {\bibinfo {title} {Quantifying the structure of controversial discussions
  with unsupervised methods: a look into the twitter climate change
  conversation},\ }\href@noop {} {\bibfield  {journal} {\bibinfo  {journal}
  {arXiv preprint arXiv:2206.14501}\ } (\bibinfo {year} {2022})}\BibitemShut
  {NoStop}%
\bibitem [{hug()}]{huggingface}%
  \BibitemOpen
  \href@noop {} {\bibinfo {title} {cardiffnlp/twitter-roberta-base-sentiment ·
  {H}ugging {F}ace --- huggingface.co}},\ \bibinfo {howpublished}
  {\url{https://huggingface.co/cardiffnlp/twitter-roberta-base-sentiment}},\
  \bibinfo {note} {[Accessed 24-03-2025]}\BibitemShut {NoStop}%
\bibitem [{\citenamefont {Barbieri}\ \emph {et~al.}(2020)\citenamefont
  {Barbieri}, \citenamefont {Camacho-Collados}, \citenamefont {Neves},\ and\
  \citenamefont {Espinosa-Anke}}]{barbieri2020tweeteval}%
  \BibitemOpen
  \bibfield  {author} {\bibinfo {author} {\bibfnamefont {F.}~\bibnamefont
  {Barbieri}}, \bibinfo {author} {\bibfnamefont {J.}~\bibnamefont
  {Camacho-Collados}}, \bibinfo {author} {\bibfnamefont {L.}~\bibnamefont
  {Neves}},\ and\ \bibinfo {author} {\bibfnamefont {L.}~\bibnamefont
  {Espinosa-Anke}},\ }\bibfield  {title} {\bibinfo {title} {Tweeteval: Unified
  benchmark and comparative evaluation for tweet classification},\ }\href@noop
  {} {\bibfield  {journal} {\bibinfo  {journal} {arXiv preprint
  arXiv:2010.12421}\ } (\bibinfo {year} {2020})}\BibitemShut {NoStop}%
\bibitem [{\citenamefont {Guess}\ \emph {et~al.}(2021)\citenamefont {Guess},
  \citenamefont {Barber{\'a}}, \citenamefont {Munzert},\ and\ \citenamefont
  {Yang}}]{guess2021consequences}%
  \BibitemOpen
  \bibfield  {author} {\bibinfo {author} {\bibfnamefont {A.~M.}\ \bibnamefont
  {Guess}}, \bibinfo {author} {\bibfnamefont {P.}~\bibnamefont {Barber{\'a}}},
  \bibinfo {author} {\bibfnamefont {S.}~\bibnamefont {Munzert}},\ and\ \bibinfo
  {author} {\bibfnamefont {J.}~\bibnamefont {Yang}},\ }\bibfield  {title}
  {\bibinfo {title} {The consequences of online partisan media},\ }\href@noop
  {} {\bibfield  {journal} {\bibinfo  {journal} {Proceedings of the National
  Academy of Sciences}\ }\textbf {\bibinfo {volume} {118}},\ \bibinfo {pages}
  {e2013464118} (\bibinfo {year} {2021})}\BibitemShut {NoStop}%
\bibitem [{\citenamefont {Meyer}\ \emph {et~al.}(2025)\citenamefont {Meyer},
  \citenamefont {Pr{\"o}schel},\ and\ \citenamefont
  {Br{\"u}ggemann}}]{meyer2025disruptive}%
  \BibitemOpen
  \bibfield  {author} {\bibinfo {author} {\bibfnamefont {H.}~\bibnamefont
  {Meyer}}, \bibinfo {author} {\bibfnamefont {L.}~\bibnamefont
  {Pr{\"o}schel}},\ and\ \bibinfo {author} {\bibfnamefont {M.}~\bibnamefont
  {Br{\"u}ggemann}},\ }\bibfield  {title} {\bibinfo {title} {From disruptive
  protests to disrupted networks? analyzing levels of polarization in the
  german twitter/x debates on “fridays for future” and “letzte
  generation”},\ }\href@noop {} {\bibfield  {journal} {\bibinfo  {journal}
  {Social Media+ Society}\ }\textbf {\bibinfo {volume} {11}},\ \bibinfo {pages}
  {20563051251337400} (\bibinfo {year} {2025})}\BibitemShut {NoStop}%
\bibitem [{\citenamefont {Eady}\ \emph {et~al.}(2019)\citenamefont {Eady},
  \citenamefont {Nagler}, \citenamefont {Guess}, \citenamefont {Zilinsky},\
  and\ \citenamefont {Tucker}}]{eady2019many}%
  \BibitemOpen
  \bibfield  {author} {\bibinfo {author} {\bibfnamefont {G.}~\bibnamefont
  {Eady}}, \bibinfo {author} {\bibfnamefont {J.}~\bibnamefont {Nagler}},
  \bibinfo {author} {\bibfnamefont {A.}~\bibnamefont {Guess}}, \bibinfo
  {author} {\bibfnamefont {J.}~\bibnamefont {Zilinsky}},\ and\ \bibinfo
  {author} {\bibfnamefont {J.~A.}\ \bibnamefont {Tucker}},\ }\bibfield  {title}
  {\bibinfo {title} {How many people live in political bubbles on social media?
  evidence from linked survey and twitter data},\ }\href@noop {} {\bibfield
  {journal} {\bibinfo  {journal} {Sage Open}\ }\textbf {\bibinfo {volume}
  {9}},\ \bibinfo {pages} {2158244019832705} (\bibinfo {year}
  {2019})}\BibitemShut {NoStop}%
\bibitem [{\citenamefont {Park}\ and\ \citenamefont
  {Park}(2024)}]{park2024filter}%
  \BibitemOpen
  \bibfield  {author} {\bibinfo {author} {\bibfnamefont {H.~W.}\ \bibnamefont
  {Park}}\ and\ \bibinfo {author} {\bibfnamefont {S.}~\bibnamefont {Park}},\
  }\bibfield  {title} {\bibinfo {title} {The filter bubble generated by
  artificial intelligence algorithms and the network dynamics of collective
  polarization on youtube: the case of south korea},\ }\href@noop {} {\bibfield
   {journal} {\bibinfo  {journal} {Asian Journal of Communication}\ }\textbf
  {\bibinfo {volume} {34}},\ \bibinfo {pages} {195} (\bibinfo {year}
  {2024})}\BibitemShut {NoStop}%
\bibitem [{\citenamefont {Liu}\ \emph {et~al.}(2025)\citenamefont {Liu},
  \citenamefont {Hu}, \citenamefont {Savas}, \citenamefont {Baum},
  \citenamefont {Berinsky}, \citenamefont {Chaney}, \citenamefont {Lucas},
  \citenamefont {Mariman}, \citenamefont {de~Benedictis-Kessner}, \citenamefont
  {Guess} \emph {et~al.}}]{liu2025short}%
  \BibitemOpen
  \bibfield  {author} {\bibinfo {author} {\bibfnamefont {N.}~\bibnamefont
  {Liu}}, \bibinfo {author} {\bibfnamefont {X.~E.}\ \bibnamefont {Hu}},
  \bibinfo {author} {\bibfnamefont {Y.}~\bibnamefont {Savas}}, \bibinfo
  {author} {\bibfnamefont {M.~A.}\ \bibnamefont {Baum}}, \bibinfo {author}
  {\bibfnamefont {A.~J.}\ \bibnamefont {Berinsky}}, \bibinfo {author}
  {\bibfnamefont {A.~J.}\ \bibnamefont {Chaney}}, \bibinfo {author}
  {\bibfnamefont {C.}~\bibnamefont {Lucas}}, \bibinfo {author} {\bibfnamefont
  {R.}~\bibnamefont {Mariman}}, \bibinfo {author} {\bibfnamefont
  {J.}~\bibnamefont {de~Benedictis-Kessner}}, \bibinfo {author} {\bibfnamefont
  {A.~M.}\ \bibnamefont {Guess}}, \emph {et~al.},\ }\bibfield  {title}
  {\bibinfo {title} {Short-term exposure to filter-bubble recommendation
  systems has limited polarization effects: Naturalistic experiments on
  youtube},\ }\href@noop {} {\bibfield  {journal} {\bibinfo  {journal}
  {Proceedings of the National Academy of Sciences}\ }\textbf {\bibinfo
  {volume} {122}},\ \bibinfo {pages} {e2318127122} (\bibinfo {year}
  {2025})}\BibitemShut {NoStop}%
\bibitem [{\citenamefont {Pacheco}\ \emph {et~al.}(2021)\citenamefont
  {Pacheco}, \citenamefont {Hui}, \citenamefont {Torres-Lugo}, \citenamefont
  {Truong}, \citenamefont {Flammini},\ and\ \citenamefont
  {Menczer}}]{pacheco2021uncovering}%
  \BibitemOpen
  \bibfield  {author} {\bibinfo {author} {\bibfnamefont {D.}~\bibnamefont
  {Pacheco}}, \bibinfo {author} {\bibfnamefont {P.-M.}\ \bibnamefont {Hui}},
  \bibinfo {author} {\bibfnamefont {C.}~\bibnamefont {Torres-Lugo}}, \bibinfo
  {author} {\bibfnamefont {B.~T.}\ \bibnamefont {Truong}}, \bibinfo {author}
  {\bibfnamefont {A.}~\bibnamefont {Flammini}},\ and\ \bibinfo {author}
  {\bibfnamefont {F.}~\bibnamefont {Menczer}},\ }\bibfield  {title} {\bibinfo
  {title} {Uncovering coordinated networks on social media: methods and case
  studies},\ }in\ \href@noop {} {\emph {\bibinfo {booktitle} {Proceedings of
  the international AAAI conference on web and social media}}},\ Vol.~\bibinfo
  {volume} {15}\ (\bibinfo {year} {2021})\ pp.\ \bibinfo {pages}
  {455--466}\BibitemShut {NoStop}%
\end{thebibliography}%

\clearpage

\onecolumngrid

\clearpage

\setcounter{figure}{0}
\setcounter{table}{0}
\setcounter{section}{0}

\renewcommand{\thefigure}{S\arabic{figure}}
\renewcommand{\thetable}{S\arabic{table}}
\renewcommand{\thesubsection}{S\arabic{section}}  
\renewcommand{\theequation}{S\arabic{equation}} 
\renewcommand{\thefigure}{S\arabic{figure}}
\renewcommand{\thetable}{S\arabic{table}}

\setcounter{equation}{0}

\appendix

\section{Dataset collection and description}

In this paper, we employed Twarc, a tool for collecting data from Twitter, to delve into user-generated content. By utilising hashtags as search queries, particularly the search and stream API, we collected Twitter conversations, targeting specific themes and topics relevant to our study (COP27, IPCC, Climate Refugees, and Do\~{n}ana). This methodological approach facilitated the systematic collection of rich and diverse data. These are the specific hashtags that we utilised to build the datasets:

\begin{itemize}
  \item \textbf{COP27}: COP27, TogetherForImplementation, COPTV, PRECOP, PRECOP27, LossAndDamage,  Egypt\_COP27, COP27Egypt, ClimateAction, ClimateCrisis, ClimateJustice.
  \item \textbf{IPCC}: ipcc, @ipcc\_ch.
  \item \textbf{Climate Refugees}: climate (refugees OR refugee), climate migration, climate displaced.
  \item \textbf{Do\~{n}ana}: Do\~{n}ana.
\end{itemize}

We also provide the detailed json queries used in the extraction in Listings \ref{lst:cop27}, \ref{lst:ipcc}, \ref{lst:climate-refugees} and \ref{lst:donana}.

\begin{lstlisting}[language=json, caption={COP27 Query Parameters}, label={lst:cop27}]
{
    "start-time": "2022-09-01",
    "end-time": "2022-11-28",
    "keywords": [
        "COP27",
        "TogetherForImplementation",
        "COPTV",
        "PRECOP",
        "PRECOP27",
        "LossAndDamage",
        "Egypt_COP27",
        "COP27Egypt",
        "ClimateAction",
        "ClimateCrisis",
        "ClimateJustice"
    ]
}
\end{lstlisting}

\begin{lstlisting}[language=json, caption={IPCC Query Parameters}, label={lst:ipcc}]
{
    "start-time": "2023-03-18",
    "end-time": "2023-03-27",
    "keywords": [
        "ipcc",
        "@ipcc_ch"
    ]
}
\end{lstlisting}

\begin{lstlisting}[language=json, caption={Climate Refugees Query Parameters}, label={lst:climate-refugees}]
[
    {
        "start-time": "2008-01-01",
        "end-time": "2023-01-01",
        "keywords": [
            "climate (refugees OR refugee)",
            "climate migration",
            "climate displaced"
        ]
    }
]
\end{lstlisting}

\begin{lstlisting}[language=json, caption={Doñana Query Parameters}, label={lst:donana}, basicstyle=\ttfamily, extendedchars=true, literate={ñ}{{\~n}}1]
{
    "start-time": "2019-01-01",
    "end-time": "2023-05-01",
    "keywords": [
        "Doñana"
    ]
}
\end{lstlisting}

The YouTube videos appearing in the Twitter dataset were filtered using specific keywords for each dataset. The keywords used were as follows: \textbf{COP27} – "COP27", "climate", and "COP"; \textbf{IPCC} – "IPCC", "report", and "climate"; \textbf{Climate Refugees} – "Refugee" and "climate"; and \textbf{Do\~{n}ana} – "Do\~{n}ana" and "clima."

\begin{figure}[!htbp]
  \begin{center}
  \includegraphics[width=0.8\textwidth]{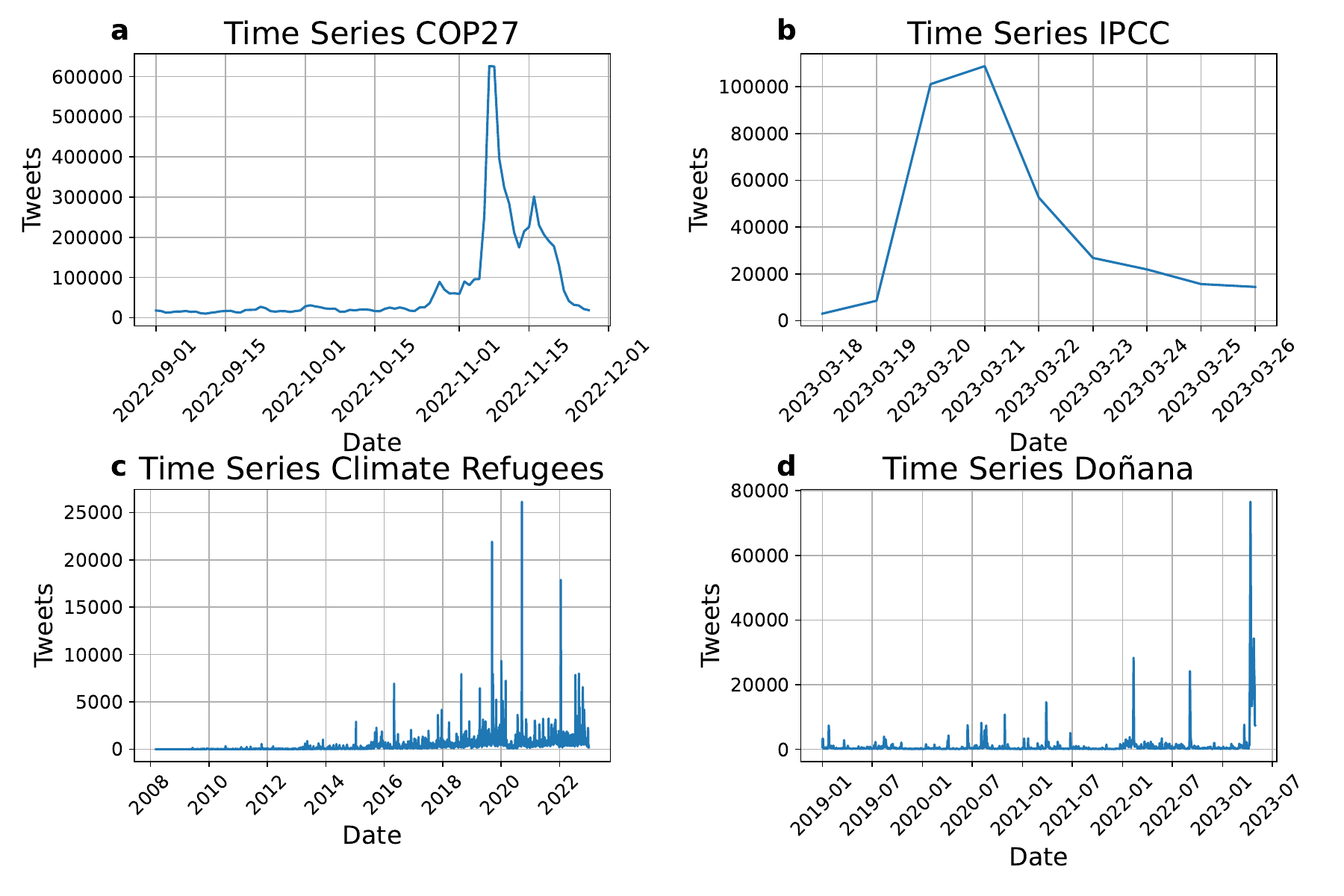}
  \end{center}
  \caption{\textbf{Timeseries of the COP27, IPCC, Climate Refugees and Do\~{n}ana datasets.} Daily timeseries of tweets in the datasets gathered for COP27, IPCC, Climate Refugees and Do\~{n}ana.} \label{timeseries_copipcc}
\end{figure}

\begin{figure}[!htbp]
  \begin{center}
  \includegraphics[width=0.8\textwidth]{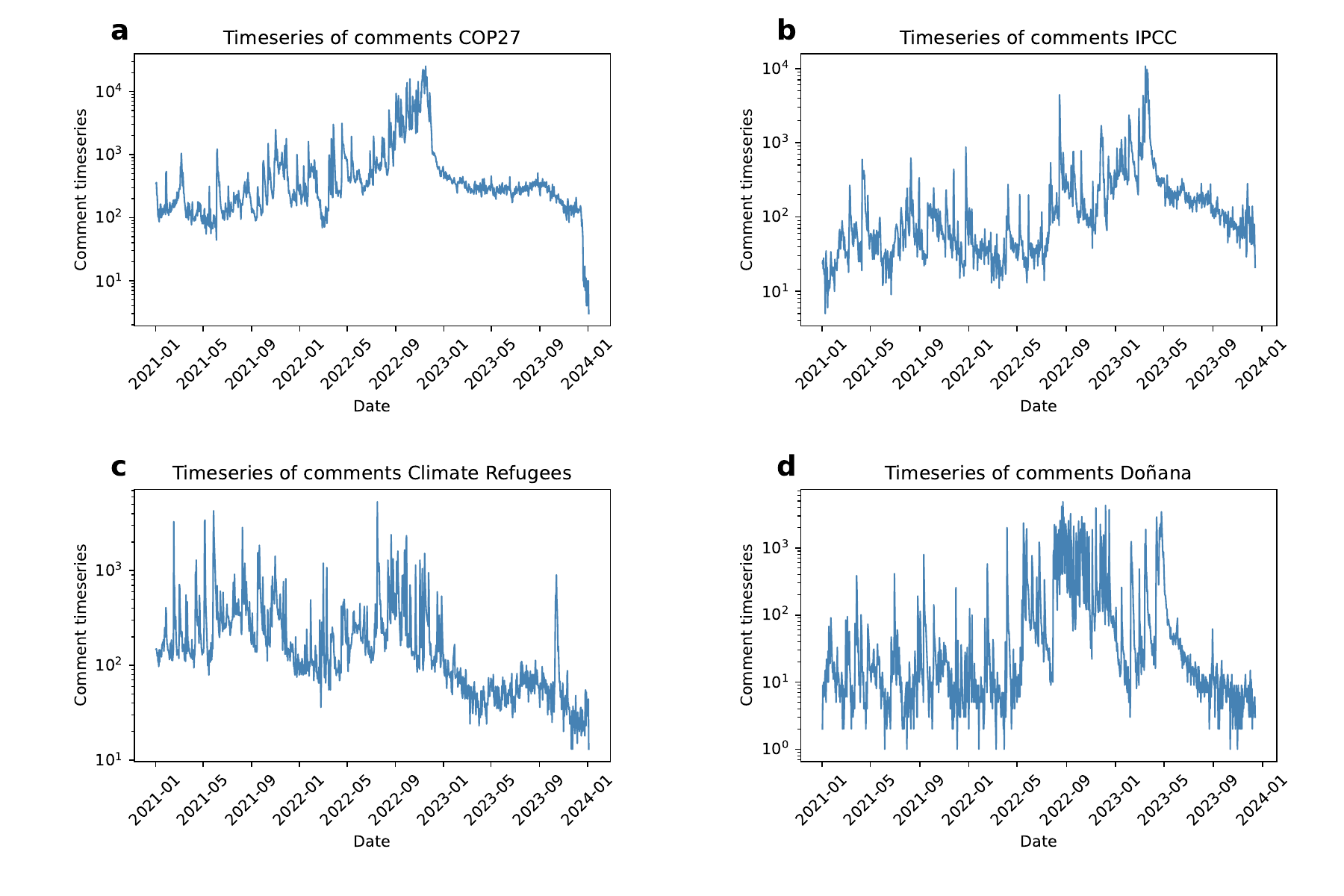}
  \end{center}
  \caption{\textbf{Timeseries of comments in the Youtube datasets.} Number of comments per date in the Youtube datasets of COP27, IPCC, Climate Refugees and Do\~{n}ana. For visualisation purposes, we show the comments after 2021.} \label{timeseries_youtube}
\end{figure}

In Fig. \ref{timeseries_copipcc} and Fig. \ref{timeseries_youtube} we show the timeseries of each dataset for Twitter and YouTube, respectively.

\clearpage

\section{Network analysis and structural polarisation}

\subsection{Influencers by community}

In this section, we identify the top influencers within different communities for each of the key datasets: COP27, IPCC, Climate Refugees, and Doñana. The influencers are ranked by their total inflow of interactions, highlighting those users who have the most significant impact within their respective communities. Tables \ref{table_users_cop27}, \ref{table_users_ipcc}, \ref{table_users_refugees}, and \ref{table_users_donana} present the top 4 most relevant users in the ten largest communities for each dataset.

\begin{table*}[h]
  \centering
  \resizebox{\textwidth}{!}{
    \begin{tabular}{lrrrrrrrr}
\toprule
User & Community & Pagerank & Inflow & In-degree & Outflow & Out-degree & Flow ratio & Degree ratio \\
\midrule
MikeHudema & 0 & 0.004604 & 126323 & 42099 & 12 & 8 & 0.999905 & 0.999810 \\
COP27P & 0 & 0.008392 & 108582 & 45664 & 380 & 142 & 0.996513 & 0.996900 \\
antonioguterres & 0 & 0.006937 & 79507 & 38380 & 5 & 5 & 0.999937 & 0.999870 \\
UN & 0 & 0.003134 & 43045 & 24757 & 211 & 56 & 0.995122 & 0.997743 \\
JamesMelville & 1 & 0.007062 & 49172 & 31076 & 8 & 5 & 0.999837 & 0.999839 \\
DiEM\_25 & 1 & 0.005087 & 28848 & 28496 & 11 & 6 & 0.999619 & 0.999789 \\
BernieSpofforth & 1 & 0.002298 & 26501 & 15218 & 21 & 10 & 0.999208 & 0.999343 \\
DrEliDavid & 1 & 0.000952 & 13377 & 11137 & 0 & 0 & 1.000000 & 1.000000 \\
kimpaim & 2 & 0.001172 & 24266 & 18407 & 2 & 2 & 0.999918 & 0.999891 \\
TerraBrasilnot & 2 & 0.001122 & 16518 & 12962 & 0 & 0 & 1.000000 & 1.000000 \\
VemPraRua\_br & 2 & 0.000899 & 16234 & 13991 & 0 & 0 & 1.000000 & 1.000000 \\
marciolabre & 2 & 0.000652 & 13147 & 13028 & 0 & 0 & 1.000000 & 1.000000 \\
LulaOficial & 3 & 0.010655 & 130064 & 38689 & 24 & 10 & 0.999816 & 0.999742 \\
choquei & 3 & 0.005587 & 25475 & 14922 & 1 & 1 & 0.999961 & 0.999933 \\
Reuters & 3 & 0.001247 & 12269 & 8240 & 67 & 13 & 0.994569 & 0.998425 \\
MarinaSilva & 3 & 0.002625 & 11904 & 8575 & 11 & 6 & 0.999077 & 0.999301 \\
CarolineLucas & 4 & 0.001137 & 33989 & 14931 & 40 & 31 & 0.998825 & 0.997928 \\
NicolaSturgeon & 4 & 0.000565 & 10778 & 4993 & 59 & 25 & 0.994556 & 0.995018 \\
\_david\_ho\_ & 4 & 0.000498 & 6706 & 6246 & 8 & 8 & 0.998808 & 0.998721 \\
johnestevens & 4 & 0.000809 & 6553 & 6154 & 2 & 2 & 0.999695 & 0.999675 \\
RishiSunak & 5 & 0.005127 & 37770 & 24640 & 3 & 3 & 0.999921 & 0.999878 \\
EmmanuelMacron & 5 & 0.003232 & 16585 & 12178 & 2 & 2 & 0.999879 & 0.999836 \\
Poulin2012 & 5 & 0.001389 & 12919 & 12735 & 3 & 3 & 0.999768 & 0.999764 \\
MickaCorreia & 5 & 0.000466 & 3869 & 3054 & 45 & 19 & 0.988503 & 0.993817 \\
POTUS & 6 & 0.004068 & 24808 & 15824 & 1 & 1 & 0.999960 & 0.999937 \\
SpeakerPelosi & 6 & 0.000971 & 9494 & 5361 & 9 & 6 & 0.999053 & 0.998882 \\
JustinTrudeau & 6 & 0.000316 & 5750 & 4247 & 2 & 2 & 0.999652 & 0.999529 \\
washingtonpost & 6 & 0.001279 & 5564 & 4655 & 14 & 2 & 0.997490 & 0.999571 \\
Novozymes & 7 & 0.000194 & 7735 & 3629 & 16 & 10 & 0.997936 & 0.997252 \\
AfDB\_Group & 7 & 0.000376 & 5991 & 2294 & 358 & 104 & 0.943613 & 0.956631 \\
RockefellerFdn & 7 & 0.000438 & 5845 & 3381 & 346 & 118 & 0.944112 & 0.966276 \\
WilliamsRuto & 7 & 0.000691 & 4941 & 3163 & 9 & 4 & 0.998182 & 0.998737 \\
Danielbricen & 8 & 0.001689 & 12626 & 7694 & 4 & 2 & 0.999683 & 0.999740 \\
IvanDuque & 8 & 0.000791 & 9105 & 6317 & 1 & 1 & 0.999890 & 0.999842 \\
alertaLatam & 8 & 0.000337 & 6695 & 4717 & 17 & 3 & 0.997467 & 0.999364 \\
marcelamvzyya & 8 & 0.000140 & 5918 & 3507 & 0 & 0 & 1.000000 & 1.000000 \\
petrogustavo & 9 & 0.005749 & 54899 & 24016 & 23 & 14 & 0.999581 & 0.999417 \\
infopresidencia & 9 & 0.001116 & 18209 & 6575 & 60 & 17 & 0.996716 & 0.997421 \\
FranciaMarquezM & 9 & 0.000399 & 10795 & 5306 & 22 & 17 & 0.997966 & 0.996806 \\
GustavoBolivar & 9 & 0.001154 & 8534 & 6545 & 7 & 2 & 0.999180 & 0.999695 \\
\bottomrule
\end{tabular}

  }
  \caption{Top 4 most relevant users by total inflow in the ten largest communities of the COP27 dataset}
  \label{table_users_cop27}
\end{table*}

\begin{table*}[h]
  \centering
  \resizebox{\textwidth}{!}{
    \begin{tabular}{lrrrrrrrr}
\toprule
User & Community & Pagerank & Inflow & In-degree & Outflow & Out-degree & Flow ratio & Degree ratio \\
\midrule
IPCC\_CH & 0 & 0.173282 & 86702 & 48491 & 22 & 12 & 0.999746 & 0.999753 \\
ed\_hawkins & 0 & 0.002570 & 3194 & 2119 & 60 & 29 & 0.981561 & 0.986499 \\
MrMatthewTodd & 0 & 0.001954 & 2766 & 2474 & 54 & 37 & 0.980851 & 0.985265 \\
CharlieJGardner & 0 & 0.002120 & 2720 & 2297 & 5 & 4 & 0.998165 & 0.998262 \\
uksciencechief & 1 & 0.004402 & 12700 & 5452 & 1 & 1 & 0.999921 & 0.999817 \\
BernieSpofforth & 1 & 0.005600 & 5735 & 4972 & 0 & 0 & 1.000000 & 1.000000 \\
fmeeus1 & 1 & 0.004689 & 3973 & 1800 & 15 & 8 & 0.996239 & 0.995575 \\
Lauratobin1 & 1 & 0.000858 & 3067 & 1287 & 4 & 3 & 0.998697 & 0.997674 \\
jasonhickel & 2 & 0.008178 & 4249 & 4120 & 2 & 2 & 0.999530 & 0.999515 \\
greenpeace\_esp & 2 & 0.003291 & 2092 & 1986 & 28 & 11 & 0.986792 & 0.994492 \\
millares & 2 & 0.001390 & 1788 & 1786 & 5 & 4 & 0.997211 & 0.997765 \\
ionebelarra & 2 & 0.000286 & 592 & 444 & 3 & 3 & 0.994958 & 0.993289 \\
GretaThunberg & 3 & 0.007815 & 8539 & 8115 & 0 & 0 & 1.000000 & 1.000000 \\
CarolineLucas & 3 & 0.003437 & 5833 & 3971 & 5 & 5 & 0.999144 & 0.998742 \\
paulpowlesland & 3 & 0.001666 & 1189 & 1179 & 0 & 0 & 1.000000 & 1.000000 \\
guardian & 3 & 0.000646 & 1016 & 915 & 0 & 0 & 1.000000 & 1.000000 \\
KHayhoe & 4 & 0.006960 & 9241 & 7626 & 82 & 39 & 0.991205 & 0.994912 \\
SenSchumer & 4 & 0.001571 & 4078 & 1973 & 2 & 2 & 0.999510 & 0.998987 \\
curious\_founder & 4 & 0.004436 & 2067 & 1779 & 9 & 5 & 0.995665 & 0.997197 \\
remblance\_erin & 4 & 0.001244 & 1380 & 1325 & 9 & 9 & 0.993521 & 0.993253 \\
rahmstorf & 5 & 0.003506 & 3135 & 2032 & 35 & 17 & 0.988959 & 0.991703 \\
micha\_bloss & 5 & 0.007133 & 2687 & 2276 & 3 & 2 & 0.998885 & 0.999122 \\
Martin\_Bethke & 5 & 0.001183 & 2124 & 1222 & 71 & 40 & 0.967654 & 0.968304 \\
HolzheuStefan & 5 & 0.001040 & 1787 & 1241 & 26 & 13 & 0.985659 & 0.989633 \\
antonioguterres & 6 & 0.015073 & 6512 & 4285 & 1 & 1 & 0.999846 & 0.999767 \\
UN & 6 & 0.007672 & 4093 & 2753 & 8 & 2 & 0.998049 & 0.999274 \\
BarackObama & 6 & 0.002001 & 1794 & 1775 & 0 & 0 & 1.000000 & 1.000000 \\
HillaryClinton & 6 & 0.001469 & 1398 & 1375 & 0 & 0 & 1.000000 & 1.000000 \\
JKSteinberger & 7 & 0.006622 & 2121 & 1930 & 76 & 26 & 0.965407 & 0.986708 \\
frantecol & 7 & 0.001867 & 2101 & 2052 & 6 & 4 & 0.997152 & 0.998054 \\
jennystojkovic & 7 & 0.000888 & 1519 & 1324 & 8 & 4 & 0.994761 & 0.996988 \\
NLRebellion & 7 & 0.000375 & 571 & 407 & 26 & 17 & 0.956449 & 0.959906 \\
MikeHudema & 8 & 0.003279 & 5639 & 3757 & 1 & 1 & 0.999823 & 0.999734 \\
s\_guilbeault & 8 & 0.001042 & 2369 & 1082 & 4 & 2 & 0.998314 & 0.998155 \\
tveitdal & 8 & 0.000368 & 561 & 409 & 35 & 7 & 0.941275 & 0.983173 \\
gmbutts & 8 & 0.000383 & 512 & 406 & 4 & 3 & 0.992248 & 0.992665 \\
DrLuetke & 9 & 0.000737 & 741 & 719 & 0 & 0 & 1.000000 & 1.000000 \\
Axel\_Bojanowski & 9 & 0.001738 & 561 & 406 & 11 & 3 & 0.980769 & 0.992665 \\
welt & 9 & 0.000674 & 543 & 471 & 2 & 1 & 0.996330 & 0.997881 \\
Schlautropf & 9 & 0.000378 & 419 & 375 & 14 & 12 & 0.967667 & 0.968992 \\
\bottomrule
\end{tabular}

  }
  \caption{Top 4 most relevant users by total inflow in the ten largest communities of the IPCC dataset}
  \label{table_users_ipcc}
\end{table*}

\begin{table*}[h]
  \centering
  \resizebox{\textwidth}{!}{
    \begin{tabular}{lrrrrrrrr}
\toprule
User & Community & Pagerank & Inflow & In-degree & Outflow & Out-degree & Flow ratio & Degree ratio \\
\midrule
AOC & 0 & 0.012350 & 35741 & 31238 & 9 & 7 & 0.999748 & 0.999776 \\
BernieSanders & 0 & 0.002378 & 12756 & 11197 & 2 & 2 & 0.999843 & 0.999821 \\
propublica & 0 & 0.003287 & 12325 & 8952 & 87 & 33 & 0.992991 & 0.996327 \\
POTUS & 0 & 0.003139 & 8389 & 7316 & 0 & 0 & 1.000000 & 1.000000 \\
MikeHudema & 1 & 0.002022 & 12798 & 10474 & 13 & 8 & 0.998985 & 0.999237 \\
AssaadRazzouk & 1 & 0.002807 & 10615 & 7388 & 44 & 31 & 0.995872 & 0.995822 \\
PaulEDawson & 1 & 0.000761 & 10559 & 5675 & 15 & 8 & 0.998581 & 0.998592 \\
ClimateBen & 1 & 0.003454 & 10327 & 9121 & 135 & 62 & 0.987096 & 0.993248 \\
zoenone0none & 2 & 0.011598 & 24876 & 24855 & 3 & 2 & 0.999879 & 0.999920 \\
studentactivism & 2 & 0.011936 & 19994 & 19861 & 2 & 2 & 0.999900 & 0.999899 \\
ajplus & 2 & 0.009282 & 6795 & 5790 & 18 & 6 & 0.997358 & 0.998965 \\
femalekissinger & 2 & 0.001781 & 5197 & 5195 & 1 & 1 & 0.999808 & 0.999808 \\
Refugees & 3 & 0.009205 & 29470 & 15697 & 197 & 86 & 0.993360 & 0.994551 \\
UN & 3 & 0.006119 & 16457 & 12513 & 91 & 33 & 0.994501 & 0.997370 \\
UNmigration & 3 & 0.001623 & 8475 & 4151 & 278 & 113 & 0.968239 & 0.973499 \\
antonioguterres & 3 & 0.002124 & 6676 & 5626 & 0 & 0 & 1.000000 & 1.000000 \\
nytimes & 4 & 0.009325 & 21229 & 18971 & 51 & 16 & 0.997603 & 0.999157 \\
guardian & 4 & 0.002638 & 7924 & 6603 & 9 & 4 & 0.998865 & 0.999395 \\
LeoDiCaprio & 4 & 0.001235 & 5447 & 4897 & 9 & 9 & 0.998350 & 0.998166 \\
NatGeo & 4 & 0.001027 & 5283 & 4718 & 0 & 0 & 1.000000 & 1.000000 \\
MrAhmednurAli & 5 & 0.002175 & 11799 & 11703 & 1 & 1 & 0.999915 & 0.999915 \\
jeremycorbyn & 5 & 0.003088 & 6753 & 5402 & 3 & 2 & 0.999556 & 0.999630 \\
LaylaMoran & 5 & 0.000665 & 3725 & 3711 & 0 & 0 & 1.000000 & 1.000000 \\
bmay & 5 & 0.000566 & 3383 & 3029 & 15 & 13 & 0.995586 & 0.995726 \\
BreitbartNews & 6 & 0.000559 & 3076 & 2481 & 0 & 0 & 1.000000 & 1.000000 \\
MaximeBernier & 6 & 0.000966 & 2799 & 2022 & 17 & 11 & 0.993963 & 0.994589 \\
PrisonPlanet & 6 & 0.001274 & 1912 & 1711 & 1 & 1 & 0.999477 & 0.999416 \\
Liz\_Wheeler & 6 & 0.000363 & 1810 & 1787 & 0 & 0 & 1.000000 & 1.000000 \\
Kon\_\_K & 7 & 0.001581 & 9741 & 6362 & 46 & 24 & 0.995300 & 0.996242 \\
bruce\_haigh & 7 & 0.000417 & 4512 & 2721 & 115 & 86 & 0.975146 & 0.969362 \\
simonahac & 7 & 0.000289 & 2775 & 2060 & 58 & 47 & 0.979527 & 0.977693 \\
abcnews & 7 & 0.000239 & 2051 & 1671 & 2 & 2 & 0.999026 & 0.998805 \\
JustinTrudeau & 8 & 0.003543 & 15733 & 15066 & 5 & 5 & 0.999682 & 0.999668 \\
ianbremmer & 8 & 0.000955 & 4529 & 4469 & 9 & 4 & 0.998017 & 0.999106 \\
cjwerleman & 8 & 0.000434 & 1052 & 1017 & 3 & 3 & 0.997156 & 0.997059 \\
MusaNV18 & 8 & 0.000294 & 931 & 877 & 4 & 3 & 0.995722 & 0.996591 \\
ndcodeine & 9 & 0.006104 & 24228 & 24205 & 0 & 0 & 1.000000 & 1.000000 \\
RoadtoMUT2022 & 9 & 0.000038 & 157 & 157 & 0 & 0 & 1.000000 & 1.000000 \\
chloeprw & 9 & 0.000016 & 71 & 71 & 0 & 0 & 1.000000 & 1.000000 \\
themomentumco & 9 & 0.000009 & 40 & 40 & 0 & 0 & 1.000000 & 1.000000 \\
\bottomrule
\end{tabular}

  }
  \caption{Top 4 most relevant users by total inflow in the ten largest communities of the Climate Refugees dataset}
  \label{table_users_refugees}
\end{table*}

\begin{table*}[h]
  \centering
  \resizebox{\textwidth}{!}{
    \begin{tabular}{lrrrrrrrr}
\toprule
User & Community & Pagerank & Inflow & In-degree & Outflow & Out-degree & Flow ratio & Degree ratio \\
\midrule
JuanMa\_Moreno & 0 & 0.014113 & 37135 & 14986 & 25 & 8 & 0.999327 & 0.999466 \\
alfonso\_ussia & 0 & 0.003464 & 17027 & 9717 & 41 & 30 & 0.997598 & 0.996922 \\
okdiario & 0 & 0.003296 & 13310 & 5792 & 44 & 11 & 0.996705 & 0.998104 \\
AndaluciaJunta & 0 & 0.011000 & 11794 & 6983 & 386 & 66 & 0.968309 & 0.990637 \\
WWFespana & 1 & 0.006797 & 20837 & 8588 & 1292 & 201 & 0.941615 & 0.977131 \\
juralde & 1 & 0.005168 & 19984 & 9507 & 363 & 94 & 0.982160 & 0.990209 \\
Santi\_MBarajas & 1 & 0.003885 & 19234 & 7772 & 270 & 65 & 0.986157 & 0.991706 \\
SEO\_BirdLife & 1 & 0.004740 & 10059 & 4557 & 499 & 138 & 0.952737 & 0.970607 \\
FonsiLoaiza & 2 & 0.009273 & 13695 & 9416 & 5 & 1 & 0.999635 & 0.999894 \\
JA\_DelgadoRamos & 2 & 0.002128 & 10446 & 5413 & 23 & 12 & 0.997803 & 0.997788 \\
Toni\_Valero & 2 & 0.002191 & 9947 & 4700 & 64 & 27 & 0.993607 & 0.994288 \\
JavierArocaA & 2 & 0.001303 & 8632 & 4994 & 84 & 30 & 0.990363 & 0.994029 \\
TeresaRodr\_ & 3 & 0.002844 & 10018 & 5616 & 209 & 73 & 0.979564 & 0.987168 \\
Ainhoasauria & 3 & 0.003481 & 5677 & 5676 & 0 & 0 & 1.000000 & 1.000000 \\
MiguelMorenatti & 3 & 0.001572 & 4024 & 3916 & 5 & 4 & 0.998759 & 0.998980 \\
lavozdelsures & 3 & 0.000936 & 3343 & 2414 & 475 & 191 & 0.875589 & 0.926679 \\
carlosromeroco & 4 & 0.016908 & 51995 & 16142 & 4210 & 302 & 0.925096 & 0.981635 \\
PartidoPACMA & 4 & 0.004616 & 10101 & 8238 & 32 & 15 & 0.996842 & 0.998182 \\
fromerofoto & 4 & 0.001280 & 4799 & 1132 & 887 & 90 & 0.844003 & 0.926350 \\
guardiacivil & 4 & 0.004166 & 3612 & 2798 & 7 & 2 & 0.998066 & 0.999286 \\
sanchezcastejon & 5 & 0.032978 & 67372 & 26401 & 2 & 2 & 0.999970 & 0.999924 \\
eldiarioes & 5 & 0.008312 & 24556 & 11979 & 205 & 32 & 0.991721 & 0.997336 \\
PSOE & 5 & 0.003014 & 22664 & 7498 & 47 & 15 & 0.997931 & 0.998003 \\
iescolar & 5 & 0.006451 & 21280 & 10952 & 6 & 3 & 0.999718 & 0.999726 \\
jcanadellb & 6 & 0.000574 & 1478 & 1084 & 15 & 8 & 0.989953 & 0.992674 \\
sninobecerra & 6 & 0.000243 & 789 & 739 & 0 & 0 & 1.000000 & 1.000000 \\
puntocriticoDH & 6 & 0.000112 & 675 & 201 & 358 & 16 & 0.653437 & 0.926267 \\
jm\_clavero & 6 & 0.000113 & 528 & 439 & 1 & 1 & 0.998110 & 0.997727 \\
ClimateBen & 7 & 0.002295 & 798 & 740 & 2 & 1 & 0.997500 & 0.998650 \\
WWF\_Deutschland & 7 & 0.000356 & 246 & 199 & 14 & 5 & 0.946154 & 0.975490 \\
iacbe & 7 & 0.000101 & 173 & 108 & 13 & 7 & 0.930108 & 0.939130 \\
AlanDaviesbirds & 7 & 0.000195 & 158 & 107 & 22 & 10 & 0.877778 & 0.914530 \\
ParlamentoAnd & 8 & 0.002065 & 2855 & 2335 & 24 & 10 & 0.991664 & 0.995736 \\
CiudadanosCs & 8 & 0.000200 & 724 & 501 & 24 & 14 & 0.967914 & 0.972816 \\
Cs\_Andalucia & 8 & 0.000254 & 557 & 440 & 67 & 32 & 0.892628 & 0.932203 \\
AsocParqueDunar & 8 & 0.000136 & 430 & 114 & 101 & 26 & 0.809793 & 0.814286 \\
GotTalentES & 9 & 0.001144 & 1618 & 1617 & 0 & 0 & 1.000000 & 1.000000 \\
proalmerienses & 9 & 0.000008 & 22 & 21 & 0 & 0 & 1.000000 & 1.000000 \\
RaholaOficial & 9 & 0.000007 & 22 & 13 & 16 & 16 & 0.578947 & 0.448276 \\
chiquisanz & 9 & 0.000001 & 10 & 5 & 308 & 60 & 0.031447 & 0.076923 \\
\bottomrule
\end{tabular}

  }
  \caption{Top 4 most relevant users by total inflow in the ten largest communities of the Do\~{n}ana dataset}
  \label{table_users_donana}
\end{table*}

\clearpage

\subsection{Language analysis}
We present the most frequently used languages across the largest communities for each dataset in Tables \ref{table_languages_cop27}, \ref{table_languages_ipcc}, \ref{table_languages_refugees}, and \ref{table_languages_donana}, which correspond to the COP27, IPCC, Climate Refugees, and Doñana datasets, respectively.

\begin{table*}[h!]
  \centering
  \resizebox{\textwidth}{!}{
    \begin{tabular}{ccccc}
\toprule
Community & Top 1 language & Top 1 language ratio & Top 2 language & Top 2 language ratio \\
\midrule
0 & en & 0.911 & de & 0.051 \\
1 & en & 0.860 & de & 0.060 \\
2 & pt & 0.980 & es & 0.011 \\
3 & pt & 0.927 & en & 0.047 \\
4 & en & 0.994 & fr & 0.001 \\
5 & fr & 0.871 & en & 0.113 \\
6 & en & 0.917 & pl & 0.021 \\
7 & en & 0.911 & fr & 0.026 \\
8 & es & 0.945 & en & 0.042 \\
9 & es & 0.926 & en & 0.061 \\
\bottomrule
\end{tabular}

  }
  \caption{Most used languages in the ten largest communities of the COP27 dataset}
  \label{table_languages_cop27}
\end{table*}

\begin{table*}[h!]
  \centering
  \resizebox{\textwidth}{!}{
    \begin{tabular}{ccccc}
\toprule
Community & Top 1 language & Top 1 language ratio & Top 2 language & Top 2 language ratio \\
\midrule
0 & en & 0.985 & de & 0.003 \\
1 & en & 0.773 & nl & 0.180 \\
2 & es & 0.699 & en & 0.261 \\
3 & en & 0.993 & de & 0.005 \\
4 & en & 0.985 & de & 0.005 \\
5 & de & 0.930 & en & 0.064 \\
6 & en & 0.986 & es & 0.004 \\
7 & en & 0.647 & nl & 0.323 \\
8 & en & 0.791 & sv & 0.192 \\
9 & de & 0.955 & en & 0.043 \\
\bottomrule
\end{tabular}

  }
  \caption{Most used languages in the ten largest communities of the IPCC dataset}
  \label{table_languages_ipcc}
\end{table*}

\newpage

\begin{table*}[h!]
  \centering
  \resizebox{\textwidth}{!}{
    \begin{tabular}{ccccc}
\toprule
Community & Top 1 language & Top 1 language ratio & Top 2 language & Top 2 language ratio \\
\midrule
0 & en & 0.998 & ro & 0.001 \\
1 & en & 0.987 & fi & 0.003 \\
2 & en & 0.990 & id & 0.009 \\
3 & en & 0.990 & es & 0.003 \\
4 & en & 0.991 & it & 0.004 \\
5 & en & 0.998 & it & 0.001 \\
6 & en & 0.990 & nl & 0.004 \\
7 & en & 0.997 & fr & 0.002 \\
8 & en & 0.996 & it & 0.001 \\
9 & th & 0.991 & en & 0.009 \\
\bottomrule
\end{tabular}

  }
  \caption{Most used languages in the ten largest communities of the Climate Refugees dataset}
  \label{table_languages_refugees}
\end{table*}

\begin{table*}[h!]
  \centering
  \resizebox{\textwidth}{!}{
    \begin{tabular}{ccccc}
\toprule
Community & Top 1 language & Top 1 language ratio & Top 2 language & Top 2 language ratio \\
\midrule
0 & es & 0.996 & pt & 0.001 \\
1 & es & 0.907 & en & 0.077 \\
2 & es & 0.993 & ca & 0.003 \\
3 & es & 0.969 & pt & 0.015 \\
4 & es & 0.952 & pt & 0.024 \\
5 & es & 0.991 & en & 0.005 \\
6 & es & 0.635 & ca & 0.295 \\
7 & en & 0.609 & de & 0.283 \\
8 & es & 0.978 & en & 0.011 \\
9 & es & 0.998 & pt & 0.001 \\
\bottomrule
\end{tabular}

  }
  \caption{Most used languages in the ten largest communities of the Do\~{n}ana dataset}
  \label{table_languages_donana}
\end{table*}

The language distribution across the largest communities in the COP27, IPCC, Climate Refugees, and Doñana datasets reveals notable patterns. These patterns not only reflect the diversity of linguistic representation within each dataset but also provide insights into the global reach.

\clearpage

\subsection{Analysis of global structural polarisation}

To assess the degree of polarisation within the online discourse surrounding key climate-related topics, we calculate the standardised polarisation in the COP27, IPCC, Climate Refugees, and Do\~{n}ana networks, following the method outlined in \cite{salloum2022separating}. The polarisation is standardised by comparing the observed polarisation \(\Phi(G)\) against the expected polarisation \(\Phi(G_{CM})\) from a corresponding null model, resulting in a z-score that accounts for network-specific variations. The formula for this calculation is given by:

\begin{equation} 
\hat{\Phi}_z(G) = \frac{\Phi(G) - \Phi(G_{CM})}{\sqrt{ \langle \Phi(G_{CM})^2 \rangle - \langle \Phi(G_{CM}) \rangle^2 }}.
\end{equation} 

\begin{figure}[!htbp]
  \begin{center}
  \includegraphics[width=0.8\textwidth]{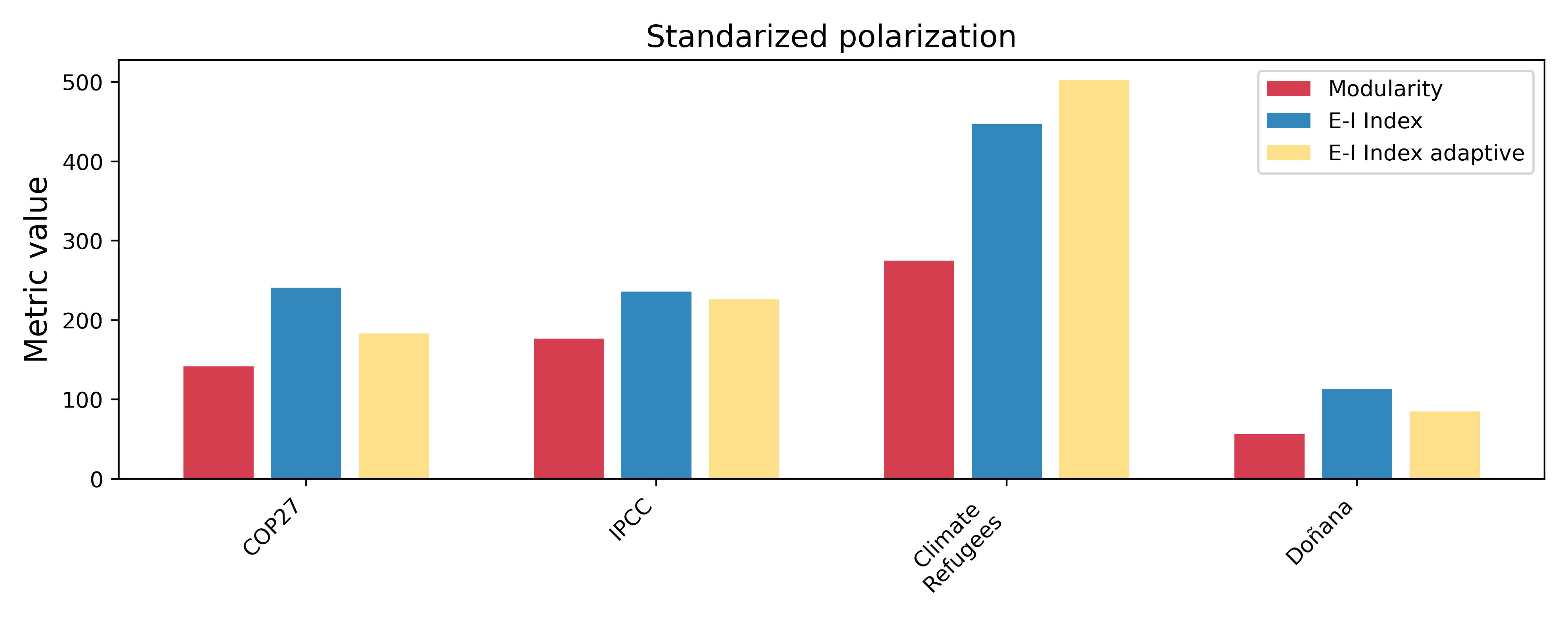}
  \end{center}
\caption{\textbf{Standardised polarisation in the Twitter dataset.} Standardized measures of polarisation in the retweet networks after the denoising and the calculation of z-scores.}
  \label{global_polarization_standarized}
\end{figure}

Figure \ref{global_polarization_standarized} presents the standardised polarisation across the retweet networks for the mentioned datasets, illustrating the extent of polarisation after denoising and z-score calculation.

\begin{figure}[!htbp]
  \begin{center}
  \includegraphics[width=0.9\textwidth]{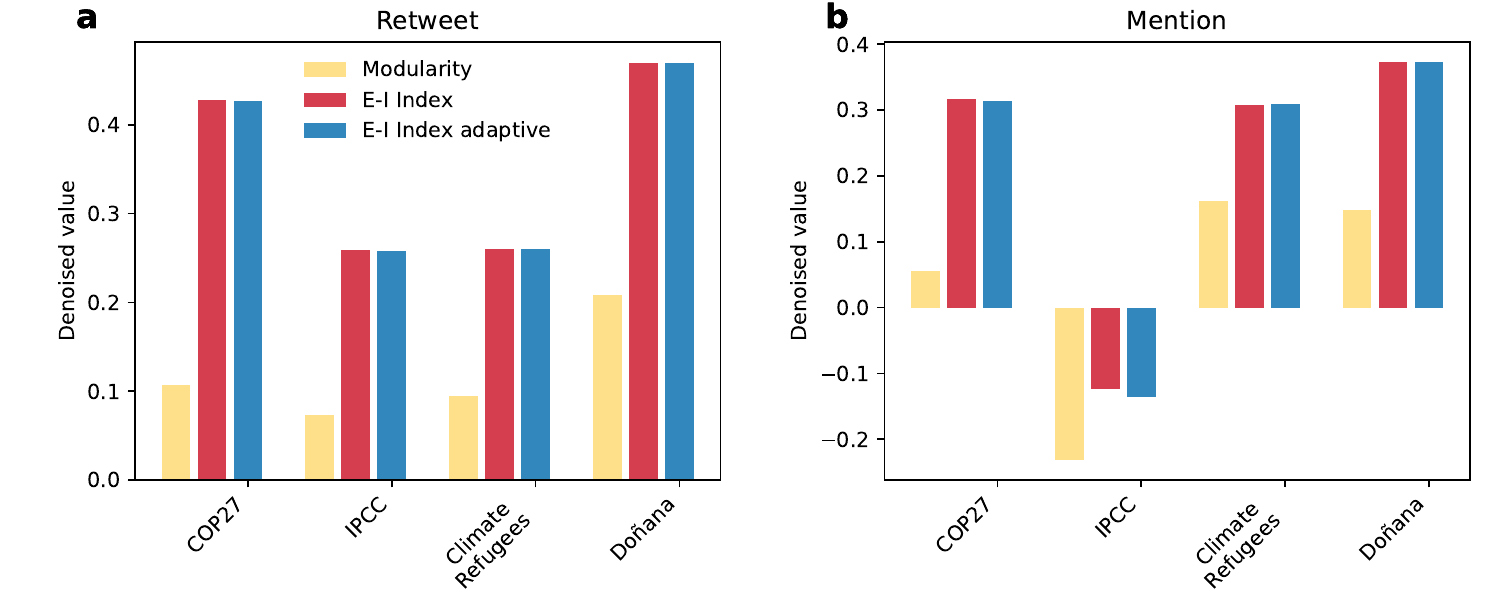}
  \end{center}
\caption{\textbf{Standarized polarisation in Twitter by interaction type.} Standardized modularity and the E-I indices in (\textbf{a}) the retweet and (\textbf{b}) mention networks. We computed the polarisation over the subnetwork of the two largest communities of opposed political bias. We have calculated the indices in the real network, in 100 realizations of the configuration model with preserved degree sequences, and standardized them.}
  \label{local_polarization_standarized}
\end{figure}

\newpage
Furthermore, we have computed in Fig. \ref{local_polarization_standarized} the polarisation scores in the subgraph between the two largest communities with opposing bias that speak the same language.

\clearpage
\subsection{Bias and reliability}

\begin{figure}[!h] 
\begin{center} 
\includegraphics[width=0.7\textwidth]{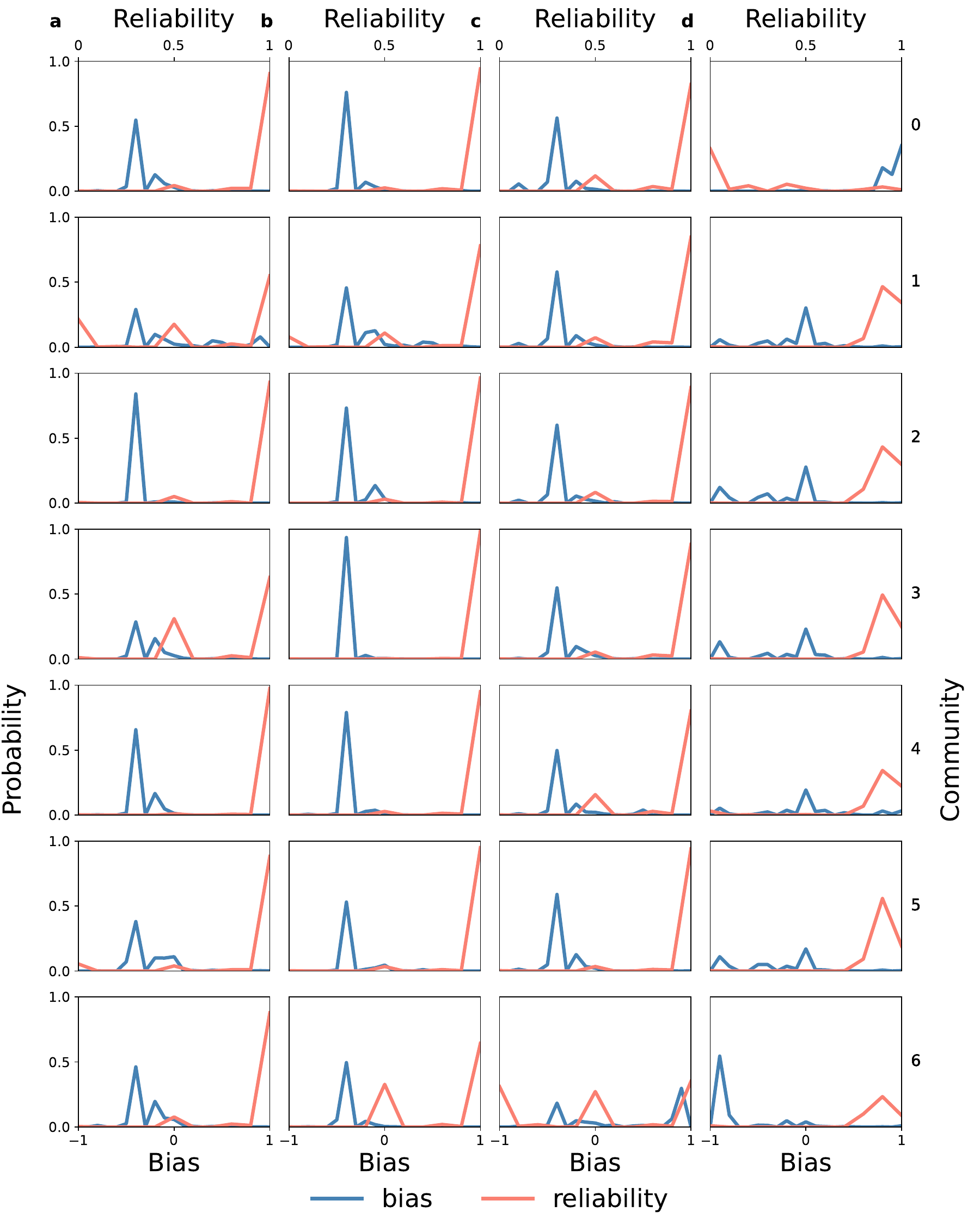} 
\end{center} 
\caption{\textbf{Distribution of bias and reliability in the largest communities.}\textbf{(a-d)} Distribution of bias and reliability in the seven largest communities for \textbf{(a)} COP27, \textbf{(b)} IPCC, \textbf{(c)} Climate Refugees and \textbf{(d)} Doñana.}
\label{bias_reliability_distribution}
 \end{figure}

\begin{figure}[!htbp]
  \begin{center}
  \includegraphics[width=0.8\textwidth]{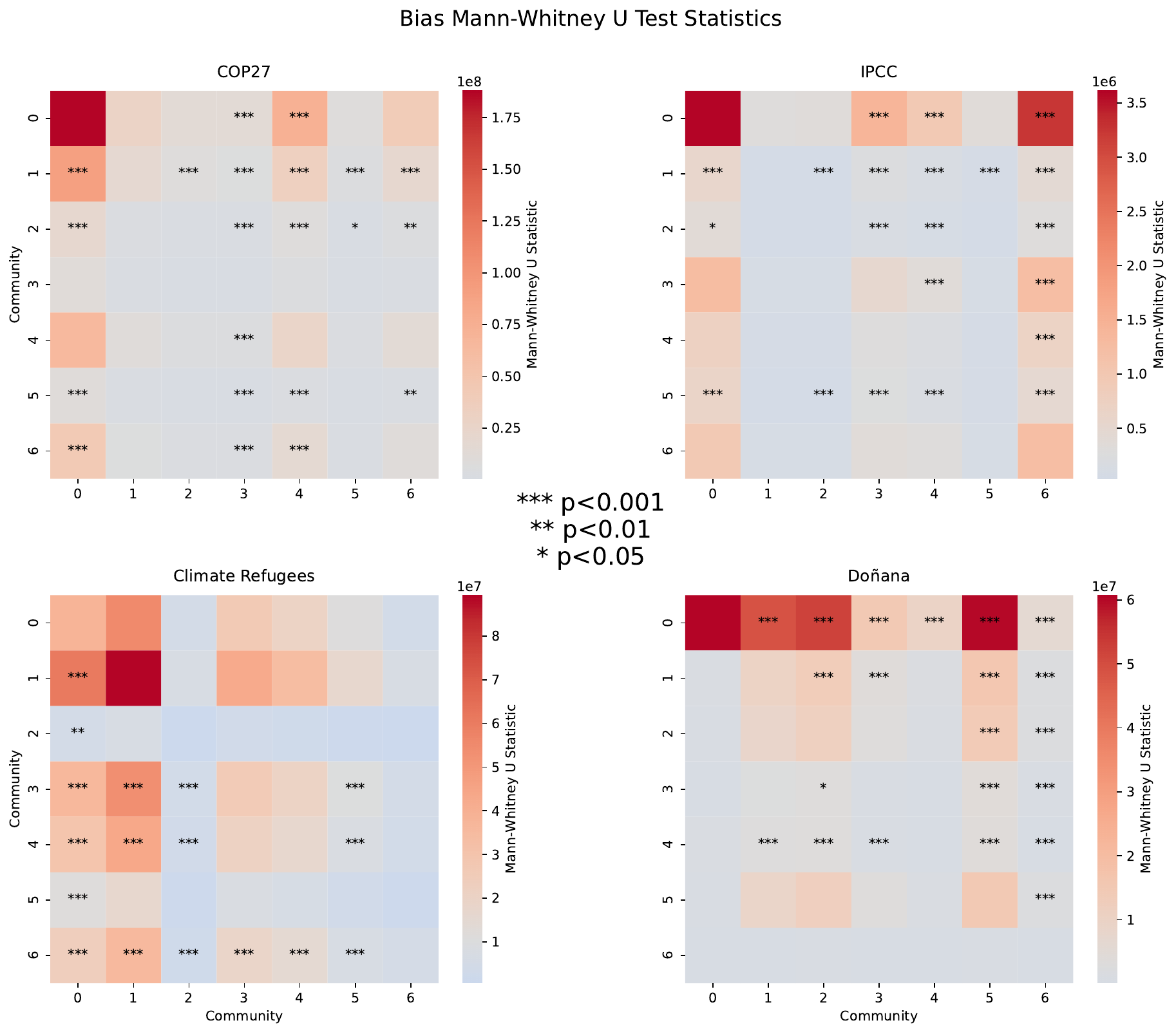}
  \end{center}
\caption{\textbf{Mann-Whitney U test between the bias of the top 7 communities.} Result of the Mann-Whitney U test between the bias of the top 7 communities. The test was conducted to assess if the community in the vertical axis is larger than the community in the horizontal axes. Asterisks report the significance of the tests (*p-value$<$0.05, **p-value$<$0.01 and ***p-value$<$0.001).}
  \label{bias_significance}
\end{figure}

\begin{figure}[!htbp]
  \begin{center}
  \includegraphics[width=0.8\textwidth]{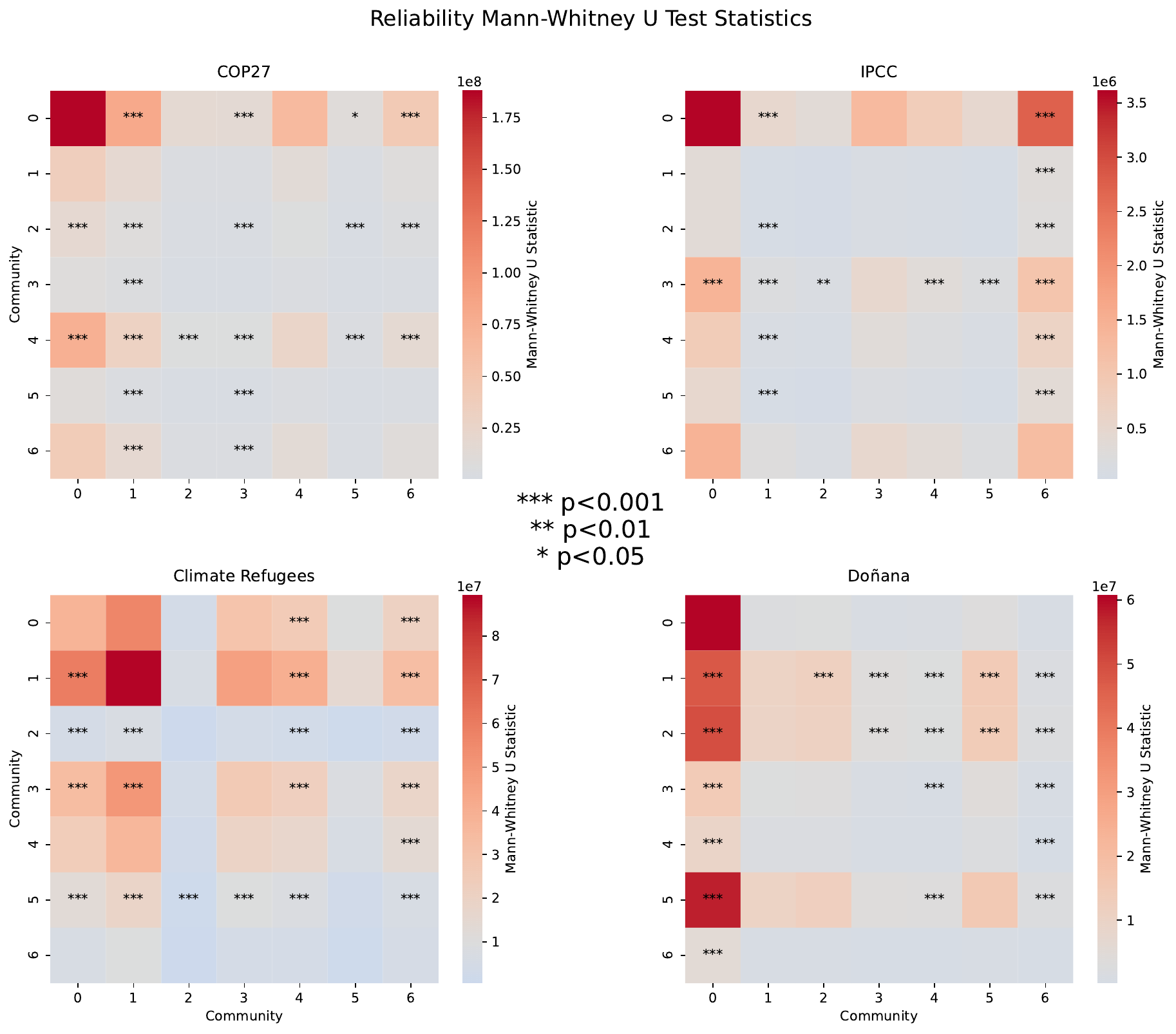}
  \end{center}
\caption{\textbf{Mann-Whitney U test between the reliability of the top 7 communities.} Result of the Mann-Whitney U test between the reliability of the top 7 communities. The test was conducted to assess if the community in the vertical axis is larger than the community in the horizontal axes. Asterisks report the significance of the tests (*p-value$<$0.05, **p-value$<$0.01 and ***p-value$<$0.001).}
  \label{reliability_significance}
\end{figure}

We tested the significance of the observed bias (Fig. \ref{bias_significance}) and reliability (Fig. \ref{reliability_significance}) to determine whether the differences identified in our analysis are statistically meaningful. In both cases, the differences between the two selected communities in each network are statistically significant.

\newpage

\clearpage
\subsection{Isolation and entropy}

\begin{figure}[!h] 
\begin{center} 
\includegraphics[width=0.7\textwidth]{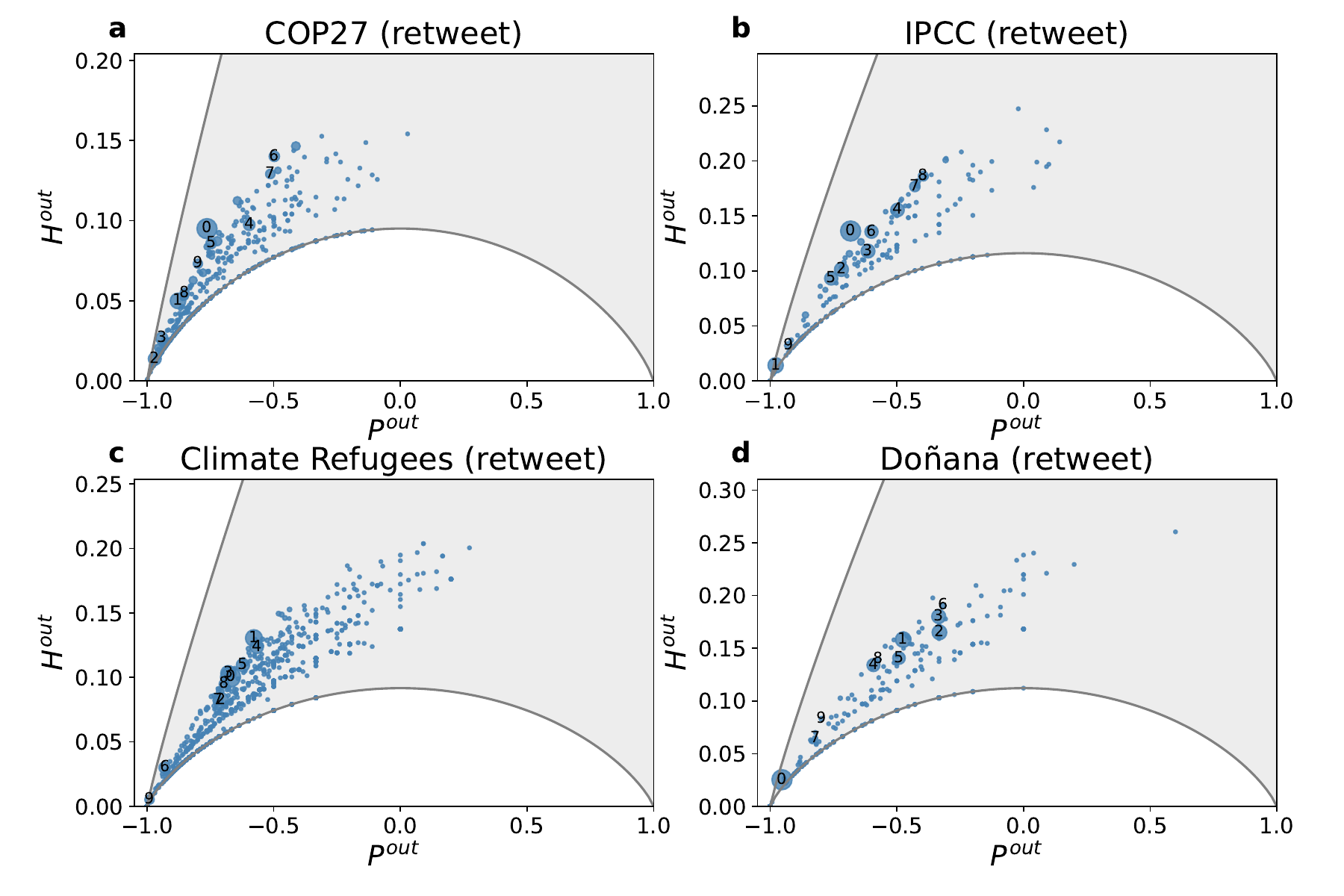} 
\end{center} 
\caption{\textbf{Community connectivity and entropy in retweet networks.} Outflow entropy as a function of the normalised outgoing isolation in the retweet networks of \textbf{(a)} COP27, \textbf{(b)} IPCC, \textbf{(c)} Climate Refugees, and \textbf{(d)} Do\~{n}ana. The grey area corresponds to the range of feasible values.} \label{polarization_entropy}
 \end{figure}

To evaluate the isolation and interaction of communities, we calculated the directed interactions between them, and we computed the normalised outflow difference given by
\begin{equation} 
\Delta F ^{out}_i=\frac{\sum_{j, j\neq i}T_{ij}-T_{ii}}{\sum_{\forall j} T_{ij}}, 
\end{equation} 
and the entropy of flows as 
\begin{equation} 
H^{out}_i=-\sum_{\forall j} p_{ij} log p_{ij}, 
\end{equation} 
where $T_{ij}$ is the number of interactions from community $i$ to community $j$ and $p_{ij}$ is the probability of having a retweet from $i$ to $j$ given by $T_{ij}/\sum_{\forall j} T_{ij}$. While the first index quantifies the ratio between external and internal links and goes from $-1$ to $1$ when all flows are external, the entropy provides information on the variety of those flows. The entropy is $0$ when there is no variety of flows and $1$ when it is maximum. In Fig. \ref{polarization_entropy}, we represent the outflow entropy as a function of the isolation for the four networks.

We computed the minimum and maximum attainable entropy values as a function of $\Delta F^{out}_i$ and $p_{ii}$ for orientation. Given a value of $\Delta F^{out}_i$ we have 
\begin{equation} 
\frac{\sum_{j, j\neq i}T_{ij}}{{\sum_{\forall j}} T_{ji}}=\Delta F_i+\frac{T_{ii}}{{\sum_{\forall j}T_{ij}}}=\Delta F_i+p_{ii}. 
\end{equation} 
The minimum value of the entropy corresponds to the situation when a community is connected only to itself and one external community (except for the extreme case $\Delta F_i=1$), which we can plug into the entropy as 
\begin{equation} H^{in}_i=- p_{ii} log p_{ii}- \Delta F_i+p_{ii} log \Delta F_i+p_{ii}. 
\end{equation} 
The maximum value of the entropy appears when a community is linked to all the others with a flow $T_{ij}/(N_c-1)$, where $N_c$ is the total number of communities in the network. Thus, it can be written as 
\begin{equation} 
H^{in}_i=- p_{ii} log p_{ii}- \sum^{N_c-1}_{j=0} \frac{\Delta F_i+p_{ii}}{N_c-1} log \frac{\Delta F_i+p_{ii}}{N_c-1}. 
\end{equation}

\newpage

\begin{figure}[!h]
  \begin{center}
  \includegraphics[width=0.7\textwidth]{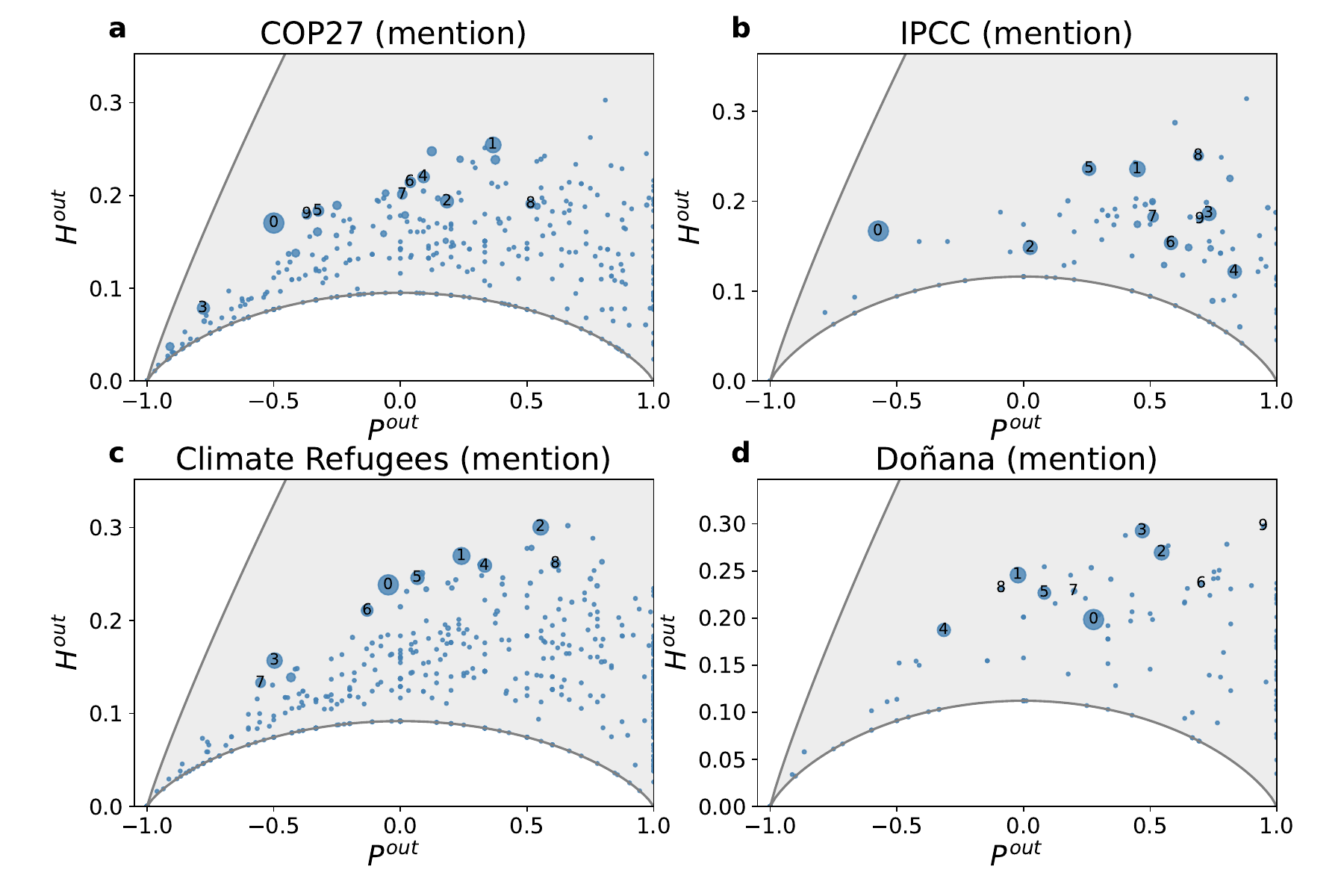}
  \end{center}
 \caption{\textbf{Community connectivity and entropy in mention networks.} Outflow entropy as a function of the normalised outgoing connectivity in the mention networks of \textbf{(a)} COP27, \textbf{(b)} IPCC, \textbf{(c)} Climate Refugees, and \textbf{(d)} Do\~{n}ana. The grey area corresponds to the range of attainable values. }\label{polarization_entropy_mention}
\end{figure}

In Figs. \ref{polarization_entropy}-\ref{polarization_entropy_mention} we show the outflow entropy as a function of the outgoing isolation for the retweet and mention networks between communities.

\newpage

\begin{figure}[!h]
  \begin{center}
  \includegraphics[width=0.6\textwidth]{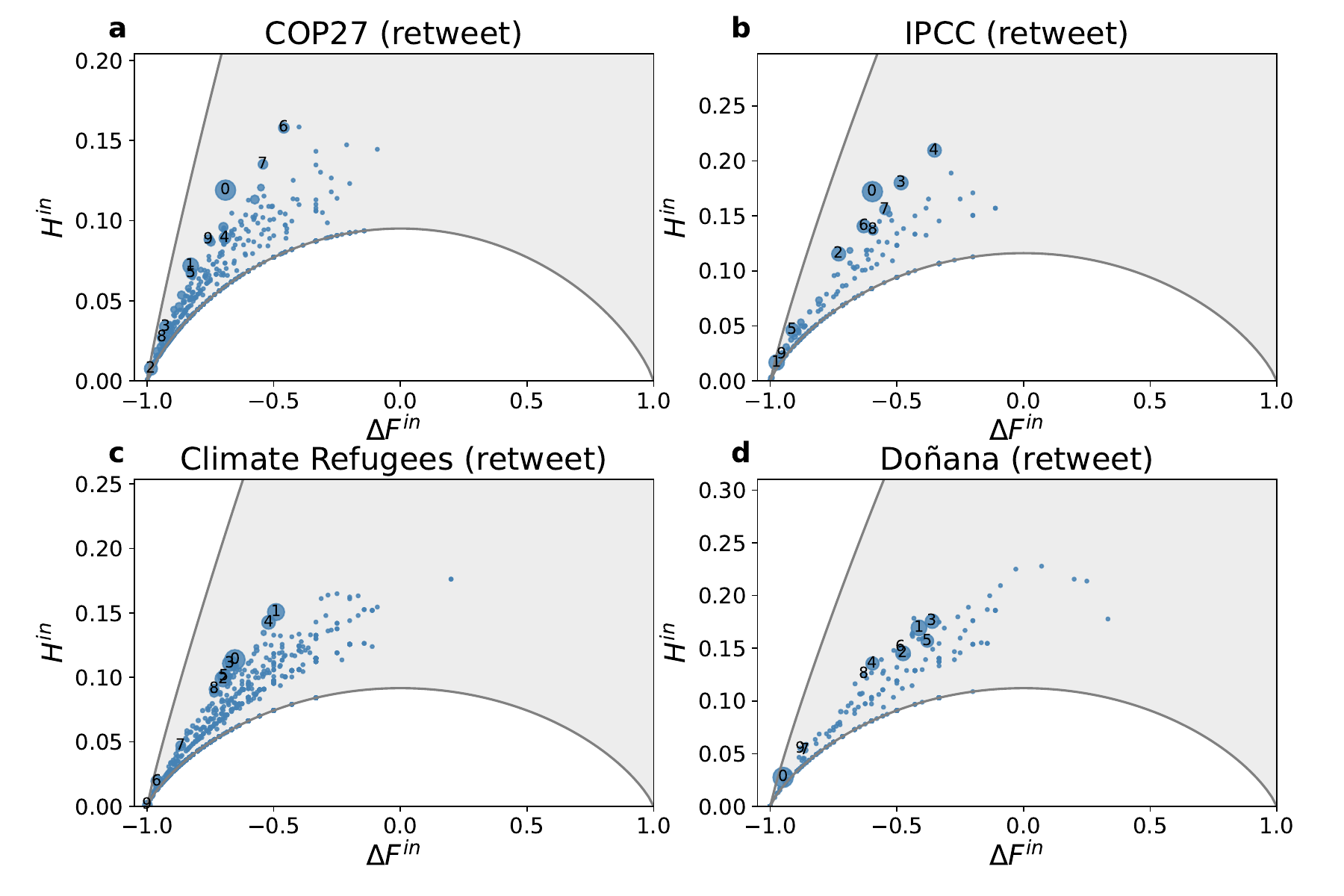}
  \end{center}
  \caption{\textbf{Community connectivity and entropy in retweet networks.} Inflow entropy as a function of the normalized incoming connectivity in the retweet networks of \textbf{(a)} COP27, \textbf{(b)} IPCC, \textbf{(c)} Climate Refugees, and \textbf{(d)} Do\~{n}ana.  The grey area corresponds to the range of attainable values.} \label{polarization_entropy_reversed}
\end{figure}

We also show in Fig. \ref{polarization_entropy_reversed} the inflow entropy as a function of the incoming isolation for the retweet network between communities. 

\newpage

We report in Tables \ref{table_polarization_out} and \ref{table_polarization_in} we report the values of isolation and entropy for the 8 largest communities in each network.

\begin{table*}[ht]
  \centering
  \resizebox{\textwidth}{!}{
    \begin{tabular}{lccrr@{\hspace{20pt}}ccrr}
\toprule
Network & Interaction & Community & $\Delta F ^{out}$ & $H^{out}$ & Interaction & Community & $\Delta F ^{out}$ & $H^{out}$ \\
\midrule
     &       & 0 & -0.68 & 0.14 &    & 0 & -0.57 & 0.17 \\
     &       & 1 & -0.98 & 0.01 &    & 1 & 0.45 & 0.24 \\
     &       & 2 & -0.72 & 0.10 &    & 2 & 0.03 & 0.15 \\
IPCC & Retweet & 3 & -0.61 & 0.12 &  Mention  & 3 & 0.73 & 0.19 \\
     &       & 4 & -0.50 & 0.16 &    & 4 & 0.83 & 0.12 \\
     &        & 5 & -0.76 & 0.09 &    & 5 & 0.26 & 0.24 \\
     &       & 6 & -0.60 & 0.14 &    & 6 & 0.58 & 0.15 \\
     &       & 7 & -0.43 & 0.18 &    & 7 & 0.51 & 0.18 \\
\midrule
      &       & 0 & -0.76 & 0.10 &    & 0 & -0.50 & 0.17 \\
      &       & 1 & -0.88 & 0.05 &    & 1 & 0.37 & 0.25 \\
      &       & 2 & -0.97 & 0.01 &    & 2 & 0.18 & 0.19 \\
COP27 & Retweet & 3 & -0.94 & 0.03 &  Mention  & 3 & -0.78 & 0.08 \\
      &       & 4 & -0.60 & 0.10 &    & 4 & 0.09 & 0.22 \\
      &       & 5 & -0.75 & 0.09 &    & 5 & -0.32 & 0.18 \\
      &       & 6 & -0.50 & 0.14 &    & 6 & 0.04 & 0.21 \\
      &       & 7 & -0.51 & 0.13 &    & 7 & 0.01 & 0.20 \\
\midrule
       &       & 0 & -0.95 & 0.03 &    & 0 & 0.28 & 0.20 \\
       &       & 1 & -0.48 & 0.16 &    & 1 & -0.02 & 0.25 \\
       &       & 2 & -0.33 & 0.17 &    & 2 & 0.55 & 0.27 \\
Doñana & Retweet & 3 & -0.33 & 0.18 &  Mention  & 3 & 0.47 & 0.29 \\
       &       & 4 & -0.59 & 0.13 &    & 4 & -0.31 & 0.19 \\
       &       & 5 & -0.49 & 0.14 &    & 5 & 0.08 & 0.23 \\
       &       & 6 & -0.32 & 0.19 &    & 6 & 0.70 & 0.24 \\
       &       & 7 & -0.82 & 0.06 &    & 7 & 0.20 & 0.23 \\
\midrule
      &       & 0 & -0.67 & 0.10 &    & 0 & -0.05 & 0.24 \\
      &       & 1 & -0.58 & 0.13 &    & 1 & 0.24 & 0.27 \\
      &       & 2 & -0.71 & 0.08 &    & 2 & 0.55 & 0.30 \\
Climate Refugees & Retweet & 3 & -0.68 & 0.10 &  Mention & 3 & -0.50 & 0.16 \\
      &       & 4 & -0.57 & 0.12 &    & 4 & 0.33 & 0.26 \\
      &       & 5 & -0.62 & 0.11 &    & 5 & 0.07 & 0.25 \\
      &       & 6 & -0.93 & 0.03 &    & 6 & -0.13 & 0.21 \\
      &       & 7 & -0.72 & 0.08 &    & 7 & -0.55 & 0.13 \\
\bottomrule
\end{tabular}

  }
\caption{Isolation and entropy metrics for the IPCC, COP27, Doñana, and Climate Refugees datasets, categorised by retweet and mention interactions for each community in terms of outflows.}  \label{table_polarization_out}
\end{table*}

\begin{table*}[ht]
  \centering
  \resizebox{\textwidth}{!}{
    \begin{tabular}{lccrr@{\hspace{20pt}}ccrr}

\toprule
Network & Interaction & Community & $\Delta F ^{in}$ & $H^{in}$ & Interaction & Community & $\Delta F ^{in}$ & $H^{in}$ \\
\midrule
 &  & 0 & -0.60 & 0.17 &   & 0 & 0.29 & 0.39 \\
 &  & 1 & -0.97 & 0.02 &   & 1 & -0.68 & 0.12 \\
 &  & 2 & -0.73 & 0.12 &   & 2 & -0.73 & 0.12 \\
IPCC & Retweet & 3 & -0.48 & 0.18 & Mention & 3 & 0.72 & 0.32 \\
 &  & 4 & -0.35 & 0.21 &   & 4 & 0.60 & 0.33 \\
 &  & 5 & -0.91 & 0.05 &   & 5 & 0.16 & 0.22 \\
 &  & 6 & -0.63 & 0.14 &   & 6 & 0.74 & 0.35 \\
 &  & 7 & -0.55 & 0.16 &   & 7 & 0.17 & 0.23 \\
\midrule
 &  & 0 & -0.69 & 0.12 &   & 0 & -0.28 & 0.25 \\
 &  & 1 & -0.83 & 0.07 &   & 1 & 0.06 & 0.22 \\
 &  & 2 & -0.98 & 0.01 &   & 2 & -0.68 & 0.08 \\
COP27 & Retweet & 3 & -0.93 & 0.03 & Mention & 3 & -0.58 & 0.13 \\
 &  & 4 & -0.69 & 0.09 &   & 4 & -0.02 & 0.23 \\
 &  & 5 & -0.82 & 0.07 &   & 5 & 0.25 & 0.29 \\
 &  & 6 & -0.46 & 0.16 &   & 6 & 0.49 & 0.33 \\
 &  & 7 & -0.54 & 0.14 &   & 7 & 0.11 & 0.25 \\
\midrule
 &  & 0 & -0.95 & 0.03 &   & 0 & 0.28 & 0.26 \\
 &  & 1 & -0.41 & 0.17 &   & 1 & -0.18 & 0.23 \\
 &  & 2 & -0.48 & 0.15 &   & 2 & 0.02 & 0.24 \\
Doñana & Retweet & 3 & -0.36 & 0.18 & Mention & 3 & 0.40 & 0.29 \\
 &  & 4 & -0.60 & 0.14 &   & 4 & -0.31 & 0.21 \\
 &  & 5 & -0.38 & 0.16 &   & 5 & 0.38 & 0.24 \\
 &  & 6 & -0.49 & 0.15 &   & 6 & 0.41 & 0.28 \\
 &  & 7 & -0.86 & 0.05 &   & 7 & 0.16 & 0.23 \\
\midrule
 &  & 0 & -0.65 & 0.11 &   & 0 & 0.15 & 0.27 \\
 &  & 1 & -0.49 & 0.15 &   & 1 & -0.02 & 0.26 \\
 &  & 2 & -0.70 & 0.10 &  & 2 & 0.40 & 0.27 \\
Climate Refugees & Retweet & 3 & -0.67 & 0.11 & Mention & 3 & -0.24 & 0.22 \\
 &  & 4 & -0.52 & 0.14 &  & 4 & 0.70 & 0.32 \\
 &  & 5 & -0.70 & 0.10 &  & 5 & -0.10 & 0.23 \\
 &  & 6 & -0.96 & 0.02 &  & 6 & -0.21 & 0.18 \\
 &  & 7 & -0.87 & 0.05 &  & 7 & -0.71 & 0.09 \\
\bottomrule
\end{tabular}
  }
\caption{Isolation and entropy metrics for the IPCC, COP27, Doñana, and Climate Refugees datasets, categorised by retweet and mention interactions for each community in terms of inflows.}
  \label{table_polarization_in}
\end{table*}

\clearpage

\subsection{Echo chambers}

\begin{figure}[!htbp]
  \begin{center}
  \includegraphics[width=0.8\textwidth]{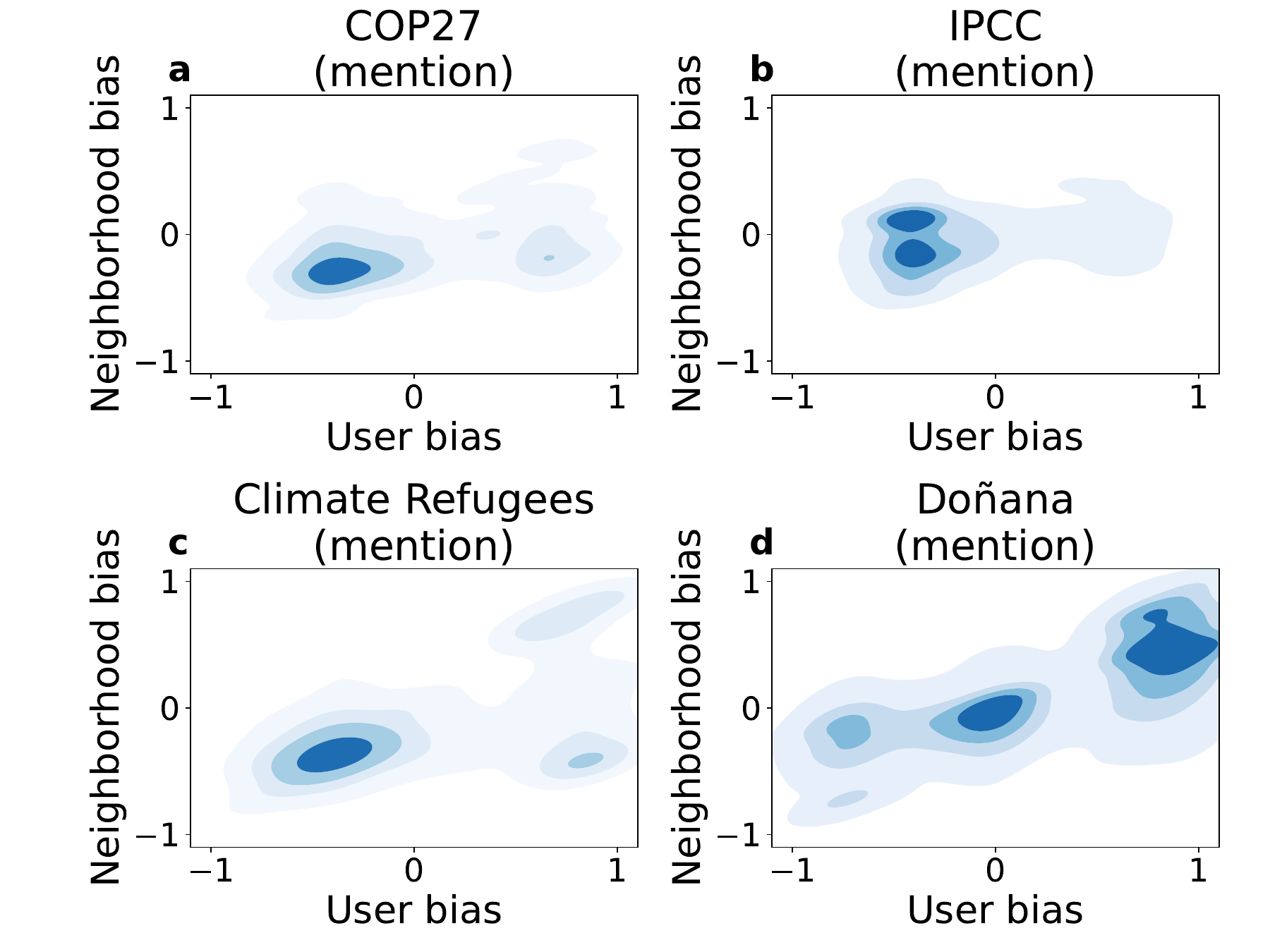}
  \end{center}
  \caption{\textbf{Detection of echo chambers in the mention network for political bias.} Political bias of the neighbourhood as a function of the bias of a user in the mention network for \textbf{(a)} COP27, \textbf{(b)} IPCC, \textbf{(c)} Climate Refugees, and \textbf{(d)} Do\~{n}ana.}\label{neighboor_bias_mention}
\end{figure}

\begin{figure}[!htbp]
  \begin{center}
  \includegraphics[width=0.8\textwidth]{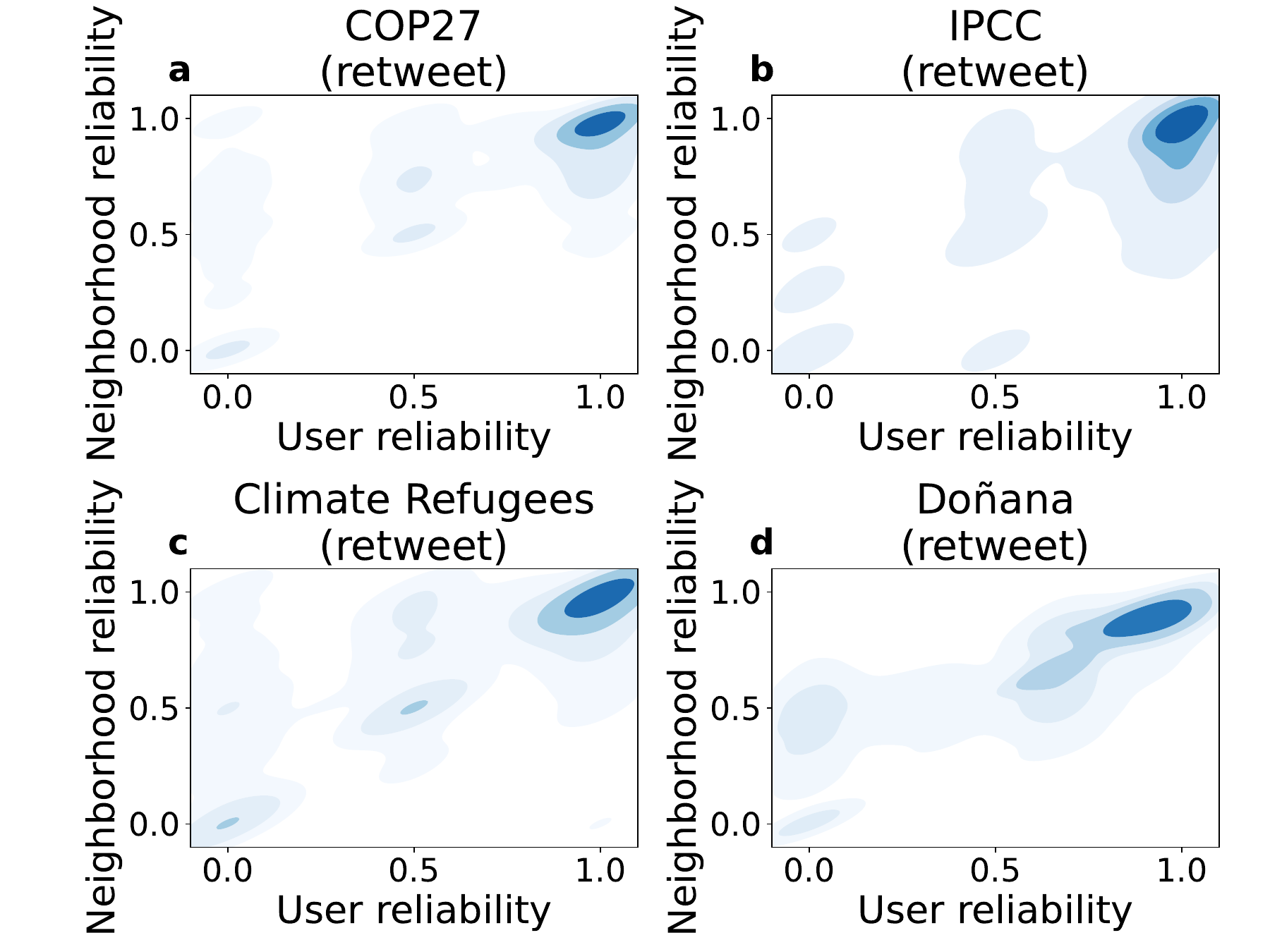}
  \end{center}
  \caption{\textbf{Detection of echo chambers in the retweet network for reliability.} Media reliability of the neighbourhood as a function of the bias of a user in \textbf{(a)} COP27, \textbf{(b)} IPCC, \textbf{(c)} Climate Refugees, and \textbf{(d)} Do\~{n}ana. studied by considering only retweets.}\label{neighboor_reliability_retweet}
\end{figure}

Echo chambers emerge in online discourse when users predominantly interact with like-minded individuals, reinforcing their existing beliefs. In political discussions, this phenomenon can manifest in different ways depending on the type of interaction network. In the mention network, users tend to engage more with others who share similar political biases, forming bias-based echo chambers. Meanwhile, in the retweet network, clustering occurs based on the reliability of the media sources that users amplify, leading to reliability-based echo chambers. These dynamics are analysed across key climate-related topics, including COP27, IPCC, Climate Refugees, and Doñana. Figures \ref{neighboor_bias_mention} and \ref{neighboor_reliability_retweet} illustrate these patterns, highlighting the structure and polarisation of online conversations.

\begin{figure}[!htbp] 
\begin{center}
 \includegraphics[width=0.8\textwidth]{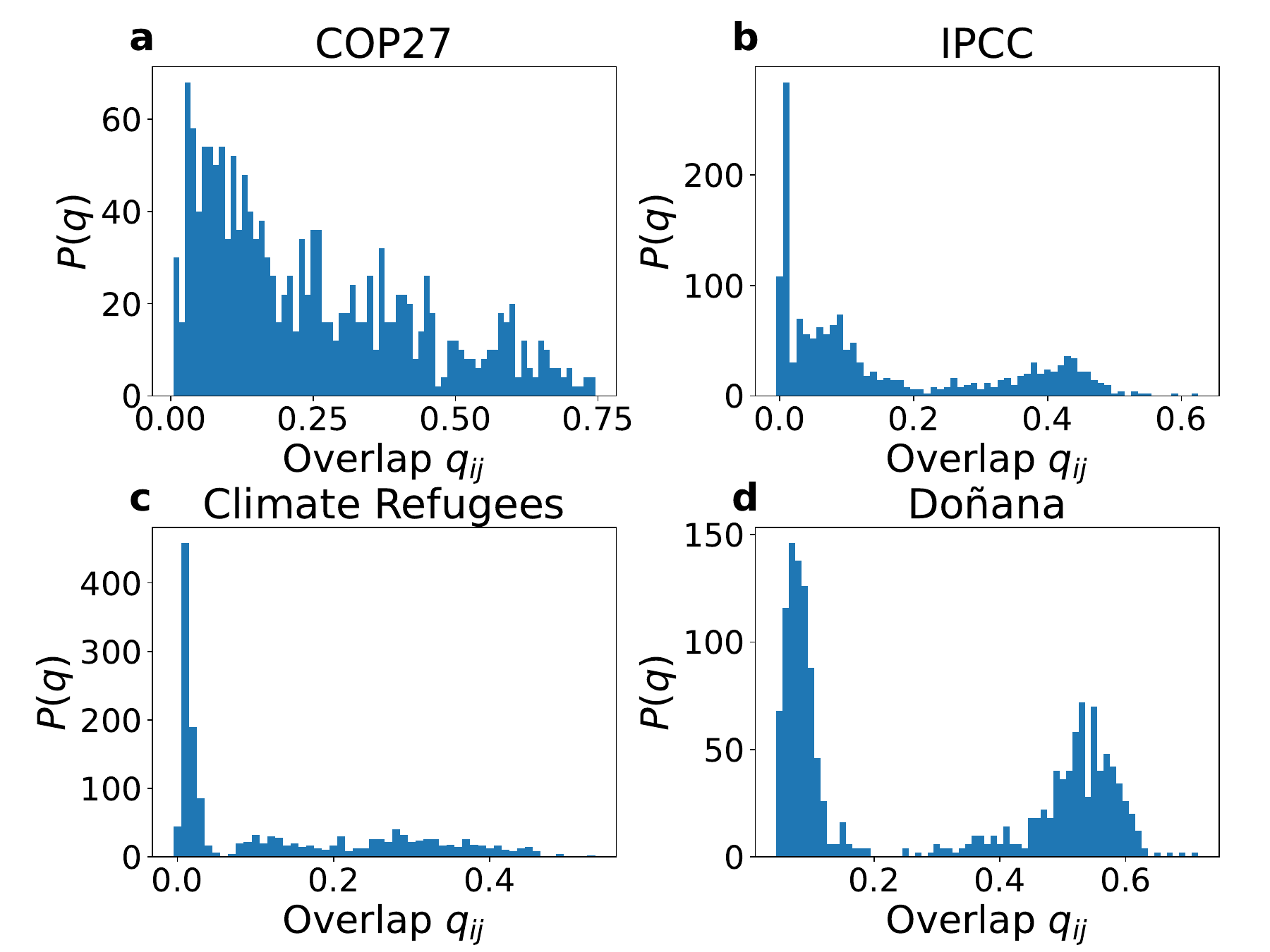} 
 \end{center}
  \caption{\textbf{Chamber overlap distribution $q_{ij}$.} Chamber overlap distribution for \textbf{(a)} COP27, \textbf{(b)} IPCC, \textbf{(c)} Climate Refugees, and \textbf{(d)} Do\~{n}ana calculated on the top $20$ users of each community.} \label{chamber_overlap_distribution} 
  \end{figure}

\begin{figure}[!htbp]
  \begin{center}
  \includegraphics[width=0.8\textwidth]{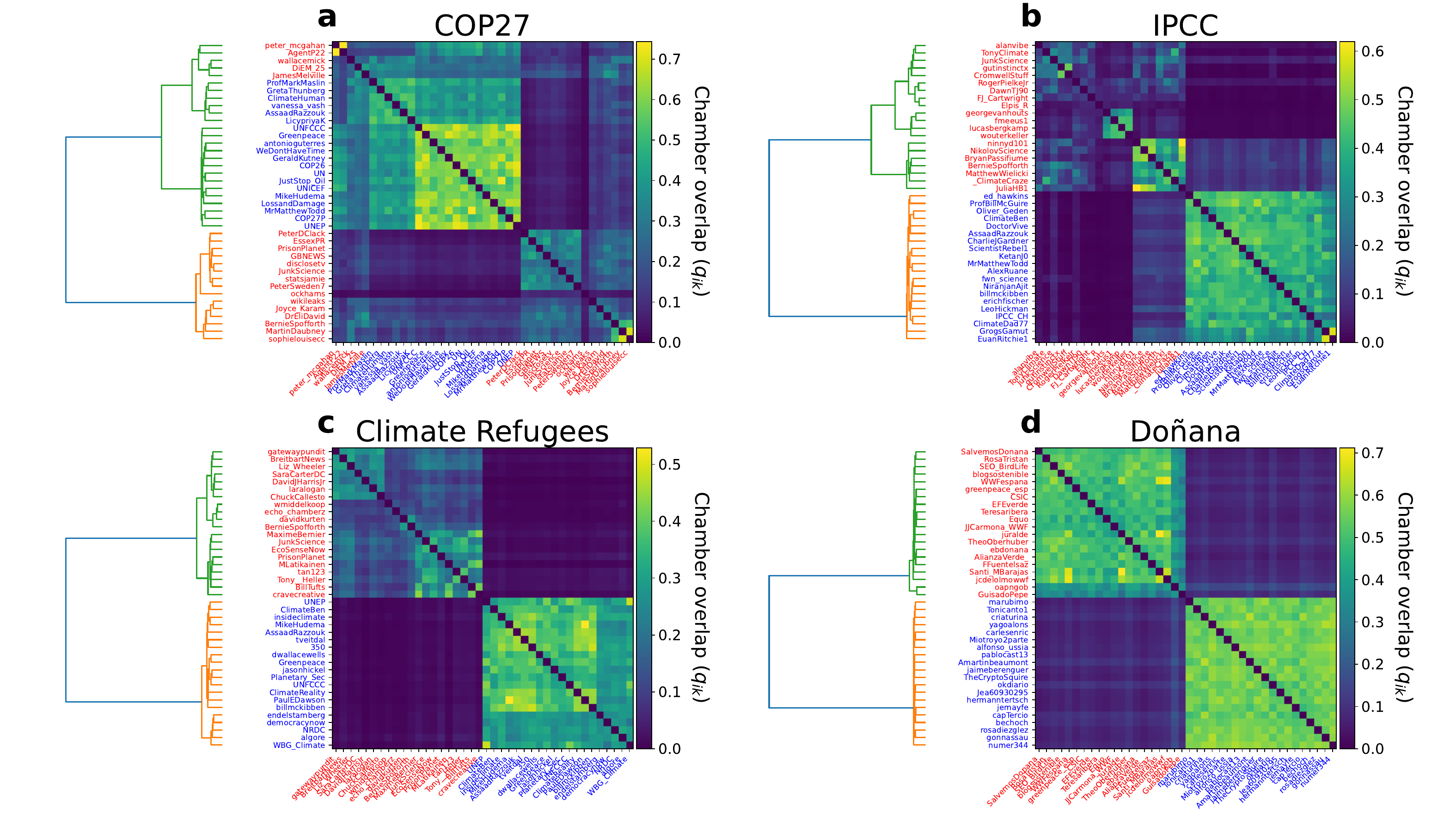}
  \end{center}
  \caption{\textbf{Chamber overlap matrix between users clustered using a hierarchical agglomerative algorithm.} Chamber overlap between the $20$ users with more retweets in the two largest communities with opposed bias in \textbf{(a)} COP27, \textbf{(b)} IPCC, \textbf{(c)} Climate Refugees, and \textbf{(d)} Do\~{n}ana.} \label{chamber_overlap_matrix}
\end{figure}

\clearpage
The chamber overlap can distinguish between users sharing similar echo chambers (Fig.\ref{chamber_overlap_matrix}). We constructed the user-to-user matrices where the values correspond to their chamber overlap. Using a hierarchical agglomerative algorithm, we clustered the distance matrix and obtained an almost perfect separation of users by communities. The separation between the users in the communities is neater in the COP27 and Do\~{n}ana datasets compared to IPCC and Climate Refugees. Nevertheless, we observe a clear trend of similar echo chambers among users within the same community. In the COP27 network, the pattern is more evident for the users in the mainstream community (in blue). The separation between communities and the matching of echo chambers suggest a clearer separation in the Do\~{n}ana dataset.

\clearpage

\section{Policy  and affective polarisation}
\subsection{Hashtag analysis}
\begin{figure}[!htbp]
  \begin{center}
  \includegraphics[width=0.8\textwidth]{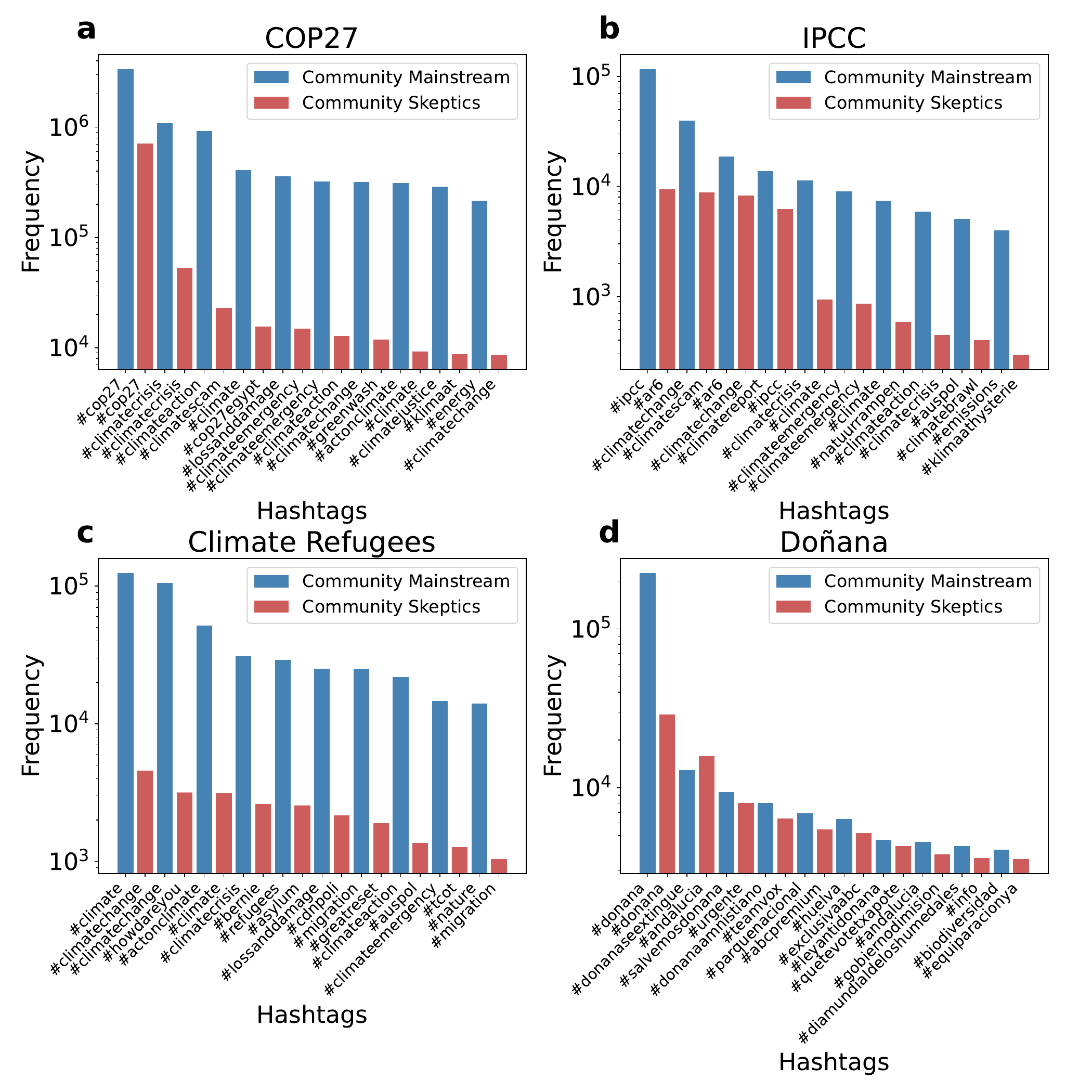}  
  \end{center}
  \caption{\textbf{Ranking of hashtags by number of interactions in the top two opposing communities.} Ranking of hashtags in the two largest communities with opposed political bias in the \textbf{(a)} COP27, \textbf{(b)} IPCC, \textbf{(c)} Climate Refugees, and \textbf{(d)} Do\~{n}ana networks.} \label{top_hashtags}
\end{figure}

\begin{figure}[!htbp]
  \begin{center}
  \includegraphics[width=0.7\textwidth]{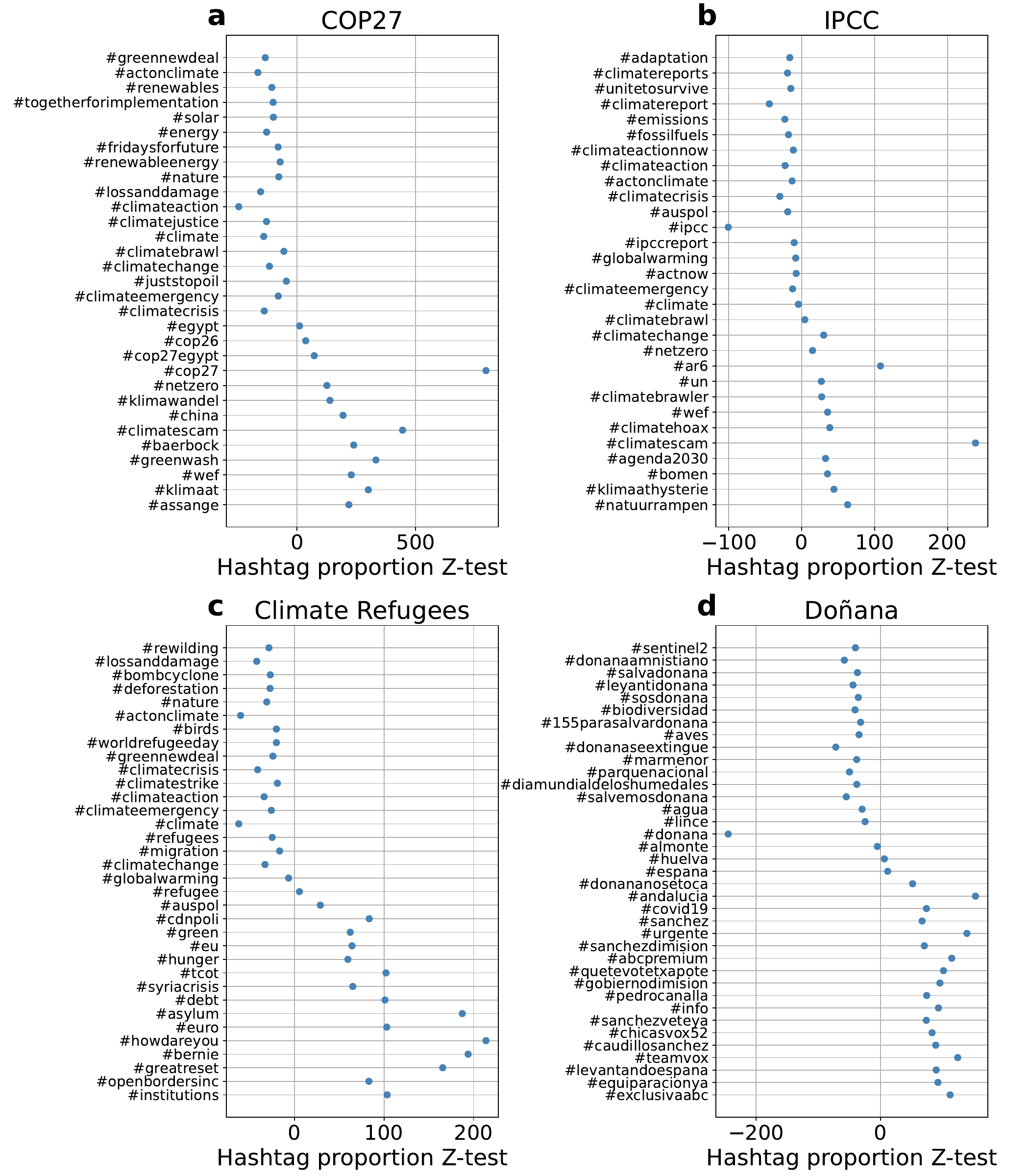}
  \end{center}
\caption{\textbf{Hashtag proportion test.} We have conducted the ztest between the proportion of hashtags in \textbf{(a)} COP27, \textbf{(b)} IPCC, \textbf{(c)} Climate Refugees, and \textbf{(d)} Do\~{n}ana networks. Blue bars indicate significance (p-value$<$0.05).}
  \label{hashtag_significance}
\end{figure}

Understanding the dynamics of hashtag usage across different communities provides valuable insights into the discourse surrounding key topics. In Fig. \ref{top_hashtags}, we present the 20 most tweeted and retweeted hashtags for the two communities analyzed in each network. Our findings reveal a substantial overlap in hashtag usage between communities, with a dominance of generalistic hashtags. To further investigate these patterns, we assess the significance of differences in hashtag presence in Fig. \ref{hashtag_significance}. The results indicate that, in all cases, the variation in hashtag abundance between communities is statistically significant, highlighting distinct engagement patterns despite the observed overlap.

\clearpage

\subsection{Topic and sentiment analysis}

\begin{figure}[!htbp] 
\begin{center} 
\includegraphics[width=\textwidth]{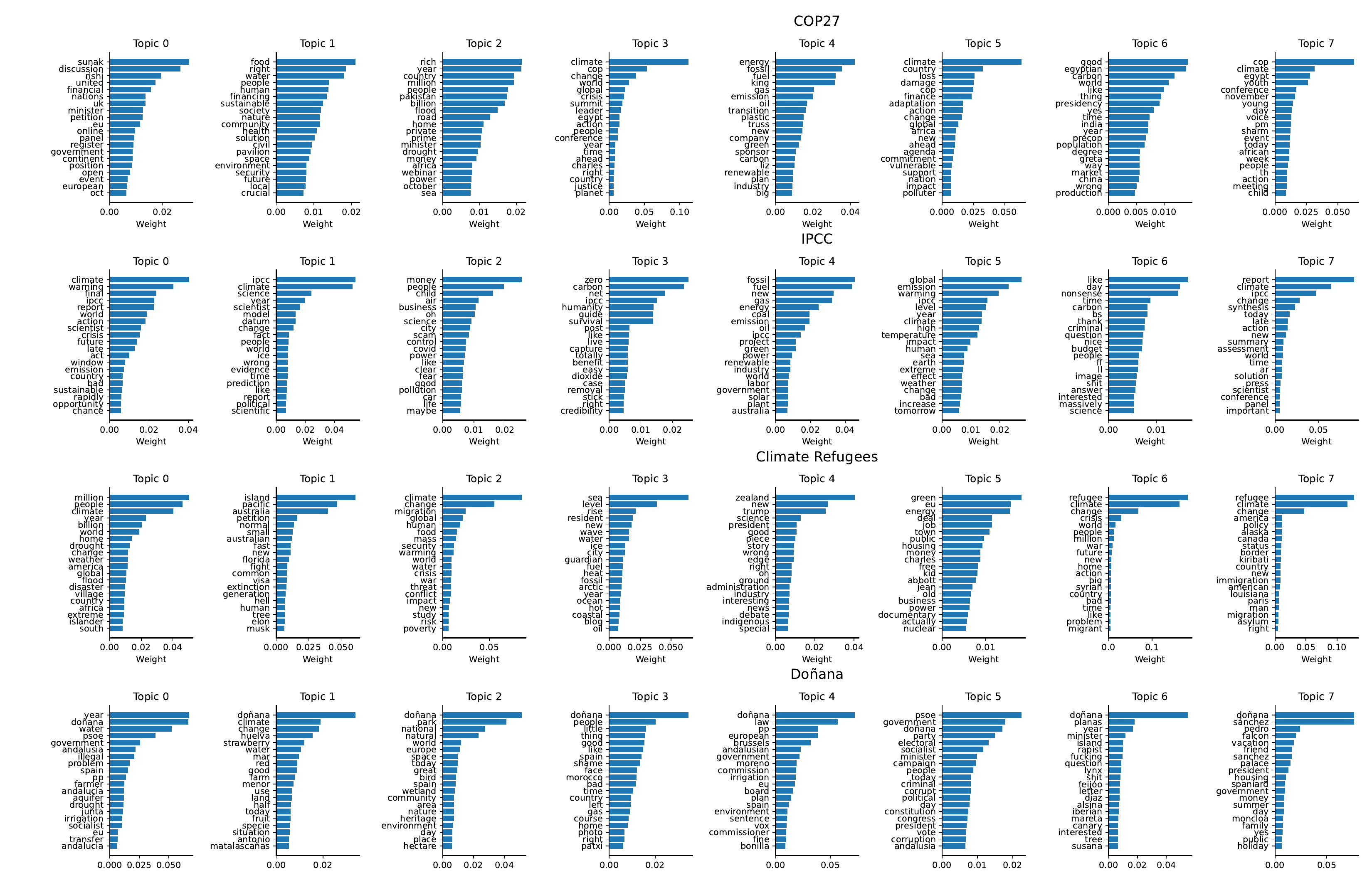} 
\end{center} 
\caption{\textbf{Topics detected in the networks analized.} Results of the Latent Dirichlet Allocation model (LDA) for eight topics in \textbf{(a)} COP27, \textbf{(b)} IPCC, \textbf{(c)} Climate Refugees, and \textbf{(d)} Do\~{n}ana networks.} \label{topic_summary} 
\end{figure}

\begin{figure}[!htbp] 
\begin{center} 
\includegraphics[width=0.9\textwidth]{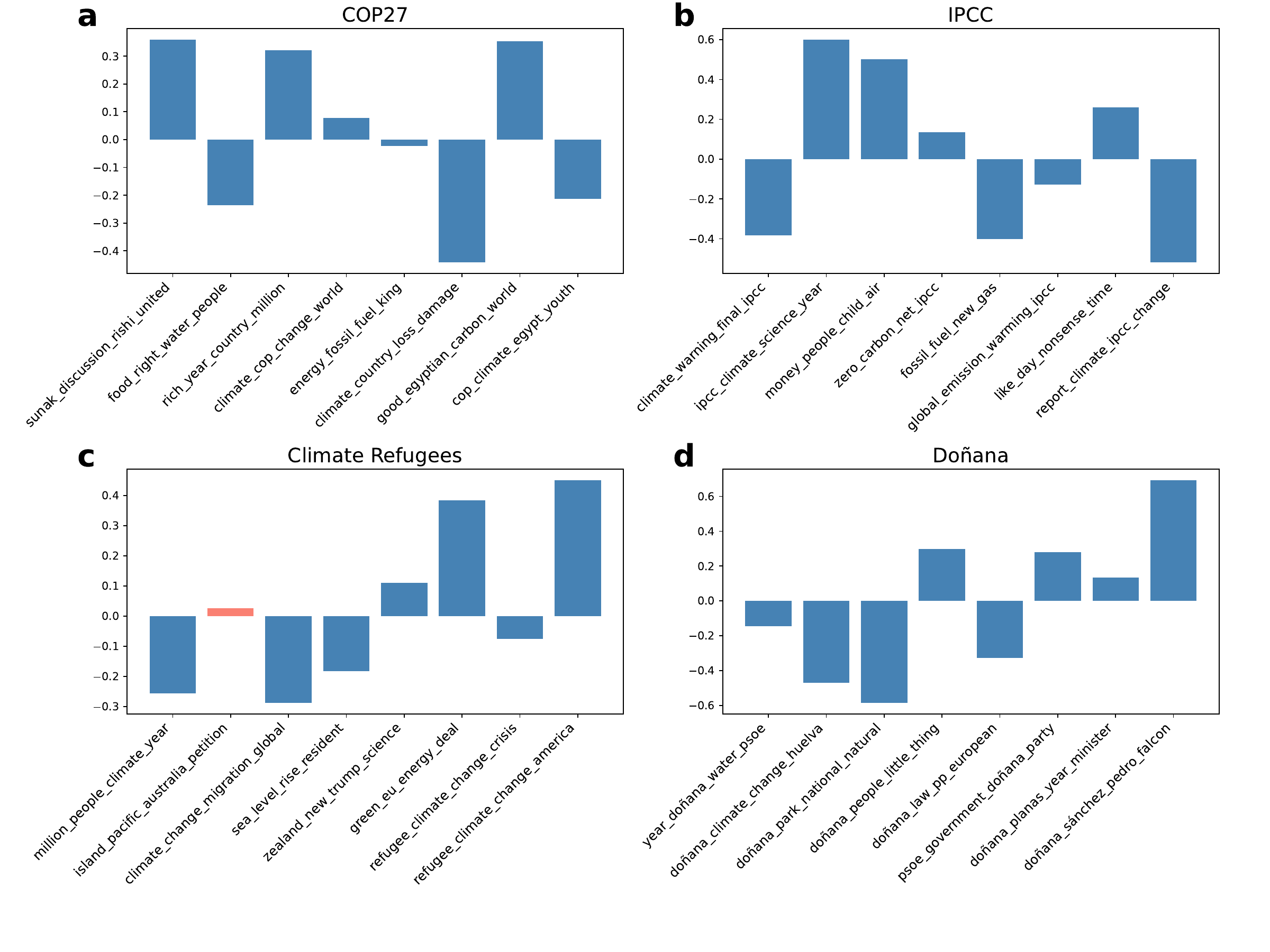} 
\end{center} 
\caption{\textbf{Distribution of the polarisation around the discussion topics.} Topics detected and polarisation in \textbf{(a)} COP27, \textbf{(b)} IPCC,\textbf{(c)} Climate Refugees, \textbf{(d)} Do\~{n}ana. Blue bars indicate that the frequency of topics between communities is statistically significant.} \label{topic_analysis} 
\end{figure}

We have implemented a Latend Dirichlet Allocation model (LDA) to uncover the main topics discussed in each of the datasets. In Fig. \ref{topic_summary} we report the most important words for each of the topics detected in the datasets. We have analyzed the polarisation around each of the topics Fig. \ref{topic_analysis}.

\begin{figure}[!htbp]
  \begin{center}
  \includegraphics[width=\textwidth]{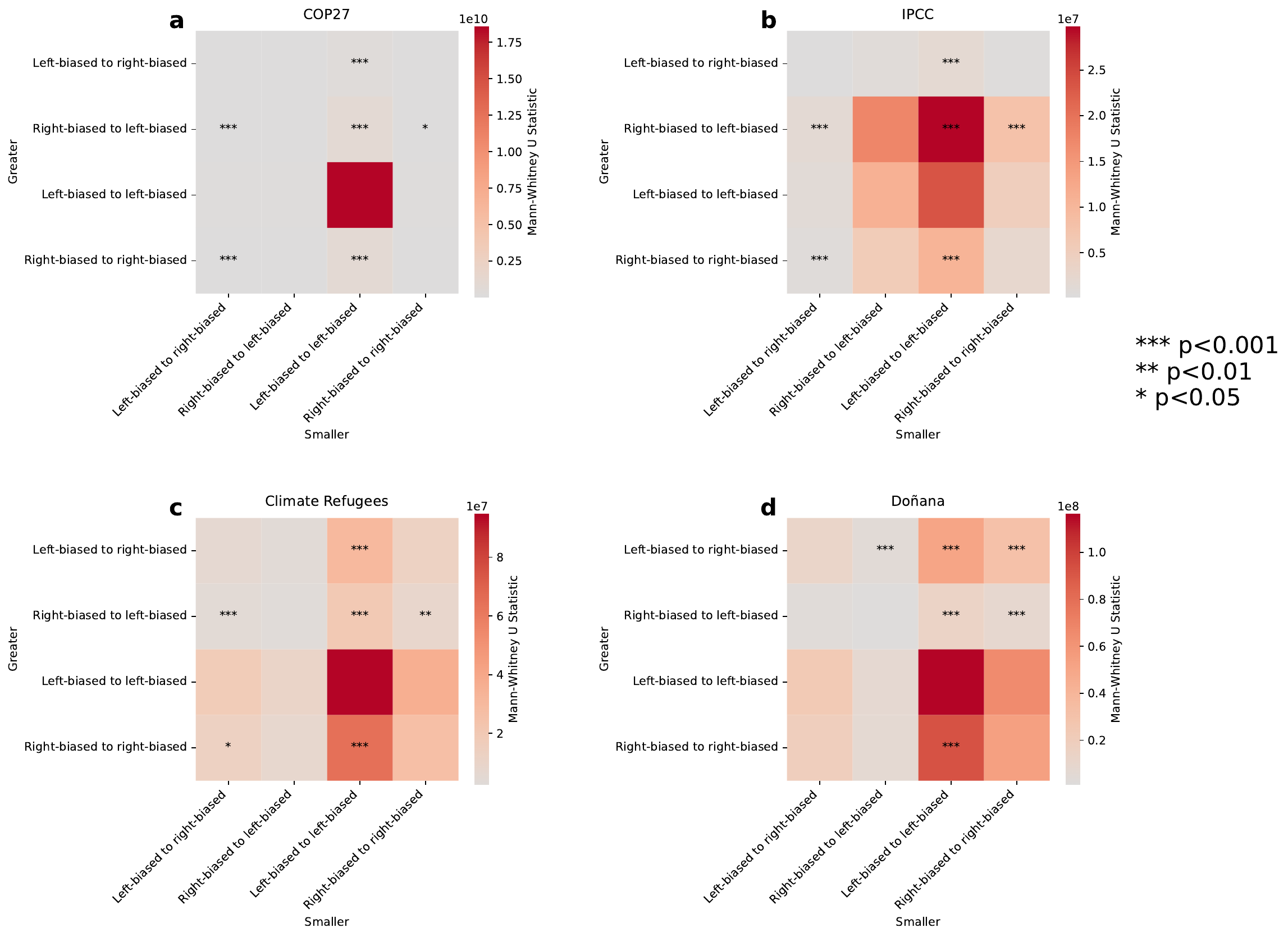}
  \end{center}
\caption{\textbf{Mann-Whitney U test in quote sentiment.} We have conducted the Mann-Whitney U test of the negative sentiment between communities in \textbf{(a)} COP27, \textbf{(b)} IPCC, \textbf{(c)} Climate Refugees, and \textbf{(d)} Do\~{n}ana networks. The test was conducted to assess if the entries in the vertical axes are greater than those in the horizontal axes.}
  \label{sentiment_significance}
\end{figure}

\clearpage
We have implemented a sentiment analysis on the four Twitter datasets using a Roberta-based model fine-tuned on around 58 million of Tweets \cite{huggingface}. We provide the Mann-Whitney U tests on the quoted messages in Fig. \ref{sentiment_significance}. 

\clearpage

\section{Cross-platform analysis}

\begin{figure}[!htbp]
  \begin{center}
  \includegraphics[width=0.9\textwidth]{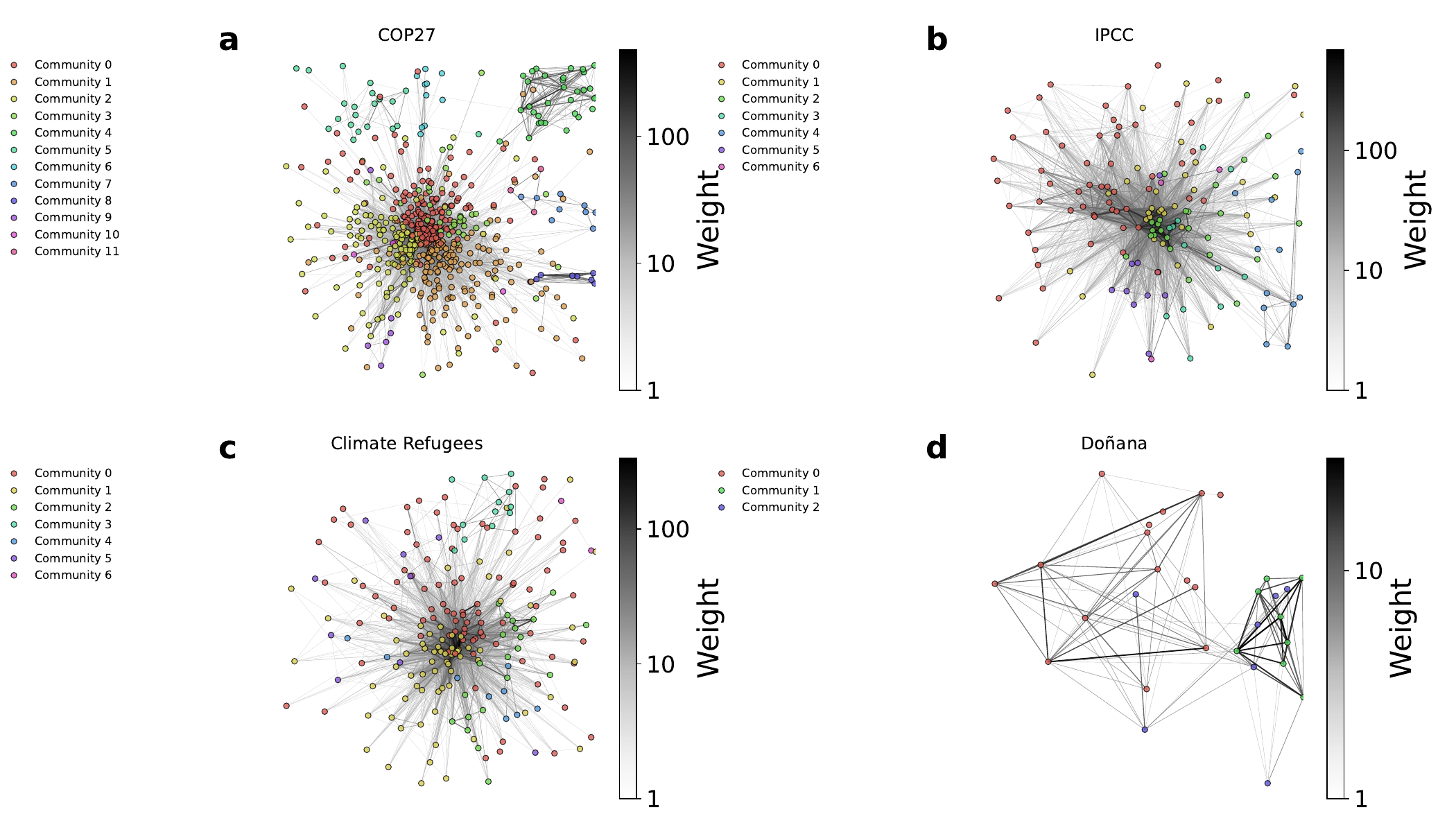}
  \end{center}
  \caption{\textbf{Network of Youtube posts extracted from the Twitter datasets.} Each node corresponds to a post and links to the number of co-commenting users. Each node is coloured according to the community calculated by optimising the modularity \cite{newman2006modularity}.} \label{youtube_network}
\end{figure}

\begin{figure}[!htbp]
  \begin{center}
  \includegraphics[width=0.9\textwidth]{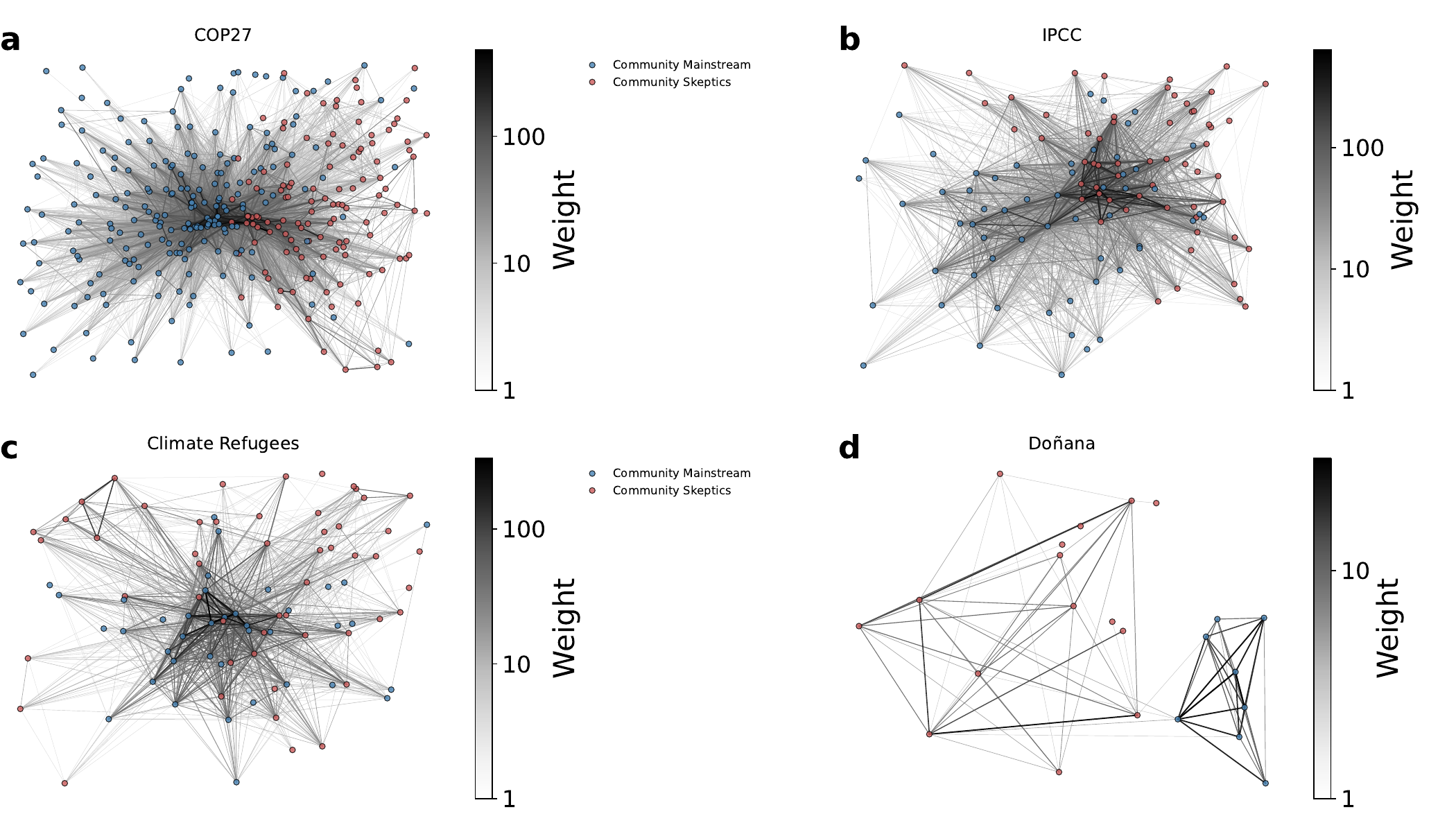}
  \end{center}
  \caption{\textbf{Network of YouTube posts after selecting the two main communities of mainstream and low-reliability content.} Interaction between the posts after selecting the larger communities with mainstream and low-reliability content. We calculated the communities by optimizing the modularity for two partitions. We identified the positioning of the communities by inspecting the videos.} \label{youtube_network_zoom}
\end{figure}

We present the network of YouTube posts extracted from the Twitter datasets, where each node represents a post and is linked based on the number of shared commenters Figure \ref{youtube_network}. Nodes are color-coded according to community structures identified through modularity optimization \cite{newman2006modularity}. To further investigate the network's structure, we analyze a subnetwork focusing on communities that predominantly spread skeptical narratives about climate change, as well as the largest community associated with mainstream content. The results of this analysis are depicted in Fig. \ref{youtube_network_zoom}.

\newpage
\clearpage

\begin{figure}[!htbp] 
\begin{center} 
\includegraphics[width=0.9\textwidth]{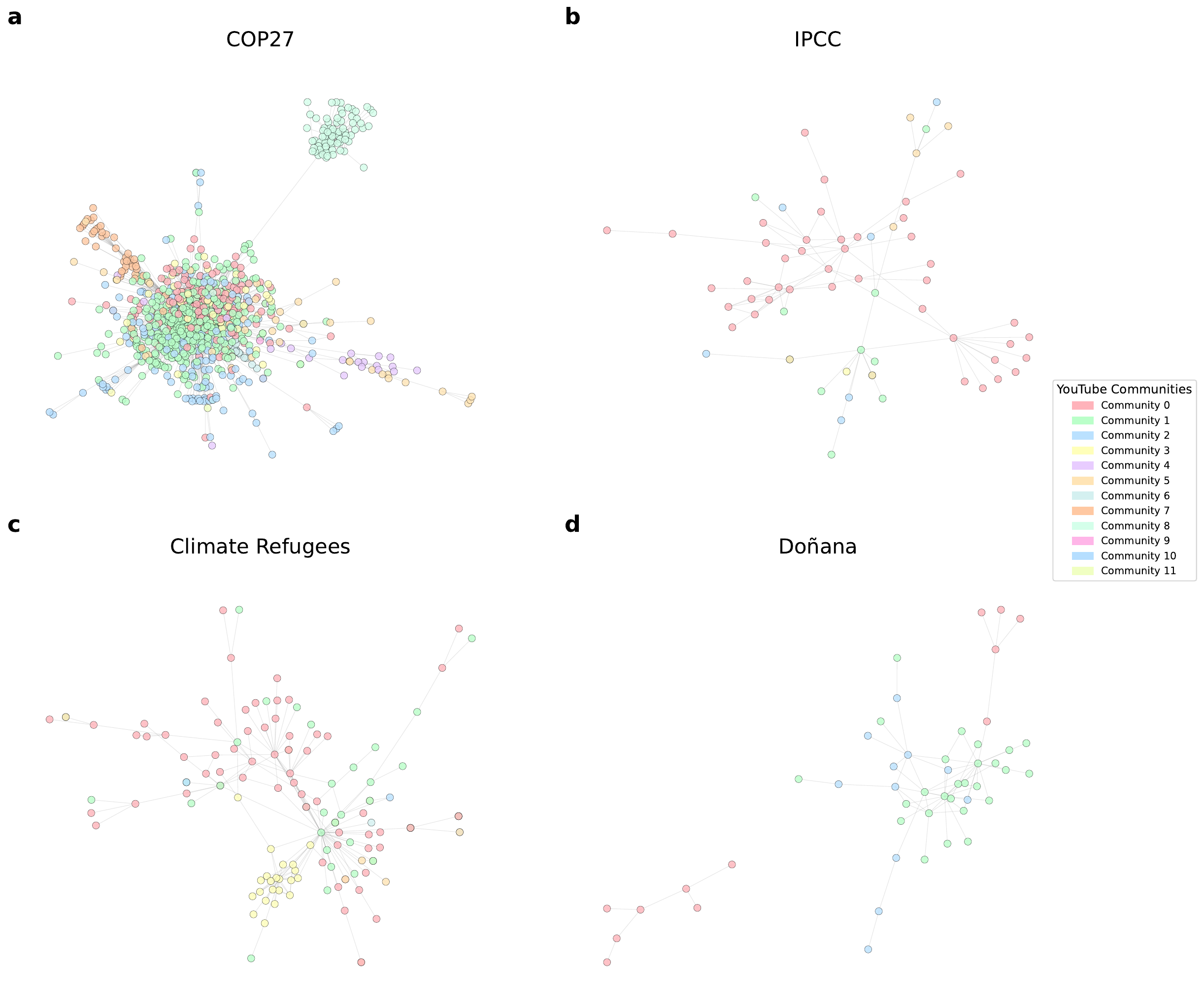} 
\end{center} 
\caption{\textbf{Network between Twitter users coloured by YouTube community.} Interactions between Twitter users that have published YouTube Content in \textbf{(a)} COP27, \textbf{(b)} IPCC, \textbf{(c)} Climate Refugees, and \textbf{(d)} Do\~{n}ana. Each dot is coloured by the content of YouTube communities they have tweeted on.} \label{twitter_youtube_network} 
\end{figure}

In Fig \ref{twitter_youtube_network} we show the Twitter subnetwork of users that shared YouTube links related to climate change coloured according to the YouTube community of the videos. Only connected components with more than 5 users are depicted.

\begin{figure}[!htbp]
  \begin{center}
  \includegraphics[width=0.9\textwidth]{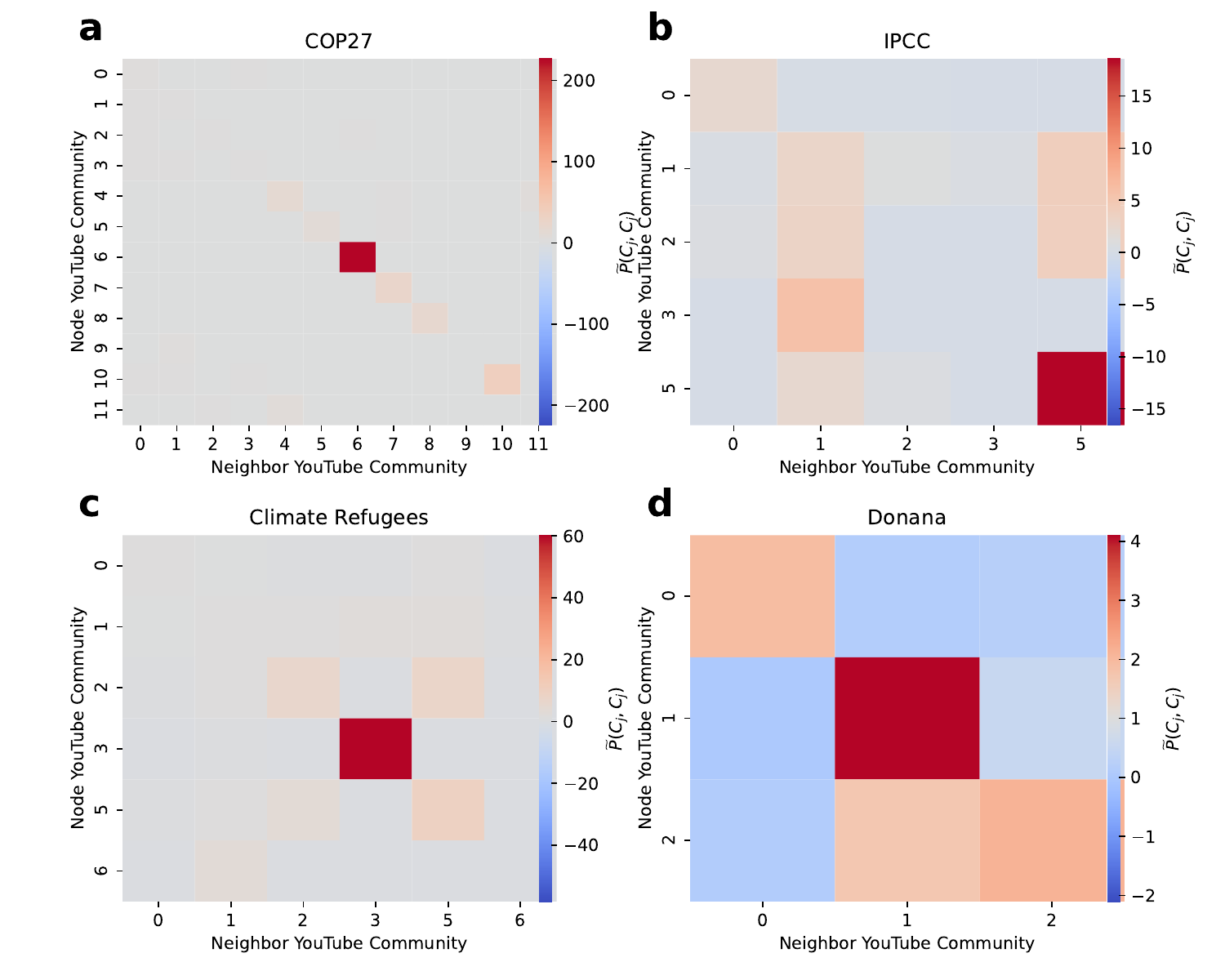}
  \end{center}
  \caption{\textbf{YouTube Community correlations in the Twitter network}. Probability of finding a user than posted a video in a YouTube community $C_i$ next to a user that posted a video in a YouTube community $C_j$. The results are normalized by 1000 random reshufflings of YouTube communities. The results correspond to \textbf{(a)} COP27, \textbf{(b)} IPCC, \textbf{(c)} Climate Refugees, and \textbf{(d)} Do\~{n}ana networks.} \label{community_correlations}
\end{figure}

We have computed the probability $P(C_i,C_j)$ of finding a user that posted a video in YouTube community $C_i$ connected to a user that posted a video in YouTube community $C_j$. To account for network effects, we have calculated $\widetilde{P}(C_i,C_j)=P(C_i,C_j)/P^{\rm null}(C_i,C_j)$  normalizing the observed probabilities by the corresponding null values over 100 random reshufflings of YouTube communities (Fig. \ref{community_correlations})

\newpage

\begin{table*}[ht]
  \centering
  \resizebox{0.65\textwidth}{!}{
    \begin{tabular}{lrlrrr}
\toprule
network & community & channel name & viewCount & subscriberCount & videoCount \\
\midrule
cop27 & 0 & ABC News & 14083290454 & 15900000 & 86239 \\
cop27 & 0 & TEDx Talks & 7712122864 & 40100000 & 210266 \\
cop27 & 0 & WION & 4701543170 & 8470000 & 136298 \\
cop27 & 0 & Al Jazeera English & 4158638169 & 12800000 & 111285 \\
cop27 & 1 & ABC News & 14083290454 & 15900000 & 86239 \\
cop27 & 1 & BBC News Hindi & 7893209075 & 17800000 & 24225 \\
cop27 & 1 & BBC News & 5099185254 & 15600000 & 20649 \\
cop27 & 1 & Hindustan Times & 4842818648 & 6530000 & 58180 \\
cop27 & 2 & WION & 4701543170 & 8470000 & 136298 \\
cop27 & 2 & Sky News Australia & 4140464784 & 3900000 & 137500 \\
cop27 & 2 & CNN-News18 & 2967286345 & 6660000 & 160062 \\
cop27 & 2 & euronews & 1368026874 & 2320000 & 99119 \\
cop27 & 3 & Democracy Now! & 672210420 & 1980000 & 15630 \\
cop27 & 3 & BreakThrough News & 138428175 & 609000 & 1403 \\
cop27 & 3 & CNBC International TV & 116672225 & 362000 & 23849 \\
cop27 & 3 & Earthling Ed & 54396530 & 452000 & 333 \\
cop27 & 4 & UOL & 3147321219 & 3880000 & 62664 \\
cop27 & 4 & CNN Brasil & 2660548853 & 4780000 & 79898 \\
cop27 & 4 & Poder360 & 441840465 & 1300000 & 19396 \\
cop27 & 4 & United Nations & 406648137 & 2870000 & 16539 \\
cop27 & 5 & DW Español & 2280313778 & 4150000 & 39964 \\
cop27 & 5 & EL PAÍS & 1525037481 & 2550000 & 32904 \\
cop27 & 5 & RTVE Noticias & 1378481206 & 2020000 & 25701 \\
cop27 & 5 & Presidencia de la República - Colombia & 92079186 & 318000 & 20699 \\
donana & 0 & Negocios TV & 961486669 & 1010000 & 51192 \\
donana & 0 & COPE & 428257253 & 680000 & 6994 \\
donana & 0 & LibertadDigital & 332062887 & 684000 & 14666 \\
donana & 0 & DISTRITOTV & 172849322 & 331000 & 17512 \\
donana & 1 & Canal Red & 40410031 & 309000 & 700 \\
donana & 1 & Pandemia Digital & 21969182 & 77300 & 1701 \\
donana & 2 & EL PAÍS & 1525037481 & 2550000 & 32904 \\
donana & 2 & RTVE Noticias & 1378481206 & 2020000 & 25701 \\
donana & 2 & elDiarioes & 172800438 & 320000 & 9763 \\
donana & 2 & Hoy por Hoy & 37690183 & 71900 & 2893 \\
ipcc & 0 & ABC News & 14083290454 & 15900000 & 86239 \\
ipcc & 0 & MSNBC & 12050180260 & 6440000 & 77964 \\
ipcc & 0 & BBC News & 5099185254 & 15600000 & 20649 \\
ipcc & 0 & Vox & 3427290494 & 11900000 & 1677 \\
ipcc & 1 & Fox News & 16126544384 & 11000000 & 101812 \\
ipcc & 1 & NBC News & 6952022416 & 9650000 & 61006 \\
ipcc & 1 & Sky News Australia & 4140464784 & 3900000 & 137500 \\
ipcc & 1 & PragerU & 1898459165 & 3210000 & 3188 \\
ipcc & 2 & Sky News Australia & 4140464784 & 3900000 & 137500 \\
ipcc & 2 & Sky News & 4139448980 & 7370000 & 44104 \\
ipcc & 2 & GBNews & 1199821224 & 1170000 & 58097 \\
ipcc & 2 & Rebel News & 716906956 & 1640000 & 21112 \\
ipcc & 3 & Fox News & 16126544384 & 11000000 & 101812 \\
ipcc & 3 & The Hill & 1462188609 & 1900000 & 30762 \\
ipcc & 3 & KTLA 5 & 1023133916 & 1020000 & 22849 \\
ipcc & 3 & WUSA9 & 337193118 & 432000 & 52569 \\
ipcc & 4 & WDR aktuell & 83165245 & 103000 & 1139 \\
ipcc & 4 & Carsten Jahn - TEAM HEIMAT & 76120963 & 179000 & 879 \\
ipcc & 4 & Stuttgarter Zeitung & 60081195 & 60000 & 3424 \\
ipcc & 4 & eingeSCHENKt.tv & 19415430 & 131000 & 544 \\
ipcc & 5 & De Nieuwe Wereld & 26340396 & 101000 & 1797 \\
ipcc & 5 & 1000frolly PhD & 10920557 & 30100 & 170 \\
ipcc & 5 & Holland Gold & 2880125 & 19800 & 181 \\
ipcc & 5 & Tom Nelson & 1587268 & 12400 & 219 \\
refugees & 0 & Fox News & 16126544384 & 11000000 & 101812 \\
refugees & 0 & MSNBC & 12050180260 & 6440000 & 77964 \\
refugees & 0 & TEDx Talks & 7712122864 & 40100000 & 210266 \\
refugees & 0 & NBC News & 6952022416 & 9650000 & 61006 \\
refugees & 1 & CNN & 15327838440 & 16000000 & 163405 \\
refugees & 1 & ABC News & 14083290454 & 15900000 & 86239 \\
refugees & 1 & NBC News & 6952022416 & 9650000 & 61006 \\
refugees & 1 & BBC News & 5099185254 & 15600000 & 20649 \\
refugees & 2 & Fox News & 16126544384 & 11000000 & 101812 \\
refugees & 2 & MSNBC & 12050180260 & 6440000 & 77964 \\
refugees & 2 & NBC News & 6952022416 & 9650000 & 61006 \\
refugees & 2 & Sky News Australia & 4140464784 & 3900000 & 137500 \\
refugees & 3 & AllatRa TV English & 2840110 & 29500 & 964 \\
refugees & 3 & Creative Society & 2664803 & 40600 & 1163 \\
refugees & 3 & ALLATRA TV International & 1507255 & 18500 & 982 \\
refugees & 4 & VICE News & 3156884335 & 8830000 & 6480 \\
refugees & 4 & Global News & 2490061489 & 4110000 & 37952 \\
refugees & 4 & DW News & 2241829454 & 5010000 & 33064 \\
refugees & 4 & CBC News & 1955874636 & 3450000 & 28005 \\
refugees & 5 & TEDx Talks & 7712122864 & 40100000 & 210266 \\
refugees & 5 & NBC News & 6952022416 & 9650000 & 61006 \\
refugees & 5 & Democracy Now! & 672210420 & 1980000 & 15630 \\
\bottomrule
\end{tabular}

  }
  \caption{Top 4 most relevant YouTube channels by view count in the 5 largest communities.}
  \label{youtube_channels}
\end{table*}

\clearpage

\end{document}